\documentclass[aps,pra,superscriptaddress,twocolumn,floatfix]{revtex4-1}
\usepackage[utf8]{inputenc}
\usepackage[T1]{fontenc}
\usepackage{amsmath}
\usepackage{amssymb}
\usepackage{mathtools}
\usepackage{graphicx}
\usepackage{overpic}
\usepackage[usenames,dvipsnames]{color}
\usepackage{bm}
\usepackage[colorlinks,urlcolor=blue,citecolor=blue,linkcolor=blue]{hyperref}
\usepackage{url}
\usepackage{subfigure}
\usepackage{slashed,bbm}
\usepackage{dsfont}
\usepackage{setspace}
\usepackage{soul}
\usepackage{braket}
\usepackage{sidecap}
\usepackage{hyperref}
\usepackage{multirow}
\usepackage{tensor}
\usepackage{ulem}

\begin{document}

\title{Understanding the three-dimensional quantum Hall effect in generic multi-Weyl semimetals}

\author{Feng Xiong}
\author{Carsten Honerkamp}
\affiliation{Institute for Theory of Statistical Physics, RWTH Aachen University, and JARA Fundamentals of Future Information Technology, 52062 Aachen, Germany}
\author{Dante M.~Kennes}
\affiliation{Institute for Theoretical Statistical Physics, RWTH Aachen University, and JARA Fundamentals of Future Information Technology, 52062 Aachen, Germany}
 \affiliation{Max Planck Institute for the Structure and Dynamics of Matter, Center for Free Electron Laser Science, 22761 Hamburg, Germany}
\author{Tanay Nag}
\affiliation{Institute for Theoretical Statistical Physics, RWTH Aachen University, and JARA Fundamentals of Future Information Technology, 52062 Aachen, Germany}

\date{\today}

\begin{abstract}

The quantum Hall effect in three-dimensional Weyl semimetal (WSM) receives significant attention for the emergence of the Fermi loop where the underlying two-dimensional Hall conductivity, namely, sheet Hall conductivity, shows quantized plateaus. In tilt multi-Weyl semimetals (mWSMs) lattice models, we systematically study Landau levels (LLs) and magneto-Hall conductivity both under the parallel and perpendicular magnetic field (referenced to the Weyl node's separation), i.e., ${\bm B}\parallel z$ and ${\bm B}\parallel x$, to explore the impact of tilting and non-linearity in the dispersion. We make use of two (single) node low-energy models to qualitatively explain the emergence of mid-gap chiral (linear crossing of chiral) LLs on the lattice for
${\bm B}\parallel z$ (${\bm B}\parallel x$). Remarkably, we find that the sheet Hall conductivity becomes quantized for ${\bm B}\parallel z$ even when two Weyl nodes project onto a single Fermi point in two opposite surfaces, forming a Fermi loop with $k_z$ as the good quantum number. On the other hand, the Fermi loop, connecting two distinct Fermi points on two opposite surfaces, with $k_x$ being the good quantum number, causes the quantization in sheet Hall conductivity for ${\bm B}\parallel x$. The quantization is almost lost (perfectly remained) in the type-II phase for ${\bm B}\parallel x$ (${\bm B}\parallel z$). Interestingly, the jump profiles between the adjacent quantized plateaus change with the topological charge for both cases. The momentum-integrated three-dimensional Hall conductivity is not quantized; however, it bears the signature of chiral LLs resulting in the linear dependence on $\mu$ for small $\mu$. The linear zone (its slope) reduces (increases) as the tilt (topological charge) of the underlying WSM increases.     

\end{abstract}

\maketitle
\section{INTRODUCTION}

With the discovery of the quantum Hall effect (QHE) \cite{Klitzing80}, the new realm of topological phases of quantum matters emerges as the central theme of research in condensed matter physics for their symmetry protected edge states \cite{Thouless82}.  
In the presence of a strong magnetic field, two-dimensional (2D) electron gases, for instance, 2D massless Dirac fermions in graphene and topological surface states, exhibit Landau levels (LLs) that further result in quantized plateaus in the integer QHE \cite{zhang2005experimental,xu2014observation}. This effect had been theoretically generalized to a three-dimensional (3D) system quite some time ago \cite{Halperin_1987, Kohmoto92}, while the experimental realization took place only recently \cite{tang2019three,Galeski2021}. Interestingly, the extra dimension along the magnetic field direction prevents the quantization of the Hall conductance in a 3D electron gas. 

Weyl semimetals (WSMs) appear suitable candidates to study 3D topological states of matter where the band structure becomes a 3D analog of graphene. However, either time-reversal symmetry or inversion symmetry, or both of these symmetries are broken in WSMs \cite{Wan11}. The WSMs host pairs of monopoles and anti-monopoles of the Berry curvature in momentum space \cite{Burkov11}, referred to as Weyl nodes (WNs) of opposite chirality. Interestingly, the  chiral-anomaly induced negative magnetoresistance \cite{Aji12,Zyuzin12,Huang15}, quantum anomalous Hall effect \cite{Yang11,Xu11}, non-local transport \cite{Parameswaran14}, and the plasmon mode \cite{Zhou15} directly reflect the topological nature of WSMs \cite{Zyuzin12, Son13}. The WSMs ideally harbor a conical spectrum with a point-like Fermi surface at the WN. This class is usually denoted as type-I WSMs. A large tilt in the WNs introduces the Lifshitz transition where the Fermi surface is no longer point-like, giving rise to the class of type-II WSMs \cite{YXu15,yan2017topological,soluyanov2015type} where the density of states at the WNs become finite. The type-I and type-II phases of WSMs have been experimentally realized in several inversion asymmetric compounds such as TaAs, MoTe$_2$, and WTe$_2$ \cite{lv2015observation,xu2015discovery}.


Interestingly it has been shown that the topological charge $m$ can be generically greater than unity in multi-WSMs (mWSMs) \cite{Xu11,Fang12,Liu17}, as compared to the conventional WSMs with $m=1$, resembling the multilayer analogs of graphene \cite{McCann06,Min08}. The non-linear anisotropic dispersion of mWSMs is extensively analyzed in low-energy as well as lattice model Hamiltonians \cite{Ahn17,Roy17}. However, the experimental discovery of mWSMs is yet to be made. Now, turning our attention to the intriguing linear responses, the electric-, thermal-, magneto-transport properties have been theoretically studied for single WSM \cite{Lundgren14,Sharma16,Spivak16,Zyuzin17,Nandy17,Nandy19} as well as mWSMs
 \cite{Chen16,Park17,Gorbar17,dantas2018magnetotransport,Nag_2020a,Das21,nag2022distinct} following 
semi-classical Boltzmann transport formalism. Meanwhile, Kubo theory is employed to study the optical responses~\cite{zyuzin2016intrinsic,Ferreiros17,Mukherjee17,Tabert16,Menon18,Nag20,Menon_2020}. On the other hand, WSMs further provide fertile ground to investigate various non-linear transport phenomena \cite{Sadhukhan21,Sadhukhan21b,Zeng21}. Apart from the theoretical studies, the topological transport properties are thoroughly investigated in several experiments with materials like, ZrTe$_5$ and TaAs \cite{Avery12,li2016chiral,Huang15,Liang17,hirschberger2016chiral,Watzman18}.

The high-field magneto-conductivity recently has acquired massive attention due to its underlying LL characteristics \cite{H-ZLu15}. Notably,  
the Fermi arcs at opposite surfaces, connected by "wormhole" tunneling through the WNs, can form a complete Fermi loop supporting the QHE \cite{Li16,Wang17,Li20}. There exist various quantum transport signatures such as field selective anomaly \cite{Udagawa16}, the magneto-thermoelectric response \cite{Ma21}, excitonic phase \cite{Chang21a}, as well as thickness-dependent magneto-conductivity \cite{Chang21b} that are associated with the LLs.
The Landau quantization in Cd$_3$As$_2$ \cite{jeon2014landau}, and TlBiSSe \cite{Novak15} has been observed experimentally using scanning tunneling microscopy. The 3D QHE has been experimentally realized in Cd$_3$As$_2$ \cite{uchida2017quantum,Schumann18,zhang2019quantum}. The QHE has been studied theoretically in the presence of interaction as well   
\cite{Qin20,RuiChen21,Peng-Lu21}. 


While much has been investigated using Boltzmann transport based on the low-energy models, we focus on studying the magneto-conductivity in the quantum limit considering a generic lattice model. Notice that the LLs have been merely extensively studied in the lattice Hamiltonian of single WSMs \cite{Udagawa16}. 
This further motivates us to contemplate the generic tilted double and triple WSMs. On the other hand, the formation of a Fermi loop via "wormhole" tunneling in the presence of the magnetic field, being perpendicular to the WN's separation, has been studied in low-energy models \cite{Wang17}. The lattice effect of such Fermi loops remains unexplored. 
We, therefore, combine the above aspects with exploring the following questions:
What are the effects of an anisotropic non-linear dispersion on the LLs for the cases with parallel and perpendicular magnetic fields? How do the Fermi loops appear and result in the quantized 2D sheet Hall conductivities?  
What are the effects of tilt in the Weyl spectrum for the above cases? What are the consequences of higher topological charges in the Hall conductivities? 
Therefore, our study is directed towards the understanding of the 3D QHE, associated with the "wormhole" tunneling,
by investigating the chiral LLs mediated magneto-transport properties
while its 2D analog has been substantially analyzed before. Our study is experimentally 
relevant in predicting the accurate response as we consider the lattice models free from any cut-off problems encountered in continuum models. Our investigations can thus become instrumental in exploring the connection of the Fermi arc surface states with the 3D QHE.


In this work, considering the generic tilted lattice model of mWSMs, we investigate the formation of LLs and quantizations in magneto-Hall conductivity when the magnetic field is parallel (${\mathbf B}=B_z \hat{z}$) and perpendicular (${\mathbf B}=B_x \hat{x}$) to the WN's separation. 
We find that for ${\bm B}||z$ (${\bm B}||x$), there exist the chiral (linear crossings of counter-propagating) LLs traversing through the WNs within the bulk gap while the number of chiral channels and their chiralities is proportional to the magnitude and sign of the topological charges of the underlying WN, respectively (see Figs.~\ref{fig:LLBz} and \ref{fig:LLBx}). These numerical findings can be explained by taking into account a two (single) node low-energy model for ${\bm B}||z$ (${\bm B}||x$). 
We show that the 2D sheet Hall conductivities, emerging from the Fermi loop construction with a good quantum number, i.e., momentum mode $k_z$ ($k_x$), can yield a quantized response for ${\bm B}||z$ (${\bm B}||x$) while their staircase-like behavior is directly connected to the filling of 
$k_x$ ($k_z$)-independent flat LLs (see Figs.~\ref{fig:hallxykz} and \ref{fig:hallyzkx}). We find that the staircase profile is maximally destroyed for the over-tilted type-II phase with ${\bm B}||x$. This is in contrast to the ${\bm B}||z$ case where the staircase nature remains preserved. On the other hand, from the behavior of the sheet Hall conductivities at small chemical potentials $\mu\to 0$, one can identify the underlying topological charge of the mWSMs. The momentum integrated Hall conductivities in 3D are found to exhibit linear $\mu$-dependence around $\mu\to 0$ indicating the crucial role of the mid-gap chiral LLs (See Figs.~\ref{fig:hallxy} and \ref{fig:hallyz}). We analytically provide a plausible explanation for this observation. The width of this $\mu$-linear zone decreases with increasing tilt while its slope gets steeper with increasing the topological charge.


This paper is organized as follows. We first discuss the generic lattice and low-energy models for mWSM in Sec.~\ref{models}. Then, we analytically compute the 
LLs in continuum models and compare them with the numerical results obtained from lattice models for ${\bm B}|| z$ and ${\bm B}|| x$ in Sec.~\ref{LL}. We next discuss the magneto-Hall conductivity in Sec.~\ref{MC} where we investigate the quantized and non-quantized structures of 2D and 3D Hall conductivities, respectively, under the magnetic field ${\bm B}|| z$ and ${\bm B}|| x$. We discuss our findings with the relevant literature in the field of 3D QHE in Sec.~\ref{literature}.
Finally we conclude in Sec.~\ref{Summary} with possible experimental connections and future directions.

\section{MODELS}
\label{models}
We consider the generic two band model of the form 
$H_{m}({\bm k})={\bm N}_m({\bm k}){\bm \cdot} {\bm \sigma}$ with ${\bm N}_m({\bm k})=(N^m_x({\bm k}),N^m_y({\bm k}),N^m_z({\bm k}))$ and ${\bm \sigma}=(\sigma_x,\sigma_y,\sigma_z)$ to describe the lattice Hamiltonian of mWSM for the topological charge $m$, where ${\bm \sigma}$ represents the pseudo-spin degrees of freedom. The individual terms are as follows for single WSM \cite{McCormick17,Roy17,Nag20} 
\begin{equation}\label{hamil1}
{\bm N}_1({\bm k})=
\begin{cases}
  N^1_x({\bm k})= t \sin k_{x} \\   
  N^1_y({\bm k})= t \sin k_{y}\\ 
  N^1_z({\bm k})= t_{z}\cos k_{z}-m_{z} + 2-\cos k_{x}-\cos k_{y}\,,
\end{cases}
\end{equation}
double WSM \cite{Fang12,yang2014classification,Roy17,Nag20}
\begin{eqnarray} \label{hamil2}
{\bm N}_2({\bm k})=
\begin{cases}
  N^2_x({\bm k})= t( \cos k_{x}-\cos k_{y} ) \\      
  N^2_y({\bm k})= t\sin k_{x}\sin k_{y}\\
  N^2_z({\bm k})= t_{z}\cos k_{z}-m_{z}+6+\cos2k_{x}+\cos2k_{y} \\
                 -4\cos k_{x}-4\cos k_{y}\,,
\end{cases}
\end{eqnarray}
and triple WSM \cite{Fang12,yang2014classification,Roy17,Nag20}
\begin{eqnarray}\label{hamil3}
{\bm N}_3({\bm k})=
\begin{cases}
  N^3_x({\bm k})= t\sin k_{x}(-2-\cos k_{x}+3\cos k_{y}) \\      
  N^3_y({\bm k})= -t\sin k_{y}(-2-\cos k_{y}+3\cos k_{x})\\
   N^3_z({\bm k})= t_{z}\cos k_{z}-m_{z}+6+\cos2k_{x}+\cos2k_{y} \\
-4\cos k_{x}-4\cos k_{y}\,.
\end{cases}
\end{eqnarray}
The above mWSM lattice Hamiltonians breaks time reversal symmetry ${\mathcal T}={\mathcal K}$ with ${\mathcal K}$ being the complex conjugation: ${\mathcal T} H_{m}(\bm k) {\mathcal T}^{-1}\ne H_{m}(-\bm k)$. We consider a universal tilt term for all the above cases $m=1,2$, and $3$ as given by $N^m_0({\bm k})=t_{0}\cos k_z$: $\mathcal{H}_{m}(\bm k)=H_{m}(\bm k)+N^m_0({\bm k}) I$. The energy eigenvalues of $\mathcal{H}_{m}({\bm k})$ are found to be $E_{m}({\bm k})=N^m_0({\bm k})\pm \sqrt{\sum_{l=x,y,z}(N^m_l({\bm k}))^2}$.
The parameters $t$ and $t_z$ denote respectively the hopping strengths between different and same pseudo-spin degrees of freedom and $m_z$ is the onsite mass term. Besides the tilt parameter $t_{0}$ tilts the energy dispersion along $k_z$ axis. For $t_{0}<t_z$ ($t_{0}>t_z$), it corresponds to type-I (type-II) WSM. Without loss of generality, parameters in the above Hamiltonians are set as $t_z=t=1$ and $m_z=0$ to locate the two WNs of opposite chiralities at $\boldsymbol{k}^{\pm}_{p}=(0,0,\pm\frac{\pi}{2})$ when solving $E_{m}({\bm k})=0$.

The low-energy effective Hamiltonian, expanding the lattice Hamiltonian of topological charge $m$ around a given WN at ${\bm k}_p^{+}$, can be written as~\cite{Roy17,Nag20} 
\begin{eqnarray}
\tilde{\mathcal{H}}_{m} \left( {\bm k} \right) &=& \alpha_{m} k^m_{\bot} \left[ \cos \left( m \phi_{k} \right) 
\sigma_{x} +\sin \left( m \phi_{k} \right) \sigma_{y} \right] + v k_z \sigma_{z}  \nonumber \\
&+& t_0 k_z\,,
\label{eq_multi1}
\end{eqnarray}
where $k_{\bot}=\sqrt{k_x^2+k_y^2}$ and $\phi_k={\rm arctan}(k_y/k_x)$. One can clearly notice the non-linear anisotropic dispersion $\tilde{E}_{m}({\bm k})=t_0 k_z \pm \sqrt{\alpha_m^2 k^{2m}_{\bot} + v^2k_z^2 }$ in mWSM as compared to the single WSM $\tilde{E}_{1}({\bm k})=t_0 k_z \pm v\sqrt{k_x^2 + k_y^2+k_z^2}$. The double (triple) WSM exhibits quadratic (cubic) dispersion along $k_{x,y}$ while keeping linear along $k_z$. The topological charge $m$ is encoded in the Berry curvature, associated with the Bloch Hamiltonian $\mathcal{H}_{m}({\bm k})$, as defined by
\begin{equation}
\Omega^{m}_{l,a} ({\bm k})= (-1)^l \frac{1}{4|{\bm N}_{{\bm k}}|^3} \epsilon_{a b c} {\bm N}_{{\bm k}} 
\cdot \left( \frac{\partial {\bm N}_{{\bm k}}}{\partial k_b} \times \frac{\partial {\bm N}_{{\bm k}}}{\partial k_c} \right)\,,
\label{bc_lattice}
\end{equation}
where $l$ denotes the band index and $a,b,c=x,y,z$. The Chern number referred to as the topological charge in this context, measures the Berry flux enclosed by the closed surface over the Brillouin zone (BZ) as given by 
\begin{equation}
\textcolor{black}{
\mathcal{C}^m_l = \dfrac{1}{2\pi}\int_{BZ} \bm{\Omega}^m_{l} (\bm{k}). d^2{\bm k}}\,.
\label{eq:Chern}
\end{equation} 
From above, one can find that $\mathcal{C}^m_{\pm}=\pm m $ \cite{dantas2018magnetotransport} for the valence ($-$) and conduction band ($+$) with the Berry curvature for the low-energy model ${\Omega}^m_{\pm}({\bm k}) =\pm  {m v \alpha_m^2 k^{2m-2}_{\bot} }( k_x, k_y, m k_z)/[2(\alpha_m^2 k^{2m}_{\bot} + v^2k_z^2)^{3/2}]$.

\begin{figure}[ht]
\begin{center}
	\includegraphics[width=1.0\linewidth]{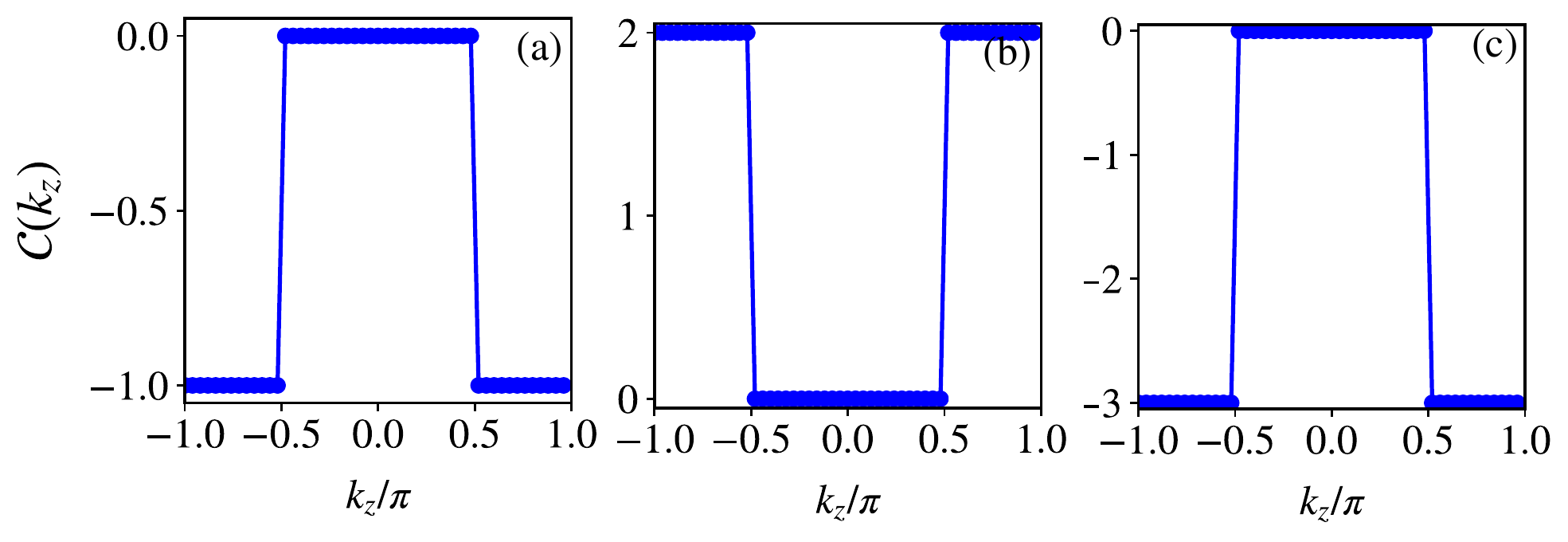}	
\end{center}
	\vspace{-0.4cm}
	\caption{Profile of 2D Chern numbers, calculated using Eq.~(\ref{eq:Chern}), as a function of $k_z$ for single (a), double (b) and triple (c) WSMs following Eqs.~(\ref{hamil1}), (\ref{hamil2}), (\ref{hamil3}), respectively. We consider $t_z=t=1$ and $m_z=0$ throughout the paper.
	} \label{fig:Chernnumber}
\end{figure}

Having discussed the notion of topological charge in the low-energy model, we now analyze the Fermi arc surface states from the lattice Hamiltonian. The WSM encompasses Fermi arc surface states connecting the projection of two WNs in the $k_y$-$k_z$ ($k_x$-$k_z$) plane with open boundary condition along $x$ ($y$)-direction. The 3D WSM conceives 2D Chern insulator plates, lying over $xy$-plane, between two WNs at $k_z=\pm \pi/2$ while the remaining region in $k_z$ consists of trivial insulator plates. Therefore, 
the 3D WSM can be regarded as stacking 2D Chern insulator layers in the direction of WNs' separation. In the present case, this can be further motivated by the fact that $\mathcal{ H}_1(k_x,k_y,k_z=\pi)$ becomes time-reversal symmetry broken quantum anomalous Hall insulator hosting
a topologically protected one-dimensional gapless chiral edge state~\cite{Slager17}. The number of Fermi arcs, interestingly, is directly given by the topological charge of mWSM~\cite{Dantas20}. Consequently, the 2D planes in between the two WNs have a Chern number given by the topological charge that is evident from Fig.~\ref{fig:Chernnumber}. To be more precise, we compute the Chern number $\mathcal{C}^m_{-}$ for the occupied valence band, following Eq.~(\ref{eq:Chern}) in $k_{x}$-$k_{y}$ plane, as a function of $k_z$ to show the underlying orientation of quantum anomalous Hall plates in the BZ. The Fermi arcs for single, double and triple WSMs Eqs.~(\ref{hamil1}), (\ref{hamil2}), and (\ref{hamil3}), are connected across the BZ between ${\bm k}_p^{\pm}$ as explicitly shown in Figs.~\ref{fig:Chernnumber} (a), (b) and (c), respectively.

\section{Landau Levels}\label{LL}

\subsection{Low energy model}\label{LowEnergyLL}
We shall now investigate the formation of LLs in the low-energy Hamiltonian. In order to capture the physics from both WNs, one can expand the Bloch Hamiltonian $\mathcal{H}_m({\bm k})$ around ${\bm \Gamma}=(0,0,0)$-point of which the low-energy Hamiltonian takes the following form 
\begin{align}
\mathcal{H}'_{m}({\bm k})&= t_{0}(1-k^2_z)\sigma_{0}+ (1-k^2)\sigma_{z}  +k^{m}_{-}\sigma_{+} +k^{m}_{+}\sigma_{-}  \,,
\label{lowHamil}
\end{align} 
where $k_{\pm }=k_x\pm ik_{y}$ and $\sigma_{\pm}=(\sigma_{x} \pm i\sigma_{y})/2$. Notice that in the above low-energy model, the WNs appears at $\boldsymbol{k}^{\pm}_{p}=(0,0,\pm 1)$ and the energy spectrum is $E'_{m}({\bm k})=(1-k^2_z) \pm \sqrt{ k_{+}^{m} k_{-}^{m} + (1-k^2)^2 }$. 

We first study the effect of a perpendicular magnetic field ${\bm B}=(0,0,B)$ in the $z$-direction that is along the separation between the WNs. For $\mu \ll \omega_c$, i.e., $\sqrt{B}\gg \mu$, the sharp LLs are formed in the strong field limit. Here, $\mu$ denotes the chemical potential and  $\omega_c=v/l_B$ refers to the cyclotron frequency where $v$, and $l_B=\frac{1}{\sqrt{eB}}$ represent, respectively, velocity and magnetic length. 
Note that thermal fluctuation, measured by the inverse of relaxation time $\tau$,  is less than the quantum fluctuation such that $\omega_c  \tau \gg 1$. By using the minimal coupling theory, the momentum ${\bm k}$ is replaced by ${\bm \Pi}= {\bm k}+e{\bm A}$ with ${\bm A}$ being the vector potential. We choose Landau gauge ${\bm A}=(-By,0,0)$ and introduce the ladder operators consistently such that $[a,a^\dagger]=1$ with $a=\frac{l_B}{\sqrt{2}}(\Pi_x -i\Pi_y)$ and $a^\dagger=\frac{l_B}{\sqrt{2}}(\Pi_x +i\Pi_y)$ for which ${\bm \Pi}=(k_x-eBy,k_y,k_z)$. Adopting the natural unit we set $e=1$. 
The low-energy Hamiltonian (Eq.~(\ref{lowHamil})) thus reduces to
\begin{align}
\mathcal{H}_{m}(k_z,a,a^\dagger)&=\begin{bmatrix}
f_{+}-\frac{2a^{\dagger}a+1}{l^2_B} & \bigg(\frac{\sqrt{2}a}{l_B}\bigg)^m \\\bigg
(\frac{\sqrt{2}a^{\dagger}}{l_B}\bigg)^m & f_{-}+\frac{2a^{\dagger}a+1}{l^2_B}
\end{bmatrix}\,,
\label{lowHamilLL}
\end{align} 
where $f_{\pm}=(t_0 \pm 1)(1-k^2_z)$. We solve the secular equation in following basis 
\begin{align}\mathcal{H}_{m}(k_z,a,a^\dagger)|\Psi^n_m\rangle=
E'_{m}(n,k_z) |\Psi^n_m\rangle  \,,
\end{align}
with the spinor part of $|\Psi^n_m\rangle$ as 
$[\chi_1|n-m\big>,~ \chi_2|n\big>]^T$ for $n\geq m$. One can solve the eigenenergies of the LLs
\begin{align}
E'_{m,\pm}(n,k_z)=\epsilon_0\pm  \sqrt{\epsilon_0^2-4\epsilon_1}\,, 
\label{analytical_LLBz}
\end{align}
where $\epsilon_0=f_{+}+ f_{-}+2m/l^2_B$ and $\epsilon_1=
f_{+} f_{-} + [(2n+1) f_{+} -(2n-2m+1)f_{-}]/l^2_B - (2n+1)(2n-2m+1)/l^4_B - \gamma^2 (2/l^2_B)^m$, and $\gamma=\sqrt{n(n-1)...(n-m+1)}$. Here, $+ $ and $-$ signs in Eq.~(\ref{analytical_LLBz}) correspond to the conduction band $n>0$ and valence band $n<0$, respectively, for the $n$-th LL. We denote $|n|$ by $n$ for valence band throughout.
The corresponding normalization factors are $\chi_{1,\pm}=(1+((E_{m,\pm}-a)/b)^2)^{-1/2}$ and $\chi_{2,\pm}=(1+((E_{m,\pm}-a)/b)^2)^{-1/2} (E_{m,\pm}-a)/b$
with $a=f_+ - 2(n-m)/l^2_B-1/l^2_B$ and $b=(\sqrt{2}/l_B)^m \gamma$. Notice that for the monolayer, bilayer, and trilayer graphene, the energies of the LLs are found proportional to $\sqrt{Bn}$, $B\sqrt{n(n-1)}$, and  $B^{3/2}\sqrt{n(n-1)(n-2)}$, respectively \cite{McCann06,yin2017landau}. The similar feature is also observed for the mWSM as visible from the last term $\gamma^2 (2/l^2_B)^m$ in $\epsilon_1$. 
To be more precise, the LL energies for single, double and triple WSMs with topological charge $m=1$, $2$ and $3$ are proportional to $\sqrt{Bn + f_1(B,k_z,n,m)}$, $\sqrt{B^2 n(n-1) + f_2(B,k_z,n,m)}$ and $\sqrt{B^3 n(n-1)(n-2) +f_3(B,k_z,n,m)}$, respectively.

It can be easily understood that LLs are independent of $k_x$ referring to their degenerate structure in $k_x$. The chiral LL for a single WSM is given by the zeroth eigenstate $|\Psi^0_1\rangle=
[0,~ |0\rangle]^T$ with energy $E'_{1,+}(0,k_z)=f_-+B$. For double WSM, zeroth and first eigenstates $|\Psi^0_2\rangle=
[0,~ |0\rangle]^T$ and $|\Psi^1_2\rangle=
[\chi_1 |0\rangle,~ \chi_2 |1\rangle]^T$ have the energies $E'_{2,+}(0,k_z)=f_-+B$ and $E'_{2,+}(1,k_z)=f_{-}+3B$, respectively. For triple WSM, zeroth, first and second eigenstates $|\Psi^0_3\rangle=
[0,~ |0\rangle]^T$, $|\Psi^1_3\rangle=
[\chi'_1 |0\rangle,~ \chi'_2 |1\rangle]^T$ and $|\Psi^2_3\rangle=
[\chi''_1 |1\rangle,~ \chi''_2 |2\rangle]^T$ have the energies $E'_{3,+}(0,k_z)=f_-+B$, $E'_{3,+}(1,k_z)=f_{-}+3B$ and $E'_{3,+}(2,k_z)=f_{-}+5B$, respectively. The normalization factors $\chi$'s can be computed thoroughly considering the above energies. One can find another set of energy solution for these LLs such as $E'_{3,-}(0,k_z)=0$, $E'_{3,-}(1,k_z)=f_{+}-B$ and $E'_{3,-}(2,k_z)=f_{+}-3B$, that we do not consider in order to maintain the notion of chirality. Importantly, the magnetic field, effectively coupled to the $\sigma_z$-term, leads to the non-degenerate chiral LLs as demonstrated above.

Having discussed the chiral structure and their associated spinor part, we now focus on the localization of these LLs as coming from their spatial part. To start with, one can consider $k_y=-i\partial_y$ while writing the low-energy Hamiltonian (Eq.~(\ref{lowHamilLL})) as follows 
\begin{align}
\mathcal{H}_{m}(k_z,\partial_y)&=\begin{bmatrix}
f_{+}-  \frac{1}{l^2_B}\bigg (1+ Z \bigg)   & \bigg(\frac{1}{l_B}(\eta +\frac{\partial}{\partial\eta})  \bigg)^m \\\bigg
( \frac{1}{l_B}(\eta -\frac{\partial}{\partial\eta}) \bigg)^m & f_{-}+\frac{1}{l^2_B}\bigg (1+Z\bigg)
\end{bmatrix}\,,
\label{lowHamilLL2}
\end{align} 
with $\eta= -y/l_B + l_B k_x= (y_0-y)/l_B$, $y_0=k_x l^2_B$. 
Considering the fact that 
$Z=(\eta^2 -\frac{\partial^2}{\partial\eta^2})$ demonstrates the harmonic oscillator,   
the eigenfunctions can be found to be 
\begin{align}
|\Psi^n_m(\eta) \rangle= \frac{e^{ik_x x +ik_z z}}{\sqrt{L_x L_z}}
\begin{bmatrix}
\chi_1 \phi_{n-m}(\eta)| n-m \rangle \\
\chi_2 \phi_{n}(\eta)| n \rangle
\end{bmatrix}\,,
\end{align}
where $\Phi_n(\eta)=\exp(-|\eta|^2/2) H_n(\eta) /\sqrt{2^n n! l_B\sqrt{\pi} }$ and $ H_n(\eta)$ represents the Hermite polynomial. The wave-functions become plane waves along $x$- and $z$-direction while localized around $y_0$ in the $y$-direction. $L_x L_z$ denotes the cross-section of the sample in the 2D plane where the electrons execute free particle motion. The Landau degeneracy is estimated to be $n_x=B L_x L_y/2\pi$ such that the cyclotron center of the electrons always remains inside the sample $y_0=2\pi n_x/L_x B < L_y$. This degeneracy is also reflected in the energy of the LLs only dependent on $k_z$, not $k_x$. Therefore, one can find for an individual LL that there exist $n_x$ number of momentum modes having the same energy with a given value of $k_z$. We discuss these issues more elaborately while connecting with the numerical results, based on the lattice models.

We now analyze the perpendicular magnetic field case where ${\bm B}=(B,0,0)$ is perpendicular to the separation of WNs. In this case, we do not need to consider the low-energy model (Eq.~(\ref{lowHamil}))
that captures the physics of two WNs simultaneously. Instead, we continue with the low-energy model at a single WN as discussed in Eq.~(\ref{eq_multi1}) . For simplicity, we choose $\alpha_m=v=1$ and the low-energy model around a given WN thus takes the form 
\begin{align}
\mathcal{H}''_{m}({\bm k})&= t_{0}k_z\sigma_{0}+ k_z\sigma_{z}  +k^{m}_{-}\sigma_{+} +k^{m}_{+}\sigma_{-}  \,.
\label{lowHamil2}
\end{align} 
One can realise that the analytical solution for LLs in mWSMs becomes way more complex and hence we have to restrict ourselves to the single WSM case with $m=1$. By employing a unitary transformation $U=\exp(i\sigma_y \pi/4 )$, $\mathcal{H}''_{1}({\bm k})$ takes a simple form allowing to continue with the analytical calculations: $U\mathcal{H}''_{1}({\bm k})U^{-1}={\tilde {\mathcal H}}^{''}_{1}({\bm k})=t_0 k_z \sigma_0 -k_z \sigma_x+ k_y\sigma_y+ k_x \sigma_z $. 
With the vector potential ${\bm A}=(0,0,By)$, 
the ladder operators become $a=\frac{l_B}{\sqrt{2}}(\Pi_z -i\Pi_y)$, $a^\dagger=\frac{l_B}{\sqrt{2}}(\Pi_z +i\Pi_y)$, ${\bm \Pi}=(k_x,k_y,k_z+By)$. As a result, the low-energy Hamiltonian can be written as
\begin{align}
\tilde{\mathcal H}^{''}_{1}(k_x,a,a^\dagger)=\begin{bmatrix}
\alpha(a+a^{\dagger}) +k_x & \beta a^{\dagger} \\
\beta a & \alpha(a+a^{\dagger}) - k_x 
\end{bmatrix}\,,
\label{lowHamilLL_Bx}
\end{align} 
where $\alpha=t_0/ \sqrt{2}l_B$ and $\beta=-\sqrt{2}/l_B$. The above 
Hamiltonian is similar to the tilted Dirac cones in presence of a perpendicular magnetic field \cite{Islam17}. The identity term proportional to $\alpha$ is analogous to a pseudo in-plane effective electric field of strength $E_{\rm eff}=\alpha\sqrt{2}/l_B$. Hence, the low-energy model can be regarded as an analog of 
monolayer graphene under a crossed electric and magnetic field, except the $k_x \sigma_z$ term \cite{Lukose07}.

One can also transform the Hamiltonian into a moving frame along $z$-direction with velocity $v=E_{\rm eff}/B=t_0$ such that the transformed electric field
vanishes and the magnetic field reduces to $B'=B \sqrt{1-t^2_0}$ \cite{Lukose07}. Therefore, in the moving frame the LLs can be obtained as $\sqrt{2Bn}(1-t^2_0)^{1/4}$ when $k_x=0$. Since we have $k_x$ in low-energy Hamiltonian, the complete expression for the energy in the moving frame is given by ${\tilde E}''_{1,{\rm MF}}(n,k_x)=\pm(1-t^2_0)^{1/4} \sqrt{2Bn+k^2_x (1-t^2_0)^{-1/2}}$. The Lorentz back transformation of momentum yields the energy of the LLs in the rest frame 
${\tilde E}''_{1}(n,k_x)=\pm (1-t^2_0)^{3/4} \sqrt{2Bn +k^2_x (1-t^2_0)^{-1/2} }$ while the argument of the wave functions becomes $\eta= \frac{(1-t^2_0)^{1/4}}{l_B} (y-k_z l^2_B + \frac{\lambda\sqrt{2n} l_B t_0}{(1-t^2_0)^{1/4}})$. 
An alternative diagonalization technique can also be employed to derive the above expression \cite{peres2007algebraic}. The chiral LL appears to be ${\tilde E}''_{1}(0,k_x)=\pm (1-t^2_0)^{1/2} k_x$. The spinor parts for the LLs are similar to the earlier case as $[\chi_1|n-1\big>,~ \chi_2|n\big>]^T$ for $n\geq 1$ and $[0,~ |1\big>]^T$ for $n= 0$, respectively. 
One can notice that the energies of the LLs are independent of $k_z$ and hence there exist a number $n_z=B L_y L_z/ 2 \pi$ of degenerate $k_z$ modes for each LL with a given $k_x$. It is noteworthy that the LLs are dependent on the tilt. Therefore, a higher magnitude of tilt essentially destroys the chiral nature of the LLs. We numerically investigate the double and triple WSM, as discussed in Eq.~(\ref{lowHamil2}), in the presence of ${\bm B}=(B,0,0)$ where we will compare with the lattice results.

Notice that it is not physically permitted to have the mid-gap LLs with opposite chiralities for a single WN. The WNs of opposite topological charges host two separate chiral LLs with positive and negative slopes of $k_x$. On the other hand, two copies of bulk LLs for two WNs merge on top to give rise to the doubly degenerate bulk LL spectrum irrespective of momentum. We anticipate that for a non-linear dispersion, such degeneracy might not appear over the entire BZ. The structure of the mid-gap chiral LLs is also expected to be non-trivially modified for the higher topological charge. We investigate the LLs for the lattice Hamiltonian to extensively verify the above predictions and tendencies obtained from the low-energy analysis.

In short, some of the generic features of the LLs that non-linear dispersion would lead to $\sqrt{{\mathcal O}((Bn)^m)}$ dependence in the energies of $n$-th LL. The above is very clearly evident when ${\bm B}$ is applied along $z$-direction. The relative spacing between two consecutive bulk LLs decreases with increasing $n$ for a given value of topological charge $m$. 
This can be observed irrespective of the choice of the magnetic fields. We note that for a linearized single WSM without the tilt $H_1({\bm k})={\bm k}\cdot {\bm \sigma}$, the bulk LLs are given by $\pm\sqrt{2Bn+k^2_{i}}$ for magnetic field along $i$-direction referring to a particle-hole symmetric nature of LL spectrum \cite{Chang21b}. Once the tilt term preserves (breaks) the particle-hole symmetry, the bulk LL spectrum, associated with the tilted WSM, is expected to preserve (break) the particle-hole symmetry.  
Interestingly, for particle-hole symmetry preserving tilt that is also 
perpendicular to the WNs' separation, one can notice the imbalance in the number of chiral modes for the magnetic field along the tilt direction \cite{Udagawa16}. We do not encounter such a situation, as evident from Eq.~(\ref{analytical_LLBz}), in the present case with particle-hole symmetry breaking tilt parallel to WNs' separation. For a higher topological charge with non-linear dispersion, the tilt can lead to richer quantum phenomena that might be absent in untilted single WSM. The analytical treatment hints at the above for parallel ${\bm B}=(0,0, B)$.
On the other hand, for perpendicular ${\bm B}=(B,0,0)$, the non-linear dispersion in mWSMs hinders a reachable analytical solution, in contrast with the case of parallel ${\bm B}$.

\subsection{Lattice model}\label{LatticeLL}
Having discussed the generation of LLs in the low-energy model, we now illustrate the formalism to execute the LLs in the tight-binding lattice Hamiltonian. With the same choice of the Landau gauge previously discussed for parallel and perpendicular magnetic fields, in lattice space, the hopping between different sites in the Hamiltonian needs to be modified by the Peierls substitution $t_{\bf{i},\bf{j}}c_{\bf{i}}^{\dagger}c_{\bf{j}}\rightarrow t_{\bf{i},\bf{j}}e^{i\int_{\bf{j}}^{\bf{i}}\boldsymbol{A}(\boldsymbol{r})d\boldsymbol{r}}c_{\bf{i}}^{\dagger}c_{\bf{j}}$. As mentioned before $k_x$ and $k_z$ are good quantum numbers.
Precisely, for both cases ${\bm B}=(0,0,B)$ and ${\bm B}=(B,0,0)$, the real space Hamiltonian takes the compact form ${\mathcal H}_m(k_x,j_y,k_z)$ so that we can equally minimize the finite size effect along $y$-direction. 
Consisting of a finite number of layers along $y$-direction, we first continue with the single WSM (Eq.~(\ref{hamil1})), as follows
\begin{align}
&{\mathcal H}_1(k_x,j_y,k_z)  =\Big\{ C_{j_{y},\boldsymbol{k}_{xz}}^{\dagger}\big[\big( \cos(k_{z}-j_{y}B_{x})+2  \nonumber \\
&-\cos(k_{x}+j_{y}B_{z})\big)\sigma^{z} + \sin(k_{x}+j_{y}B_{z})\sigma^{x} \nonumber \\
& +(t_{0} \cos (k_z-j_y B_x)-\mu)\sigma^{0}\big]  \nonumber\\
& +\big[\frac{1}{2i}(C_{j_{y}-1,\boldsymbol{k}_{xz}}^{\dagger}-C_{j_{y}+1,\boldsymbol{k}_{xz}}^{\dagger})\sigma^{y} \nonumber\\
&-\frac{1}{2}(C_{j_{y}-1,\boldsymbol{k}_{xz}}^{\dagger}+C_{j_{y}+1,\boldsymbol{k}_{xz}}^{\dagger})\sigma^{z}\big]\Big\} C_{j_{y},\boldsymbol{k}_{xz}}\,.
\label{sWSM_lattice}
\end{align}
The Hamiltonian for double WSM (Eq.~(\ref{hamil2})) can be written as 
\begin{align} 
&{\mathcal H}_2(k_x,j_y,k_z)  =\Big\{ C_{j_{y},\boldsymbol{k}_{xz}}^{\dagger}\big[\big(t_{z}\cos(k_{z}-j_{y}B_{x})+6 \nonumber \\
& + \cos2(k_{x}+j_{y}B_{z})-4\cos(k_{x}+j_{y}B_{z})\big)\sigma^{z} \nonumber \\
& + \cos(k_{x}+j_{y}B_{z})\sigma^{x}+(t_{0} \cos (k_z-j_y B_x)-\mu)\sigma^{0}\big]  \nonumber \\
& + \big[-\frac{1}{2}(C_{j_{y}-1,\boldsymbol{k}_{xz}}^{\dagger}+C_{j_{y}+1,\boldsymbol{k}_{xz}}^{\dagger})\sigma^{x}  \nonumber \\
&+\frac{1}{2i}\sin(k_{x}+j_{y}B_{z})(C_{j_{y}-1,\boldsymbol{k}_{xz}}^{\dagger}-C_{j_{y}+1,\boldsymbol{k}_{xz}}^{\dagger})\sigma^{y}  \nonumber \\
 &-2(C_{j_{y}-1,\boldsymbol{k}_{xz}}^{\dagger}+C_{j_{y}+1,\boldsymbol{k}_{xz}}^{\dagger})\sigma^{z} \nonumber \\
 & + \frac{1}{2}(C_{j_{y}-2,\boldsymbol{k}_{xz}}^{\dagger}+C_{j_{y}+2,\boldsymbol{k}_{xz}}^{\dagger})\sigma^{z}\big]\Big\} C_{j_{y},\boldsymbol{k}_{xz}}\,.
 \label{dWSM_lattice}
\end{align}
Finally, the triple WSM (Eq.~(\ref{hamil3})) takes the form 
\begin{align}
&{\mathcal H}_3(k_x,j_y,k_z)  =\Big\{ C_{j_{y},\boldsymbol{k}_{xz}}^{\dagger}\big[\big(t_{z}\cos(k_{z}-j_{y}B_{x})+6  \nonumber \\
&+\cos2(k_{x}+j_{y}B_{z})-4\cos(k_{x}+j_{y}B_{z})\big)\sigma^{z}  \nonumber \\
&+\sin(k_{x}+j_{y}B_{z})(-2-\cos(k_{x}+j_{y}B_{z}))\sigma^{x}\nonumber \\
&+(t_{0} \cos (k_z-j_y B_x)-\mu)\sigma^{0}\big]  \nonumber \\
& +\big[\frac{3}{2}\sin(k_{x}+j_{y}B_{z})(C_{j_{y}-1,\boldsymbol{k}_{xz}}^{\dagger}+C_{j_{y}+1,\boldsymbol{k}_{xz}}^{\dagger})\sigma^{x}\nonumber \\
&-\frac{1}{2i}(-2+3\cos(k_{x}+j_{y}B_{z}))(C_{j_{y}-1,\boldsymbol{k}_{xz}}^{\dagger}-C_{j_{y}+1,\boldsymbol{k}_{xz}}^{\dagger})\sigma^{y} \nonumber \\
&-2(C_{j_{y}-1,\boldsymbol{k}_{xz}}^{\dagger}+C_{j_{y}+1,\boldsymbol{k}_{xz}}^{\dagger})\sigma^{z} \big] \nonumber \\
&+\big[ \frac{1}{4i}(C_{j_{y}-2,\boldsymbol{k}_{xz}}^{\dagger}-C_{j_{y}+2,\boldsymbol{k}_{xz}}^{\dagger})\sigma^{y} \nonumber \\
&+\frac{1}{2}(C_{j_{y}-2,\boldsymbol{k}_{xz}}^{\dagger}+C_{j_{y}+2,\boldsymbol{k}_{xz}}^{\dagger})\sigma^{z}\big]\Big\} C_{j_{y},\boldsymbol{k}_{xz}}\,.
\label{tWSM_lattice}
\end{align}
Therefore, we can cast all three Hamiltonians in $2L_y\times 2L_y$ matrices. For the numerical evaluation, we consider the sample size $L_{y}=100$ with a periodic boundary condition in the $y$-direction. To satisfy the $y$-direction periodicity, the magnetic field can only be chosen as $\frac{2\pi}{Q}$ with $Q$ commensurate with $L_{y}$ so that $Q=L_y/n$ reduces to an integer only. Otherwise explicitly mentioned, all the following numerical results are performed under the magnetic field of amplitude $B=\frac{2\pi}{L_y}$ \cite{Udagawa16,abdulla2021time}.

\subsubsection{$\bf{B}\parallel z$}\label{LLBz}

\begin{figure}[ht]
\begin{center}
\includegraphics[width=1.0\linewidth]{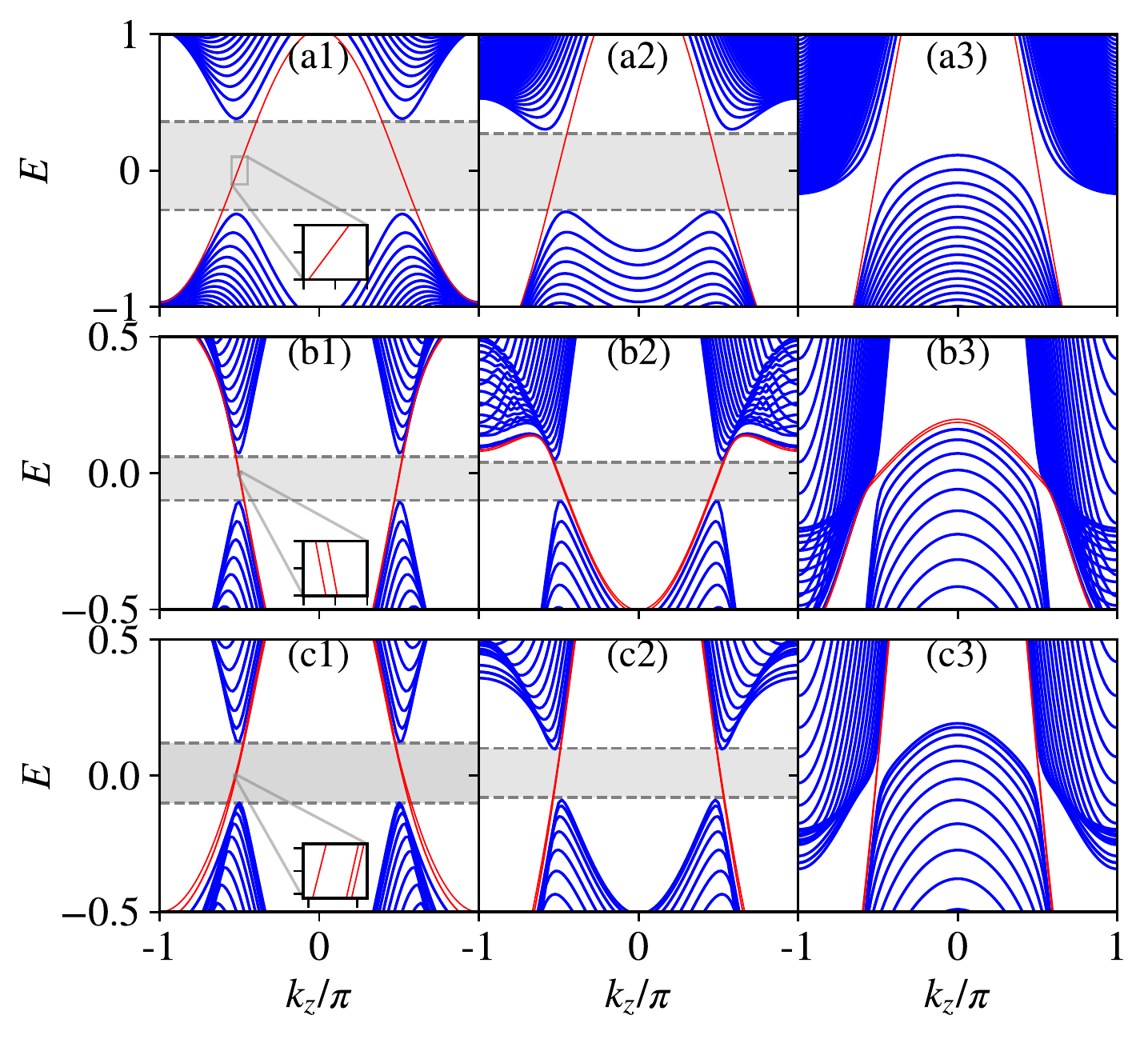}
\end{center}
	\vspace{-0.4cm}
	\caption{LL energies as a function of $k_z$ under ${\bm B} \parallel z$, shown for a single WSM with $m=1$ (Eq.~(\ref{sWSM_lattice})), a double WSM with $m=2$ (Eq.~(\ref{dWSM_lattice})), and a triple WSM with $m=3$ (Eq.~(\ref{tWSM_lattice})) in (a1, a2, a3), (b1, b2, b3), and (c1, c2, c3), respectively, keeping $k_x=0$ fixed. We consider $t_0=0$ for (a1, b1, c1), $t_0=0.5$ for (a2, b2, c2) and $t_0=1.2$ for (a3, b3, c3).  
	The mid-gap chiral LLs, traversing through the WNs at ${\bm k}^{\pm}_p=(0,0,\pm\pi/2)$, exist both in type-I and type-II phases as depicted in red color.
	These numerical results on the lattice are qualitatively consistent with analytical LL energies calculated using Eq.~(\ref{analytical_LLBz}). The insets depict the number of non-degenerate chiral Landau levels, proportional to the topological charge $m$, inside the bulk gap as designated by the grey shaded region. 
	} \label{fig:LLBz}
\end{figure}

We depict the evolution of LLs for a single WSM in Figs.~\ref{fig:LLBz} (a1, a2, a3) with three different values of the tilt parameter $t_0=0$, $0.5$ and $1.2$, denoting type-I untilted, type-I and type-II tilted WSM, respectively. The exact order of LLs for double and triple WSMs are shown in Figs.~\ref{fig:LLBz} (b1, b2, b3) and (c1, c2, c3), respectively.
By starting from the untilted case with $t_0=0$, the non-linear structure of the LLs as a function of $k_z$ is visible. A gap exists between the bulk LLs with $n=\pm 1$, $n=\pm 2$, and $n=\pm 3$ for single WSM, double WSM, and triple WSM, respectively. The size of this gap can be analytically calculated as $\Delta E'_{m,\pm}(n,k_z=\pm 1)=2 \sqrt{\epsilon_0^2-4\epsilon_1}$ from Eq.~(\ref{analytical_LLBz}) revealing the fact that the gap size depends on the magnetic field and topological charge. The non-degenerate chiral LLs are visible inside the gap (see the insets of Figs.~\ref{fig:LLBz} (a1, b1, c1)) as predicted by the analytical analysis in Sec.~\ref{LowEnergyLL}. The sign of the associated topological charge of a given WN determines the chirality of the mid-gap LLs traversing across the WNs at $k_z=\pm \pi/2$. Apart from the fact that the number of chiral modes is determined by the topological charge $m$, these modes are robust even under a larger tilt.    
The conservation of topological charge actively results in pairs of positive and negative chiral modes. However, the gap between the bulk LLs vanishes at $t_0=1$ when the semimetallic nature emerges. In the over tilted case $t_0 >1$, bulk LLs for positive (negative) values can appear below (above) zero energy.    

A close inspection of Fig.~\ref{fig:LLBz} suggests that the LLs for single and triple WSMs qualitatively follow $-E'_{m,+}(n,k_z)$ (Eq.~(\ref{analytical_LLBz})) for $m=1,~3$. The analytical solutions, given by $E'_{m,+}(n,k_z)$, thus correctly indicate the above profile except the sign.  
On the other hand, for double WSM, LLs obtained analytically with $E'_{2,+}(n,k_z)$ can describe the numerical findings. The apparent chirality reversal for the 
mid-gap chiral LLs for $m=1,3$ between analytical and numerical calculations might originate from the following lattice effect. The sign of the topological charge, associated with a given WN, changes for double WSM compared to that for the single and triple WSM. Based on the low-energy model, the analytical solution can not accurately capture the distribution of the topological charge of the WNs in the BZ. Another mismatch is that for $m=3$ lattice calculations, the chiral LLs are irregularly spaced (see the inset of Fig.~\ref{fig:LLBz} (c1)) in contrast to equally spaced analytically obtained for low-energy one.

\subsubsection{$\bf{B}\parallel x$}\label{LLBx}


 \begin{figure}[ht]
\begin{center}
	\includegraphics[width=1.0\linewidth]{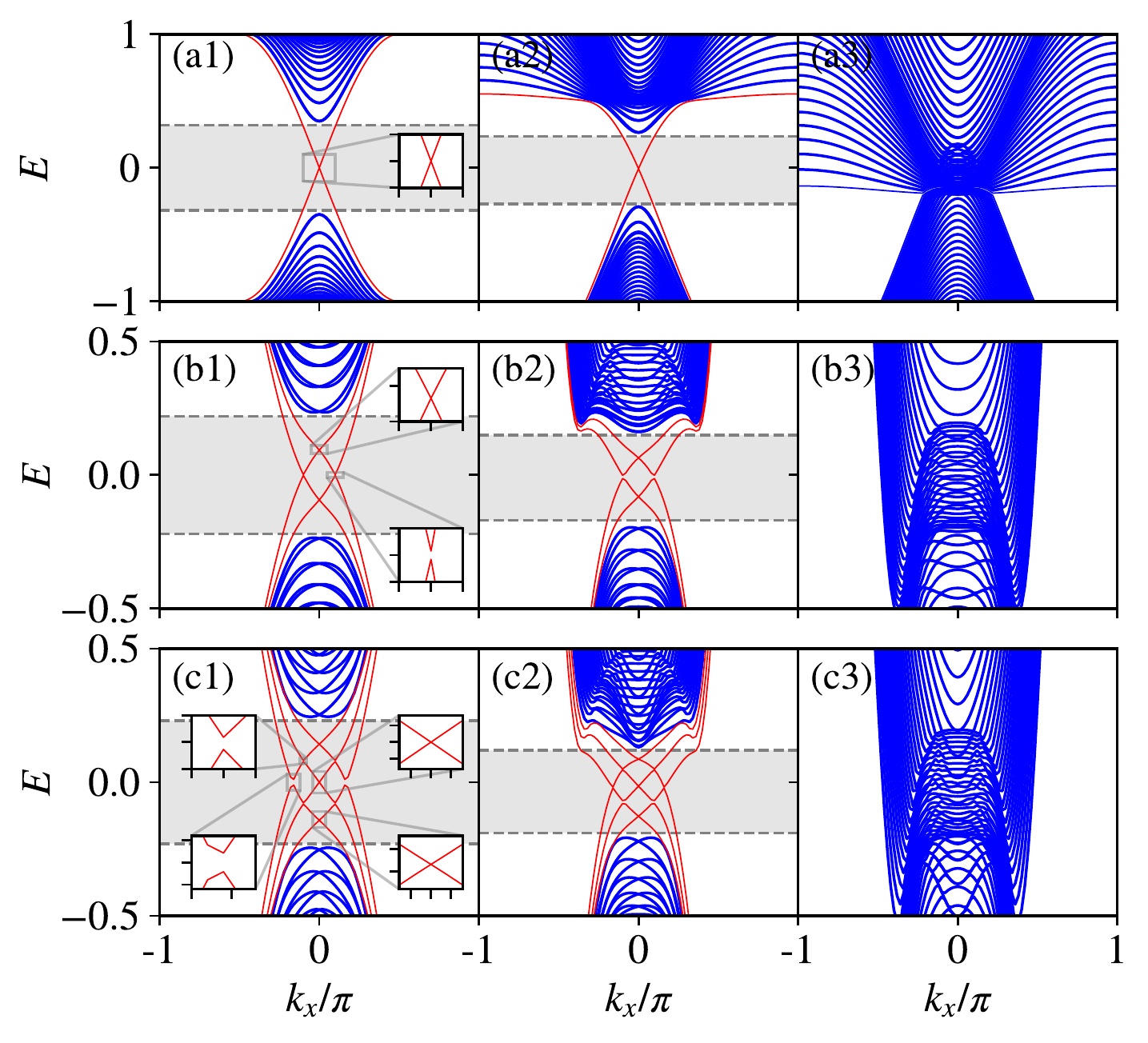}
\end{center}
	\vspace{-0.4cm}
	\caption{The variation of LL energies, computed from Eqs.~(\ref{sWSM_lattice}), (\ref{dWSM_lattice}) and (\ref{tWSM_lattice}),as a function of $k_x$ for ${\bm B} \parallel x$ while following the format given in Fig.~\ref{fig:LLBz}. 
    The mid-gap chiral LLs, depicted in red color, linearly cross each other at $k_x=0$ (displayed in the insets) only for the type-I phase, while for the type-II, these chiral modes become indistinguishable from the bulk LLs. The chiral LLs are always gapped except at $k_x$=0. The number of linear crossings associated with chiral LLs at $k_x=0$ can directly determine the topological charge of the underlying WSMs. Each of the bulk LLs is doubly degenerate for untilted single WSM. By contrast, for untilted double and triple WSMs, the bulk LLs only degenerate at $k_x=0$. Similar to Fig.~\ref{fig:LLBz}, the bulk gap is represented by the grey shaded region.
	} \label{fig:LLBx}
\end{figure}

Now, coming to a situation where the magnetic field is along $x$-direction, the LLs for single, double, and triple WSMs are plotted in Figs.~\ref{fig:LLBx} (a1, a2, a3), (b1, b2, b3)
and (c1, c2, c3), respectively as a function of $k_x$. For $t_0=0$, we find that the counter-propagating chiral LLs linearly cross each other at $k_x=0$ within the bulk gap as noticed for a single WSM. The bulk gap between $n=\pm 1$ can be estimated 
by the analytical expression $\Delta {\tilde E}''_{1}(1,0)=2(1-t^2_0)^{3/4} \sqrt{2B}$. Interestingly, both the WNs of opposite topological charges have their common projection at $k_x=0$, causing the crossing of positive and negative chiral LLs. One can find two and three such linear crossings, respectively, at $k_x=0$ for double and triple WSMs (see the insets of Figs.~\ref{fig:LLBx} (a1,b1,c1)). Small gaps exist at $k_x \neq 0$ where chiral LLs exhibit avoided level crossing. A higher topological charge WSM hosts richer microscopic variation of the mid-gap chiral LLs.

Different from the parallel magnetic field, the chirality of these LLs can only be meaningful at $k_x=0$ irrespective of the value of $m$. 
One can understand that the topological charge imprints its signature by the number of degenerate points at $k_x=0$. The number of chiral LLs passing through $k_x=0$ is twice the topological charge of the underlying WSMs. 
With increasing tilt, the gap of the avoided crossing increases within the bulk gap, and it can no longer preserve the particle-hole symmetry in the LL spectrum. For sufficiently large tilt in the type-II phase, the chiral structure of LLs at $k_x=0$ becomes wholly dissolved into the bulk. This is in stark contrast to the case $\bm{B}\parallel z$. Another crucial difference is that for untilted single WSM with perpendicular ${\bm B}$, the bulk LLs are doubly degenerate irrespective of momentum $k_x$ while lifting the degeneracy for the tilted one. For double and triple WSMs, non-linear anisotropic dispersion lifts the degeneracy everywhere except at $k_x=0$.

Bearing in mind that the analytical solution $E''_1(n,k_x)$ qualitatively explains the numerical findings, for $m=2$ and $3$, the numerical lattice results are consistent with numerical LL energies calculated in the low-energy model Eq.~(\ref{lowHamil2}) with appropriate gauge choice: $k_{\pm }=k_x \pm \frac{i}{\sqrt{2}l_B}(a+a^{\dagger})$ and $k_z=\frac{i}{\sqrt{2}l_B}(a-a^{\dagger})$. Unlike the parallel magnetic field, the linear crossings of chiral LLs, obtained from the lattice model are correctly captured by the low-energy model. The distribution of individual topological charges in BZ is not crucial for the perpendicular magnetic field. For completeness, we comment that the LL spectrum in the lattice model for Fig.~\ref{fig:LLBx} can also be qualitatively explained by the numerical result incorporating perpendicular ${\bm B}$ with the same gauge choice mentioned above in Eq.~(\ref{lowHamil}). However, it would be challenging to tackle the problem analytically even for a single WSM, and that is why we probe the perpendicular ${\bm B}$ case at least for the single WSM of the low-energy model in Eq.~(\ref{lowHamil2}).

\section{Magnetoconductivity}\label{MC}

We now focus on a highly relevant physical observable, namely the magneto-Hall conductivity, that has been extensively studied in 2D systems \cite{Peeters92,Charbonneau82,Vasilopoulos12,Islam_2018}. However, here we will investigate the 3D system and highlight the intriguing outcomes as compared to the 2D systems. The non-diagonal Hall conductivity with $i\neq j$, following the Kubo linear-response theory, is expressed as follows 
\begin{align}
\sigma_{ij}&=\frac{ie^{2}}{N}\sum_{k_x,k_z}\sum_{\alpha,\beta\ne\alpha}\frac{(f_{\alpha}-f_{\beta})}{\varepsilon_{\alpha}-\varepsilon_{\beta}}\frac{\big<\alpha|\mathcal{V}_{k_{i}}|\beta\big>\big<\beta|\mathcal{V}_{k_{j}}|\alpha\big>}{(\varepsilon_{\alpha}-\varepsilon_{\beta})+i\eta}   \nonumber\\
&=\frac{ie^{2}}{N}\sum_{k_x,k_z}\sum_{\alpha,\beta\ne\alpha}(f_{\alpha}-f_{\beta})\frac{\big<\alpha|\mathcal{V}_{k_{i}}|\beta\big>\big<\beta|\mathcal{V}_{k_{j}}|\alpha\big>}{(\varepsilon_{\alpha}-\varepsilon_{\beta})^{2}+\eta^{2}}\,,
\label{hall}
\end{align}
where $\varepsilon_{\alpha}$ is the eigenvalue, associated with the state $|\alpha\big>$ for the underlying Hamiltonian ${\mathcal H}_m(\bm k)$ and $\eta \to 0$ in a clean system. The velocity matrix is given by $\mathcal{V}_{k_{i}}=\frac{\partial {\mathcal H}_{m}(\bm k)}{\partial k_{i}}$. We compute ${\mathcal V}_{k_i}(k_x,j_y,k_z)$ by doing the partial derivative of the Hamiltonians ${\mathcal H}_{m}(k_x,j_y,k_z)$ (Eqs.~(\ref{hamil1}), (\ref{hamil2}), and (\ref{hamil3})); see Appendix \ref{LatticeVelocity} for more details. Here, $f_{\alpha}$ denotes the zero-temperature Fermi-Dirac distribution function and the overall normalization is given by $N= n_x n_z$. We are mainly interested in $\sigma_{xy}$ ($\sigma_{yz}$), that is $\sigma_{ij}$ with the $i=x(y)$ and $j=y(z)$ component for ${\bm B}$ along $z$ ($x$)-direction.

In this paper, we deal with 3D systems and, therefore, need to be careful with the summation of momentum modes to capture the essential physics. For the 2D problem, one can only encounter the summation over the good quantum number, i.e., momentum, resulting in the $l^{-2}_B$ factor in the normalization $N$. In the 3D case, the normalization incorporates a length scale in addition to the above factor. Usually, 
the normalization $N$ for the 3D case refers to the slab's volume, while for the 2D case, it represents the surfaces that host the Fermi arcs. Normally, $\sigma^{2D}_{ij}$ has the dimensionality $e^2/h$ and in 3D, $\sigma_{ij}$ becomes $e^2/h$ over length $\sigma_{ij}=\sigma^{2D}_{ij}/L$ \cite{Wang17}. 
To understand the behavior of Hall conductivity, we compute the 2D sheet Hall conductivity $\sigma^{2D}_{xy}(k_z)$ while summing the degenerate energy levels only over $k_x$ for ${\bm B}=(0,0, B)$. Henceforth, we will refer to the 2D sheet Hall conductivity as 2D Hall conductivity.
Similarly, for ${\bm B}=(B,0,0)$, we examine $\sigma^{2D}_{yz}(k_x)$ while summing the degenerate energy levels only over $k_z$. The analysis is motivated by the fact that the LL spectrum is independent of $k_x(k_z)$ for ${\bm B}$ along $z(x)$-direction. Therefore, the 3D Hall conductivity takes the form $\sigma_{ij}=\sum_{k_l} \sigma^{2D}_{ij}(k_l)/n_l$ with $i\neq j\neq l$ where $n_l$ has the length dimension along $l$-direction such that $k_l= 2\pi p/n_l $ ($p$ denotes integer number). To be precise, $n_x=BL_x L_y/2\pi$ and $n_z=L_z$ ($n_z=BL_z L_y/2\pi$ and $n_x=L_x$) for $\sigma_{xy}$ ($\sigma_{yz}$). 
The above discussion resembles Halperin's argument that for Fermi energy lying within this gap, the 3D Hall conductivity $\sigma_{ij}$ is given by $(e^2/h)\sum_k \epsilon_{ijk} G_k$ where ${\bm G}$ is reciprocal lattice vector of an internal potential \cite{Halperin_1987}.

In order to acquire an idea about the possible quantization in $\sigma^{2D}_{xy}(k_z)$ at the outset, one can continue with $\mathcal{V}_{k_{x,y}}=m k^{m-1}_{\bot} ( \cos \left( (m-1) \phi_{k} \right) \sigma_x \pm \sin \left( (m-1) \phi_{k} \right) \sigma_y )-2 k_{x,y} \sigma_z$ from low-energy Hamiltonian (Eq.~(\ref{eq_multi1})) and $|\alpha\big>=[|n-m\rangle~|n\rangle ]^T$. This results in $\big<\alpha|\mathcal{V}_{k_{x}}|\beta\big>\big<\beta|\mathcal{V}_{k_{y}}|\alpha\big>=m^2 k^{2m-2}_{\bot}(\delta_{n'-m,n} + \delta_{n',n-m} )$. The energy denominator in Eq.~(\ref{hall}) thus can be accordingly selected with $n'=n\pm 1$, $n'=n\pm 2$, and $n'=n \pm 3$ for single, double and triple WSM, respectively, for a given set of $(k_x,k_z)$. Following the above argument, the quantized plateaus are expected to show jumps by topological charge $m$ for 2D Hall conductivity. 
However, the argument is oversimplified compared with lattice models.

One more aspect that we would like to discuss is the density of states (DOS). It is intimately connected to the Hall conductivity, as we will analyze below. Since the LLs are discrete, the DOS can be expressed as the sum of a series of delta functions given by
\begin{align}
D(\mu)=\frac{1}{N}\sum_{\alpha} \delta(\mu- \varepsilon_\alpha)\,, 
\label{eq_dos}
\end{align}
where the normalization factor $N$ changes accordingly with the dimensionality of the underlying problem and the LL energy $\varepsilon_\alpha$ are obtained after diagonalizing ${\mathcal H}_{m}(k_x,j_y,k_z)$. In order to understand $\sigma^{2D}_{ij}(k_l)$, we compute partial density of state (PDOS) for a given momentum mode $k_l$, defined by $D_{2D}(\mu,k_l)$ with normalization $N=n_{p}$, $l\ne p$. The 3D DOS, expressed as  $D(\mu)=\sum_{k_l} D_{2D}(\mu,k_l)/n_l$ turns out to be relevant while analyzing the integrated response $\sigma_{ij}$. In the 3D case, one can find that $N=n_p n_l$ using Eq.~(\ref{eq_dos}). We refer to the PDOS as $D_{2D}(k_l)$ similar to $\sigma^{2D}_{ij}(k_l)$
for convenience. We numerically execute the $\delta$-function by a Lorentzian i.e., $\delta(\mu- \varepsilon_\alpha)=\eta/[(\mu- \varepsilon_\alpha)^2 +\eta^2]$ with $\eta$ being the broadening parameter to mimic disorder effects in experiment.

\subsection{$\bf{B} \parallel z$}

\begin{figure*}[ht]
\begin{center}
	\includegraphics[width=1\linewidth]{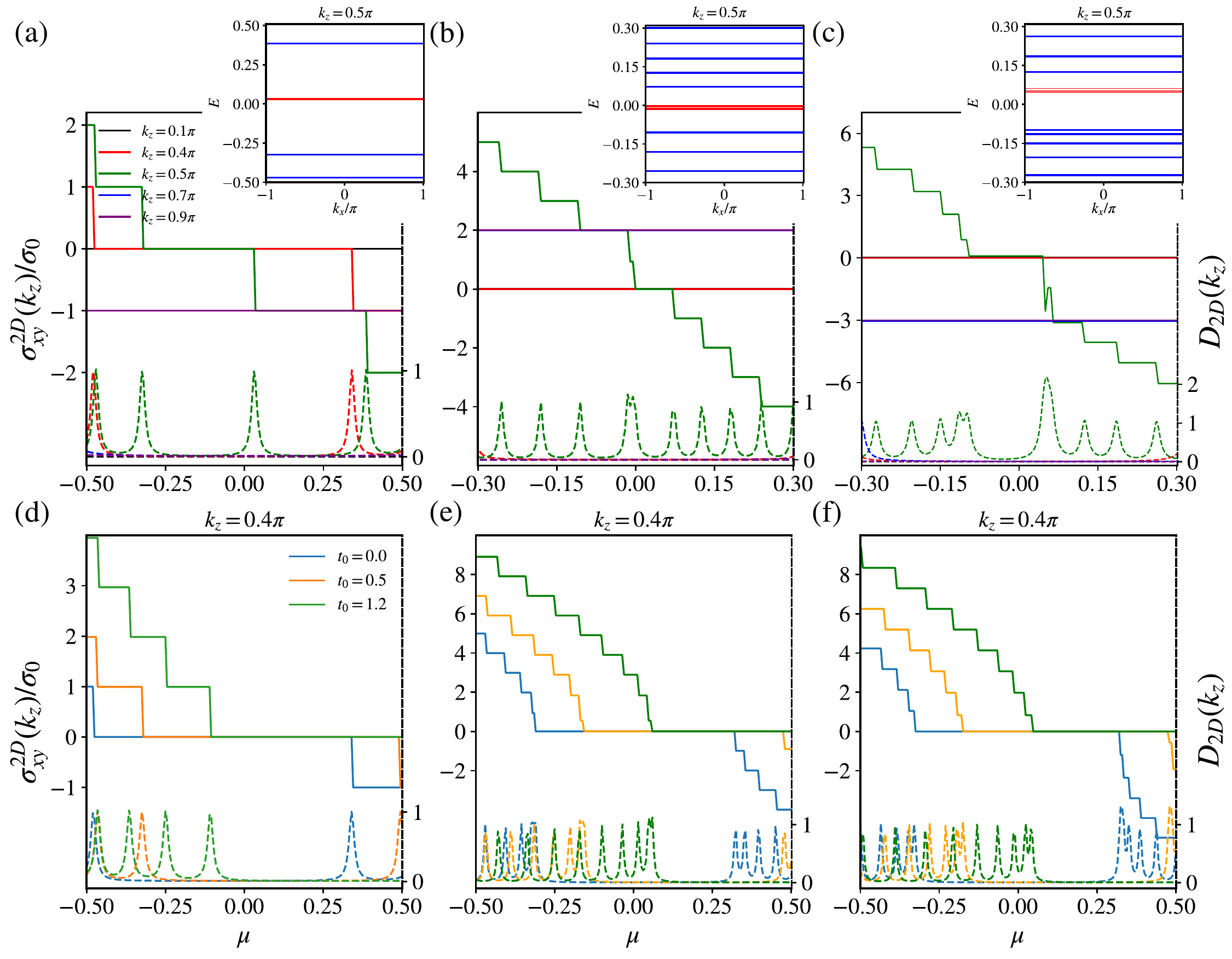}	
\end{center}
	\vspace{-0.4cm}
	\caption{Normalized 2D Hall conductivity $\sigma^{2D}_{xy}(k_z)/\sigma_0$ (shown in left axis), following Eq.~(\ref{hall}), and PDOS $D_{2D}(k_z)$ (shown in right axis), following Eq.~(\ref{eq_dos}), 
	as a function of $\mu$, keeping the tilt term fixed at $t_0=0$, for $m=1$ in (a), $m=2$ in (b), and $m=3$ in (c) when ${\bm B}\parallel z$.  
	The upper-right insets show the flat $k_x$-independent LLs at $k_z=\pi/2$.
	One can notice that $\sigma^{2D}_{xy}(k_z)$ jumps by unity
	to the next plateau at certain $\mu$ when there exists a flat LL with the energy given by $\mu$. The jump profiles for different values of $k_z$ are in accordance with the LL spectrum illustrated in Fig.~\ref{fig:LLBz}.  
	We repeat (a), (b) and (c) in (d), (e), and (f) with $\sigma^{2D}_{xy}(k_z=0.4\pi)$ for the
	tilt $t_0=0,~0.5,~1.2$, respectively. 
	The staircase-like structure continues to exist for type-II phases as well. We set $\eta=0.01$.  The 2D sheet Hall conductivity  is measured in the unit of $e^2/h$. We follow this convention throughout.
	} 
\label{fig:hallxykz}
\end{figure*}



In Fig. \ref{fig:hallxykz} (a), (b) and (c), we present $\sigma^{2D}_{xy}(k_z)$ for untilted single, double and triple WSMs, respectively with ${\bm B}=(0,0,B)$. Focusing on the untilted mWSM, the $\sigma^{2D}_{xy}(k_z)$s for single, double and triple WSM are quantized to $-1$, $2$ and $-3$ respectively when $|k_z|>\pi/2$ even though their corresponding PDOSs signal no density of electrons. This can be understood from the fact that the nontrivial 2D Chern insulator plates are stacked in the region for $|k_z|>\pi/2$ while constructing the 3D mWSM as shown in Fig.~\ref{fig:Chernnumber}(a)-(c) without external magnetic fields.
While for $|k_z|<\pi/2$, we remarkably notice the unit jumps, i.e., quantization changes by unity, in $\sigma^{2D}_{xy}(k_z)$ wherever $\mu$ crosses the $k_x$-independent one flat LL as shown in the upper-right insets. This is accurately captured by the peaks in the PDOS associated with the jumps in $\sigma^{2D}_{xy}(k_z)$.
This suggests that the staircase profile can also emerge for $|k_z|>\pi/2$ when $|\mu|$ is sufficiently large to pass through flat LLs lying far away from zero energy.
Note that we are restricting ourselves within the bulk gap where the chiral LLs are purely observed. We do not find degenerate chiral LLs, and as a result, we always find $\pm 1$ jumps in 2D Hall conductivity irrespective of the topological charge of the underlying WSMs. However, for triple WSM with $k_z=\pi/2$ in Fig.~\ref{fig:Chernnumber} (c), due to finer spacing gaps between the chiral LLs below the numerical resolution $\eta=0.01$, we observe the jumps are not perfectly quantized around $\mu\simeq 0.04$.

The flat LL picture is evident in the momentum zone $|k_z|<\pi/2$ where the 2D-layered insulator behaves trivially without the magnetic field. This is the reason that the 2D Hall conductivity $\sigma^{2D}_{xy}(|k_z|<\pi/2)$ vanishes for $m=1$, $2$ and $3$ as shown in Figs.~\ref{fig:hallxykz} (a), (b) and (c), respectively. On the other hand, the topological nature of Chern insulator plates, in the residual momentum zone $|k_z|>\pi/2$, remains unaltered with the magnetic field as long as flat LLs do not appear within the $\mu$ window of interest. We hence observe quantized plateau given by the topological charge $\sigma^{2D}_{xy}(k_z)=\mathcal{C}(k_z)$ for $|k_z|>\pi/2$. This global unity jump feature of 2D Hall conductivity is consistent with the non-degeneracy of LLs as obtained (understood)
from the lattice (low-energy) model.
Besides, the width of the plateau is determined by the gap size between two consecutive flat LLs. The plateau is maximally stretched for a single WSM as the non-linearity in the dispersion for higher charger mWSMs might reduce the relative gap between two consecutive LLs.

Now we turn our attention to the effect of tilt as shown in Figs.~\ref{fig:hallxykz} (d), (e), and (f) for single, double, and triple WSMs, respectively, with a fixed value of $k_z=0.4\pi$. With increasing tilt, more bulk LLs come within a given range of $\mu$ as the bulk gap reduces. This results in the reduction in the width of a plateau for higher tilt values. Moreover, due to the particle-hole asymmetry in the LL spectrum, the number of jumps above zero and below zero are not equal. Therefore, the underlying 2D conductivity can react to the varying tilt; however, the jump magnitude is already settled by the non-degenerate $k_x$ independent flat LLs within the concerned window of $\mu$. It is to be noted 
that the indirect nature of gap for LL spectrum with $t_0>1$ in Fig.~\ref{fig:LLBz} can preserve the staircase-like profile of 2D Hall conductivity.

\begin{figure}
\begin{center}
	\includegraphics[width=1\linewidth]{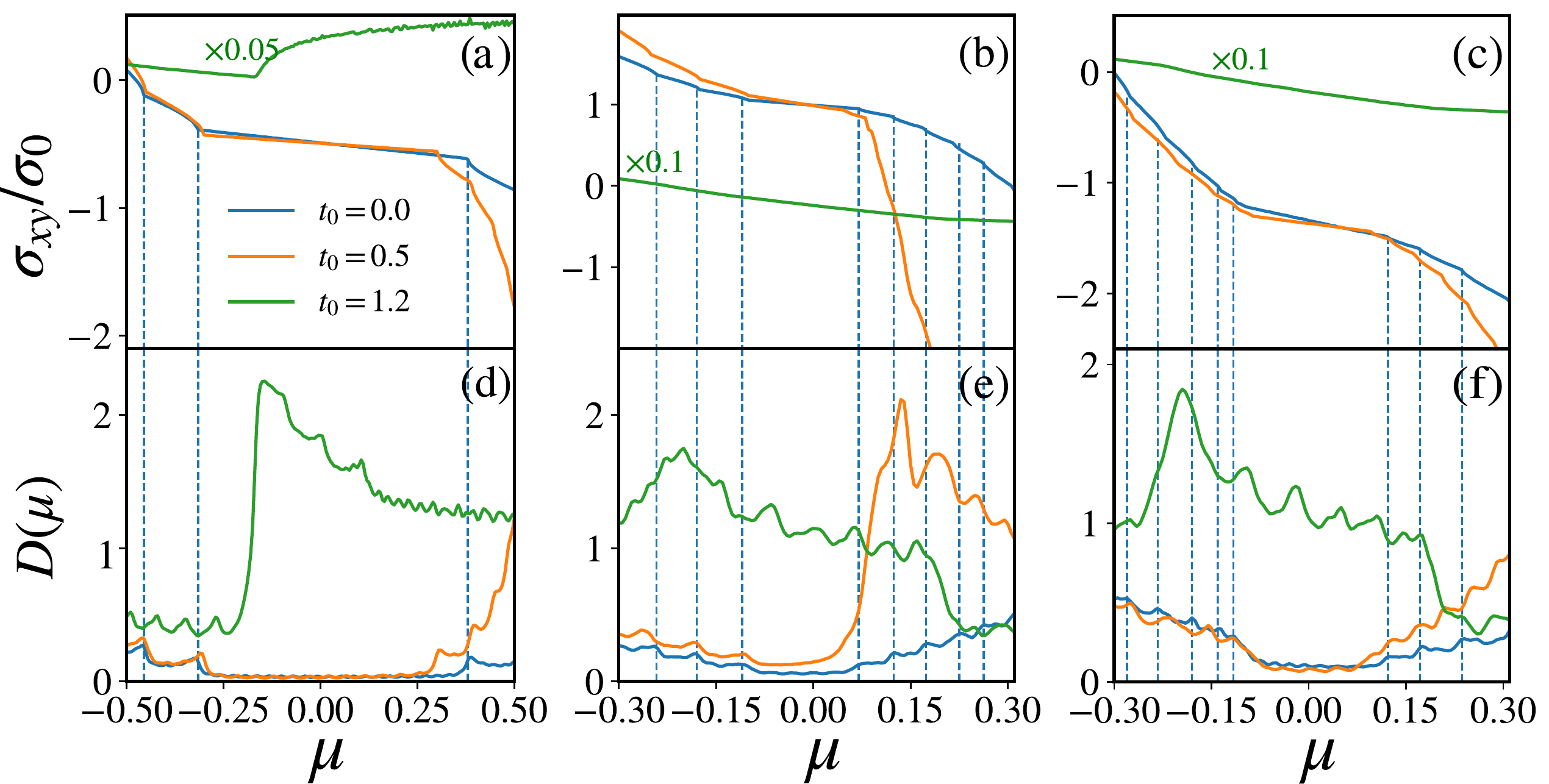} 
\end{center}
	\vspace{-0.4cm}
	\caption{Normalized 3D magneto-Hall conductivities $\sigma_{xy}/\sigma_0$, defined in Eq.~(\ref{hall}), for ${\bm B} \parallel z$ as a function of $\mu$ for $m=1$ in (a), $m=2$ in (b), and $m=3$ in (c) while varying $t_0=0,~0.5,~1.2$. The DOSs are computed from Eq.~(\ref{eq_dos}), in (d), (e) and (f) with the same parameter set as used for (a), (b) and (c). The flat regions in DOS, observed for type-I phase only, cause the $\mu$-linear regions in $\sigma_{xy}$ where the chiral LLs contribute maximally. We associate the peaks in the DOS with the change in the slope of $\sigma_{xy}$ for $t_0=0$ with the blue dashed lines to emphasize the connection between the above two quantities. The  3D Hall conductivity  is measured in the unit of $(e^2/h)/L$.  We follow this convention throughout.} 
\label{fig:hallxy}
\end{figure}

Having discussed the 2D structure of the quantized conductivity, we then investigate the 3D Hall conductivity $\sigma_{xy}=\sum_{k_z} \sigma^{2D}_{xy}(k_z)/n_z$ in Fig.~\ref{fig:hallxy}. The different Chern insulator plates, with quantized 2D conductivity along $z$-direction, would combine to yield the 3D Hall conductivity. Therefore, the Hall conductivity no longer exhibits quantized structure as observed for $\sigma^{2D}_{xy}(k_z)$. Let us first focus on the type-I single WSM. 
Interestingly, we find that $\sigma_{xy}$ varies linearly with $\mu$ when there exist the chiral LLs only inside the bulk gap. One can understand this behavior because there are $n_x$ degenerate LLs associated with each perpendicular momentum mode $k_z$. Therefore, it shows a continuous distribution of flat LLs in $k_x$ while $k_z$ is varied. After the summation over the perpendicular momentum $k_z$, the quantization is missing due to the interference among various $\sigma^{2D}_{xy}(k_z)$ profiles. Notice that the occupied bulk LLs below $\mu$ add up destructively to wash out the quantized signal even though $\mu$ stays inside the gap.   
When $\mu$ is varied outside the bulk gap, we find non-linear $\mu$ dependence with additional bulk LLs.

We can appreciate the 3D phenomena by investigating the structure of DOS with $\mu$ following the similar line of argument presented for the 2D case. We find that inside the bulk gap where only chiral LLs exist, the DOS demonstrates a non-zero flat structure.   The linear $\mu$ dependence of Hall conductivity gets destroyed as long as bulk LLs start contributing.   The slope of 3D Hall conductivity changes discontinuously when a peak exists in the DOS profile at a certain $\mu$. 
Now coming to the case of a higher topological charge, the width of the flat region in DOS decreases, and so does the linear area in the Hall conductivity, as depicted in Fig.~\ref{fig:hallxy}(b) and (c).
Interestingly, with increasing tilt, more bulk LLs come into the picture, and the contributions from the chiral LLs become insufficient to yield the linear behavior of Hall conductivity with $\mu$. It is worth mentioning that the slope of the $\mu$-linear regime increases with a larger topological charge. The responses of $\sigma_{xy}$ at $\mu=0$ for single, double, and triple WSM are approximately related: $|\sigma^{m=1}_{xy}| \approx |\frac{\sigma^{m=2}_{xy}}{2}|\approx|\frac{\sigma^{m=3}_{xy}}{3}|$, where the denominator matches with number of chiral LLs given by $m$.

The linear dependence on $\mu$ in $\sigma_{xy}$ for $\mu\to 0$ can plausibly be explained considering that LLs are observed only in high magnetic fields. To be precise, in order to experience the chiral LLs with $n<m$, one has to consider small carrier density such that $\sqrt{B}/\mu\gg1$. This essentially allows to cast the $f_{\alpha}-f_{\beta}$ in terms of Taylor series expansion around the energies of the LLs: $f_{\alpha}(\mu+E'_m(\alpha,k_z))-f_{\beta}(\mu+E'_m(\beta,k_z))\simeq f_{\alpha}(E'_m(\alpha,k_z))- f_{\beta}(E'_m(\beta,k_z)) + \mu (f'_{\alpha}-f'_{\beta})$ where $\alpha,~\beta<m$  refers to the chiral LLs within the bulk gap and $f'=\partial f(x)/\partial x$.  
Therefore, the relative occupancy factor $f_{\alpha}-f_{\beta}$ can yield the linear $\mu$ dependence for $\mu \to 0$ while such analysis is not accurate for $\mu$ far away from $0$. From this assumption, the leading order term in $f_{\alpha}-f_{\beta}$ is $\mu$ independent, which resembles the $\mu$ independent behavior of quantized Hall conductivity $\sigma^{2D}_{xy}(k_z)$ between two adjacent jumps. However, as discussed above, the summation over $k_z$ destroys the quantization leaving the linear $\mu$ behavior in $\sigma_{xy}$. We here comment that 3D anomalous Hall conductivity  for WSM in absence of any magnetic field is not expected to be quantized as $\sigma_{xy} \approx \mathcal{C}k_0$  where $k_0$ denotes the separation between two WNs \cite{Steiner17,Burkov14_PRL}.

\begin{figure}
\begin{center}
	\includegraphics[width=1.\linewidth]{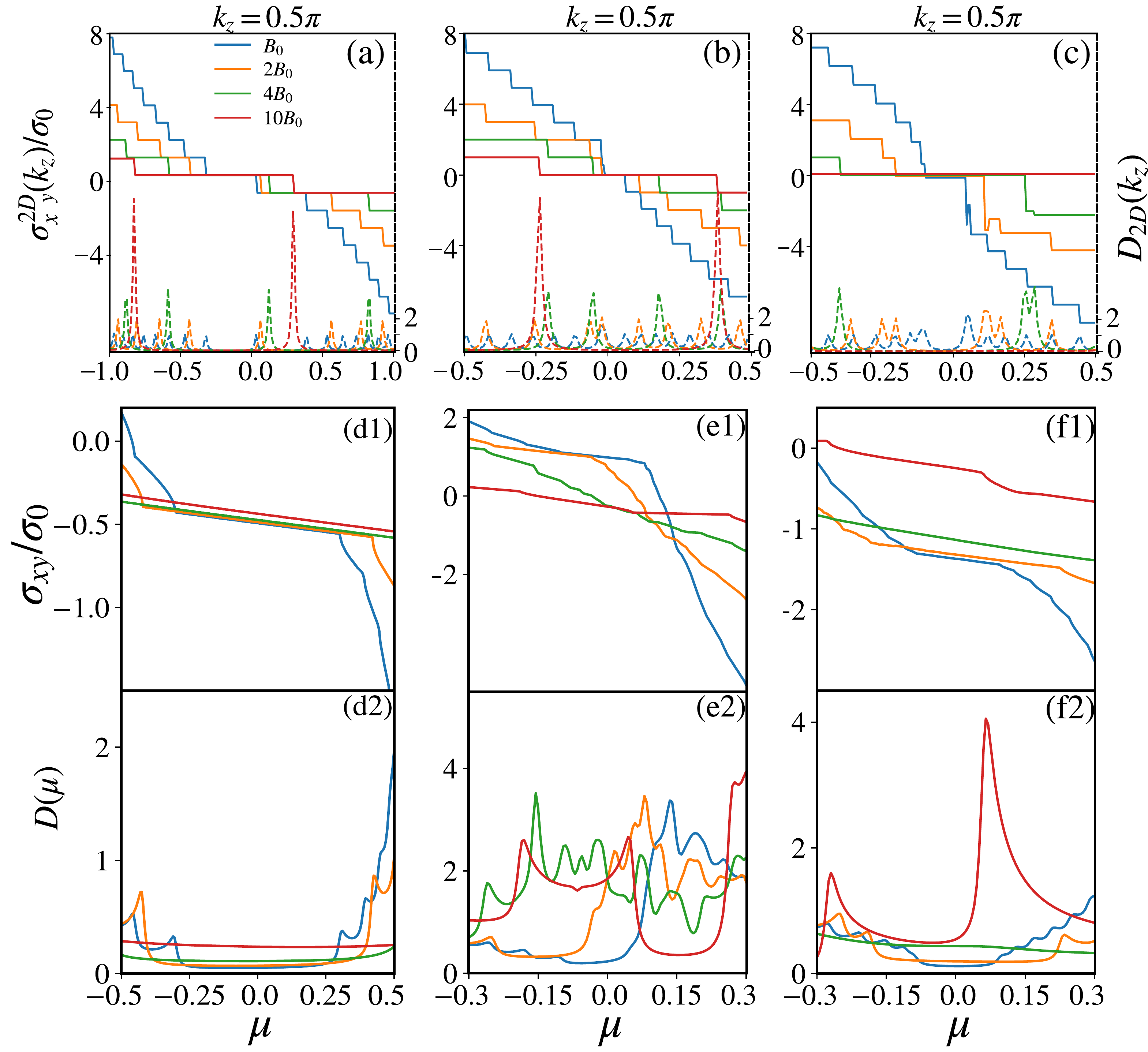} 
\end{center}
	\vspace{-0.4cm}
	\caption{Normalized 2D Hall conductivity
	$\sigma^{2D}_{xy}(k_z=\pi/2)/\sigma_0$ (shown in left axis), following  Eq.~(\ref{hall}), and PDOS $D_{2D}(k_z=\pi/2)$ (shown in right axis), using Eq.~(\ref{eq_dos}), as a function of $\mu$ for $m=1$ in (a), $m=2$ in (b), and $m=3$ in (c),  by varying the strength of the magnetic field $B=B_0$, $2B_0$, $4B_0$, and $10B_0$ keeping  $t_0=0.5$ fixed. We show (d1), (e1), and (f1), [(d2), (e2), and (f2)] for the 3D magneto-Hall conductivity $\sigma_{xy}$ [DOS $D(\mu)$] as a function of $\mu$ with the above set of parameters. The degeneracy of the LLs increases with increasing $B$. This is reflected in the increased width of the quantized plateau [$\mu$-linear region] shown in (a-c) [(d1-f1)].
	} 
\label{fig:hallxyBvary}
\end{figure}


We have investigated the effect of tilt under a constant magnetic field. We now focus on the response of 2D and 3D Hall conductivities concerning the variation of magnetic fields as shown in Figs.~\ref{fig:hallxyBvary}(a)-(c), and (d1)-(f1), respectively. The width of the quantized Hall plateau in $\sigma^{2D}_{xy}(k_z=\pi/2)$ increases with $B$. Because of the degeneracy of each LL linearly proportional to $B$, it takes a higher value of the magnetic field to fill up one LL before the electrons jump into the next empty one. 
The staircase-like structure is in complete agreement with the PDOS pattern. In the case of triple WSM, we find a jump with a higher magnitude possibly caused by the irregular spacing between the chiral LLs around $\mu\simeq 0$. 
The resulting 3D Hall conductivity $\sigma_{xy}$ after adding the contributions from all $k_z$ modes show prominent linear $\mu$ dependence when $\mu$ remains in the vicinity of chiral LLs. The change in slope can be well explained by the DOS structure as demonstrated in Figs.~\ref{fig:hallxyBvary} (d2)-(f2). It is noteworthy that the bulk gap in the LL spectrum $E_{m}(n,k_z)$ increases with the magnetic field. In DOS, the flat region, capturing chiral LLs further confirms this for single WSMs when $B$ increases. This picture qualitatively holds for mWSMs but quantitatively changes for a higher topological charge.

\subsection{$\bf{B}\parallel x$}

We shall now investigate magneto-Hall conductivity in the presence of perpendicular magnetic field ${\bm B}=(B,0,0)$, i.e., perpendicular to the WN's separation. We reiterate that the LLs at $k_x=0$ are doubly degenerate, as clearly observed from numerical findings (see Fig.~\ref{fig:LLBx}), irrespective of the topological charge of the WSM. This is in contrast to the parallel magnetic field ${\bm B}=(0,0,B)$ case where the LLs are non-degenerate for all values of $k_z$ (see Fig.~\ref{fig:LLBz}). At the outset, we comment that a perpendicular magnetic field would lead to distinct response characteristics compared to a parallel magnetic field. We below extensively analyze the effect of tilt and the amplitude of the magnetic field as well.  


\begin{figure*}[ht]
\begin{center}
	\includegraphics[width=0.95\linewidth]{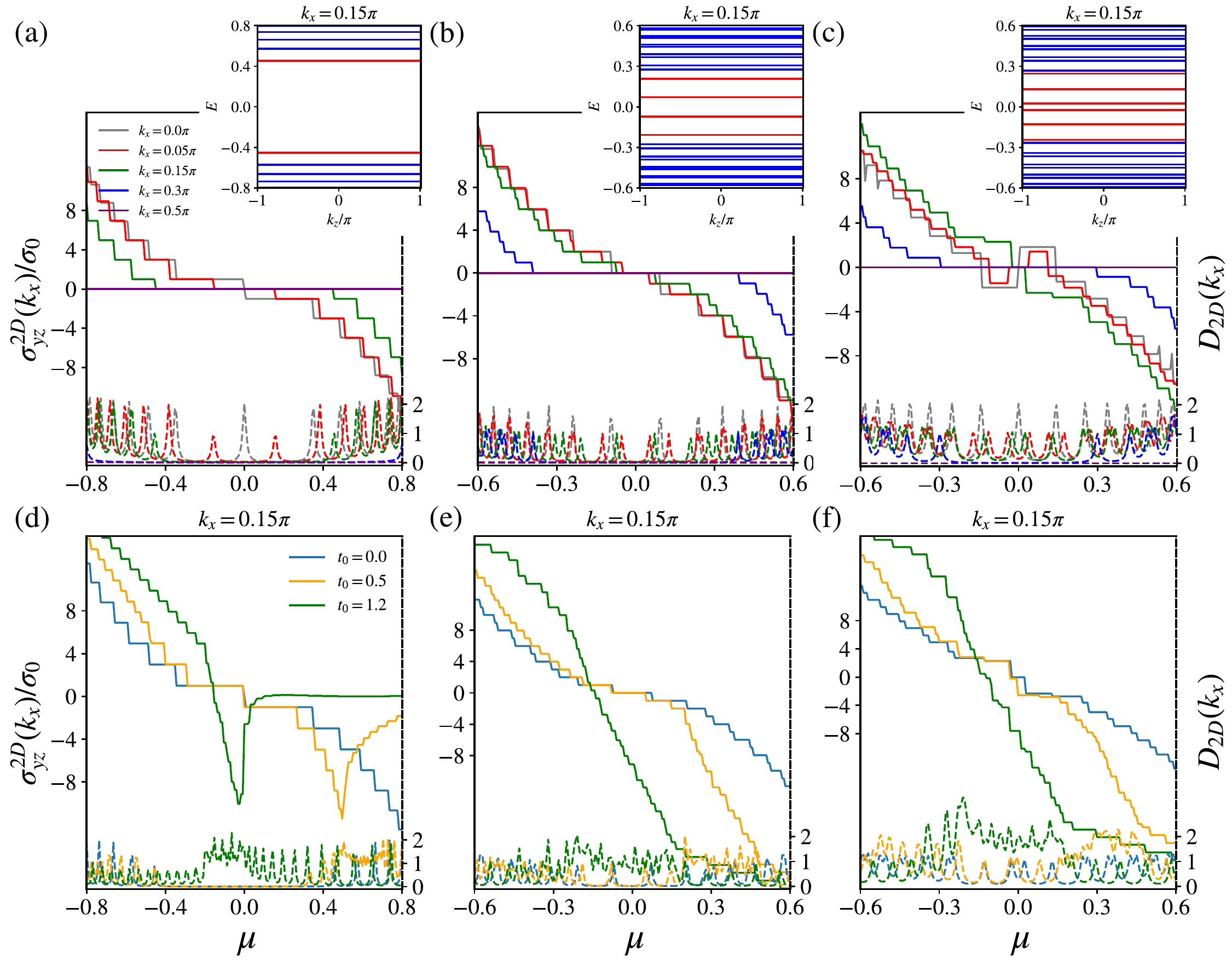}	
\end{center}
	\vspace{-0.4cm}
	\caption{Normalized 2D sheet Hall conductivity $\sigma^{2D}_{yz}(k_x)/\sigma_0$, computed from Eq.~(\ref{hall}), and PDOS $D_{2D}(k_x)$, evaluated from Eq.~(\ref{eq_dos}), as a function of $\mu$ with tilt term set at $t_0=0$, for $m=1$ in (a), $m=2$ in (b), and $m=3$ in (c) when ${\bm B}\parallel x$. The insets show the  $k_z$-independent flat LLs at $k_x=0.15\pi$. One can notice that $\sigma^{2D}_{yz}(k_x)$ jumps to the next plateau at certain $\mu$ when there exists a flat LL within the energy window scanned by $\mu$.
	We find double jumps, i.e., quantization changes by two, in $\sigma^{2D}_{yz}(k_x=0)$ for all of the WSMs, while the triple WSM additionally exhibits a non-monotonic profile following the LL spectrum illustrated in Fig.~\ref{fig:LLBx}.  
	We repeat (a), (b) and (c), respectively, in (d), (e), and (f) with $\sigma^{2D}_{yz}(k_x=0.15\pi)$ for $t_0=0,~0.5,~1.2$. The clean staircase-like structure almost vanishes in type-II phase while it exists in type-I phase.
	} 
\label{fig:hallyzkx}
\end{figure*}

\begin{figure}
\begin{center}
	\includegraphics[width=1\linewidth]{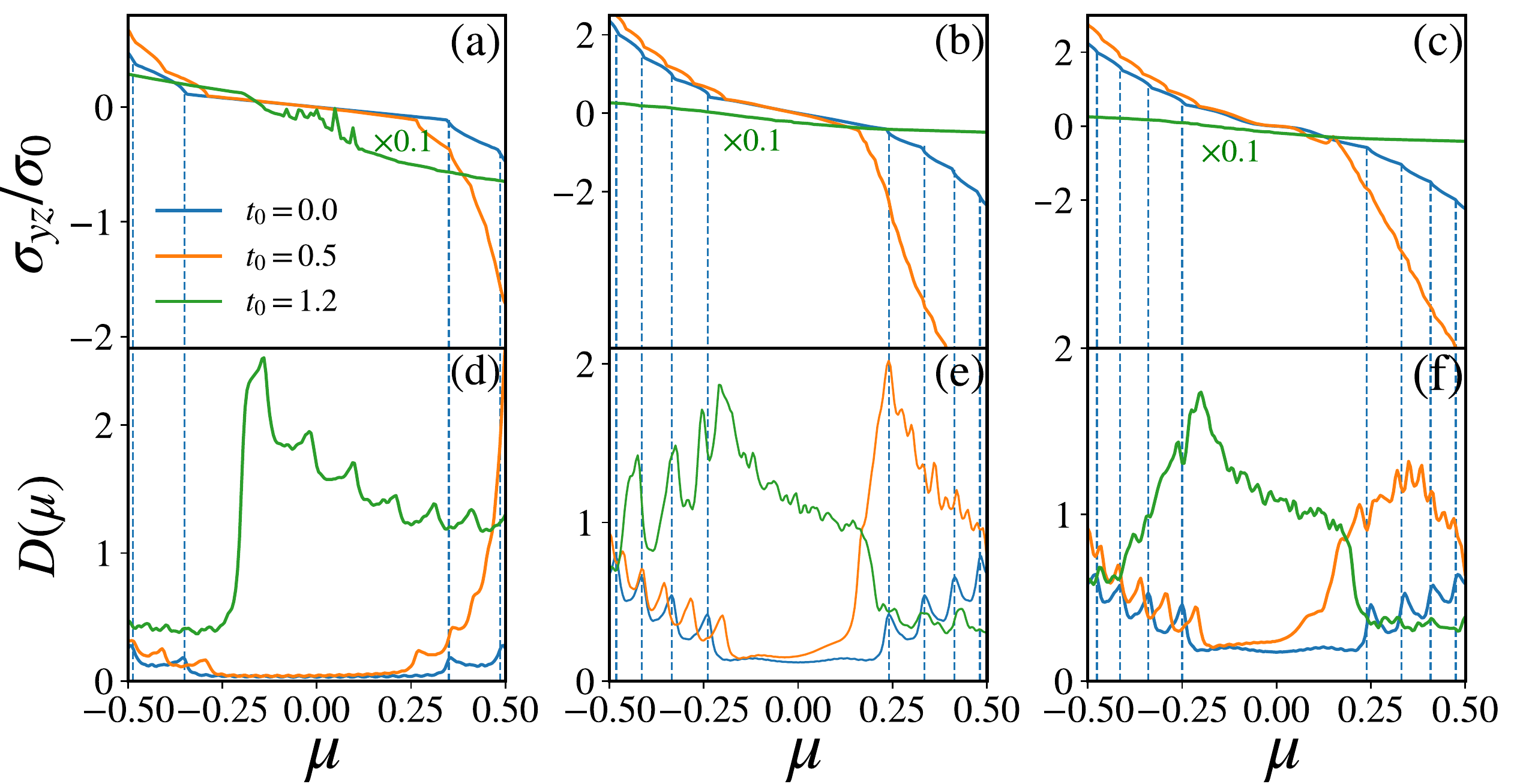}
\end{center}
	\vspace{-0.4cm}
	\caption{We repeat Fig.~\ref{fig:hallxy} for ${\bm B} \parallel x$. The flat regions in DOS, observed for type-I phase only, similar to Fig.~\ref{fig:hallxy} results in the $\mu$-linear regions in $\sigma_{yz}$ with the maximum contribution coming from the mid-gap chiral LLs. Similar to Fig.~\ref{fig:hallxy}, the slope change in $\sigma_{yz}$ is mediated by the peak in the DOS as designated by vertical blue dashed line for $t_0=0$. 
	} 
\label{fig:hallyz}
\end{figure}


\begin{figure}
\begin{center}
	\includegraphics[width=1.0\linewidth]{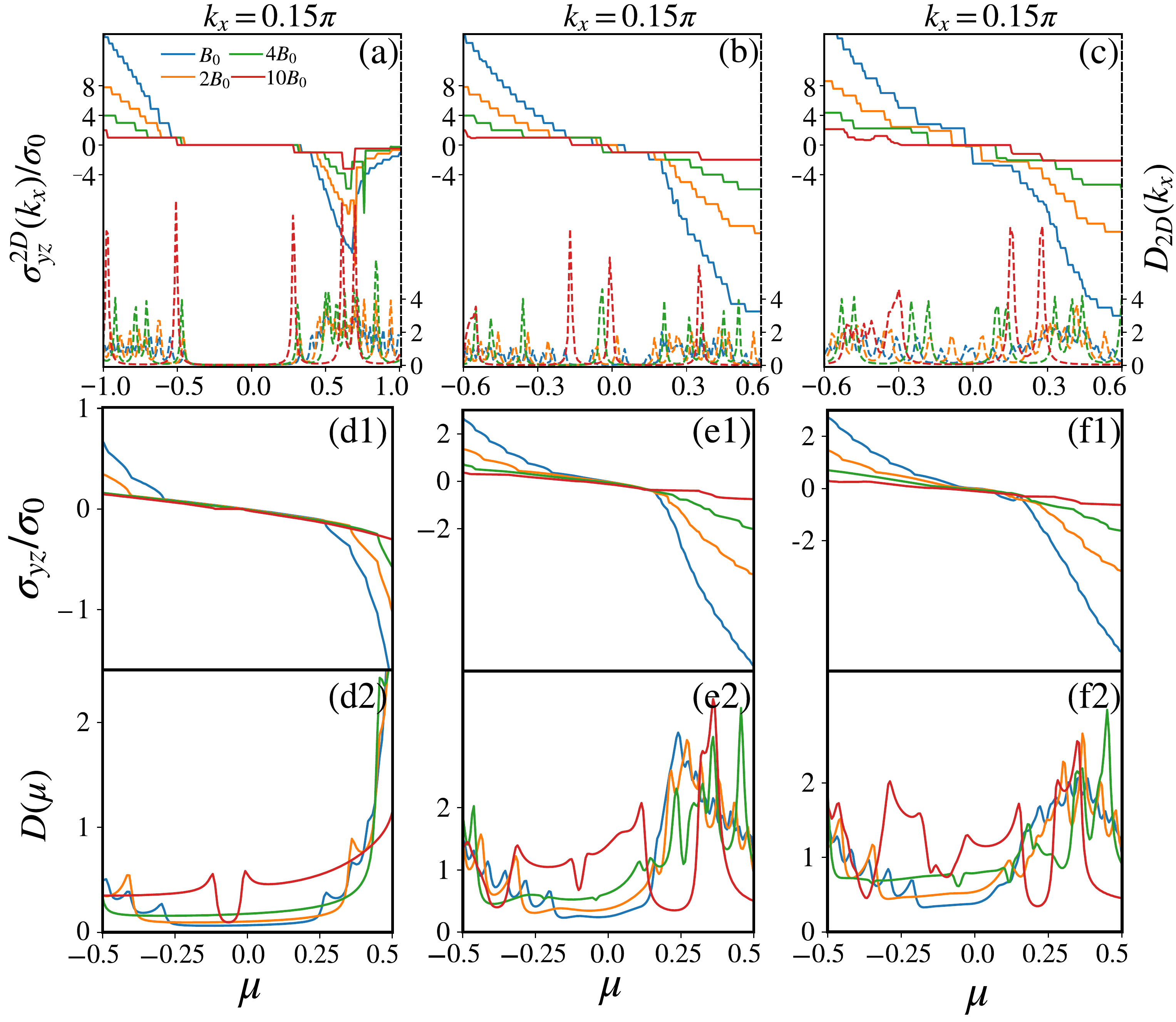} 
\end{center}
	\vspace{-0.4cm}
	\caption{Normalized 2D Hall conductivity $\sigma^{2D}_{yz}(k_x=3\pi/20)/\sigma_0$ and PDOS $D_{2D}(k_x=3\pi/20)$ as a function of $\mu$ for $m=1$ in (a), $m=2$ in (b), and $m=3$ in (c), by varying $B=B_0$, $2B_0$, $4B_0$ and $10B_0$ at $t_0=0.5$. 
	We show (d1), (e1), and (f1), [(d2), (e2), and (f2)] for the 3D magneto-Hall conductivity $\sigma_{yz}$ [DOS $D(\mu)$] as a function of $\mu$ with the above set of parameters. The degeneracy of LLs increases with $B$. This is apparently reflected in the increased width of the quantized plateau shown in (a)-(c) as well as $\mu$-linear region shown in (d1)-(f1).
	A stronger enough magnetic field $B=10 B_0$ can lead to a non-monotonic profile of $\sigma^{2D}_{yz}(k_x=3\pi/20)$ and a deviation from $\mu$-linear behavior in $\sigma_{yz}$.}
\label{fig:hallyzBvary}
\end{figure}

Let us first concentrate on the momentum labeled 2D Hall conductivity $\sigma^{2D}_{yz}(k_x)$ as displayed in Figs.~\ref{fig:hallyzkx}(a), (b) and (c) for untilted single, double and triple WSMs, respectively.
For a single WSM, $\sigma^{2D}_{yz}(k_x=0)$ always exhibits a jump by two throughout the range of $\mu$, including $\mu=0$, as there exists $k_z$-independent doubly degenerate flat LLs. For  
$\sigma^{2D}_{yz}(k_x\ne 0)$, we do not find any jump at $\mu=0$ due to the absence of LLs. We find jumps by unity at $\mu\simeq  \pm 0.45 $ for $k_x=3\pi/20$ that is consistent with the flat $k_z$-independent LLs (see insets in Fig.~\ref{fig:hallyzkx}(a)). Interestingly, the remaining flat LLs are doubly degenerate, leading to the double jump, i.e., quantization changes by two, in $\sigma^{2D}_{yz}(k_x)$ except when $\mu$ crosses the mid-gap chiral LLs for certain values of $k_x$.
Due to the particle-hole symmetry of the flat LL spectrum, the 2D Hall conductivity is an odd function of $\mu$:
$\sigma^{2D}_{yz}(k_x,\mu)=-\sigma^{2D}_{yz}(k_x,-\mu)$. In the case of double WSM, $\sigma^{2D}_{yz}(k_x=0)$ shows double jumps except for $\mu=0$ as the LLs at zero energy are not degenerate for $k_x=0$ (see Fig.~\ref{fig:LLBx} and Fig.~\ref{fig:hallyzkx}(b)). On the other hand, $\sigma^{2D}_{yz}(k_x\ne 0)$ always exhibits single jump i.e., quantization changes by unity, as none of the LLs are doubly degenerate. Although we sometime get double jump for $k_x=\pi/20$ due to numerical artifact where energy difference between two consecutive flat LLs is less than the resolution $\eta=0.01$ considered in numerical analysis.

Last for triple WSM, $\sigma^{2D}_{yz}(k_x=0)$ indeed represents counter-intuitive behavior as we find non-monotonic jump profile with respect to $\mu$ (see Fig.~\ref{fig:hallyzkx}(c)). A close inspection suggests that multiple degenerate LLs at $\mu=\pm 0.15,~0$ yield jumps by more than $\pm 2$. The non-monotonicity at $\mu=0$ might relate to the underlying chirality of LLs in the vicinity of the above values of $\mu$. The chiral nature of the mid-gap LLs for the crossing at $\mu=0$, $k_x=0$ is opposite to that for the crossing at $\mu=\pm 0.15$, $k_x=0$ (see Fig.~\ref{fig:LLBx}(c1)).
However, the unequal jump magnitude is hard to understand, while the rest of the uniform double jumps directly connect to the double degeneracy of LLs at $k_x=0$. Now for $\sigma^{2D}_{yz}(k_x \ne 0)$, the single jump pattern is not visible for closely spaced LLs as indicated by the PDOS structure, similar to the previous case of double WSM. The non-monotonicity around $\mu=0$ gets suppressed as $k_x$ staying from $0$, i.e., shown as $k_x=2\pi/30$, as the mid-gap LLs do not reverse their chiralities through the linear crossings. Importantly, due to particle-hole symmetry in the $k_z$-independent flat LL spectra, $\sigma^{2D}_{yz}(k_x)$ vanishes identically at $\mu=0$ for all $m=1,2,3$ irrespective of the values of $k_x$. This zero Hall conductance at $\mu=0$ is because the electrons with opposite chirality cancel out each other's contribution at $\mu=0$.

Next, coming to the tilt mediated complex behavior of $\sigma^{2D}_{yz}(k_x=0)$, as displayed in Figs.~\ref{fig:hallyzkx} (d)-(f), we find that the staircase-like structure becomes distorted and eventually almost disappears around $\mu=0$ for sufficiently large tilt strength. Notice that $\sigma^{2D}_{yz}(k_x)$ no longer behaves like an odd function of $\mu$, as a consequence of the breaking of particle-hole symmetry in the presence of the tilt term. For type-II mWSMs, the chiral LLs are entirely dissolved into the bulk, and hence it exhibits a substantially deformed staircase
(with highly irregular width of the plateau and non-uniform  jump) structure instead of the 
clean staircase  (with almost regular  width and uniform jump)
profile. This is in sharp contrast to the earlier case of the parallel magnetic field ${\bm B}|| z$ where the type-II WSMs still exhibit the staircase-like structure (see Figs.~\ref{fig:hallxykz} (d)-(f)). 
The metallic nature of the LL spectrum with $t_0>1$ in Fig.~\ref{fig:LLBx} can in principle destroy the staircase-like profile of 2D Hall conductivity.

Having explained the 2D Hall conductivity, we now analyse the 3D Hall conductivity by summing over all $k_x$ in BZ: $\sigma_{yz}=\sum_{k_x}\sigma^{2D}_{yz}(k_x)/n_x$ as shown in  Figs.~\ref{fig:hallyz}(a)-(c). Noticeably, 
$\sigma_{yz}$ continues showing linear dependence on $\mu$ when DOS is roughly flat, indicating that the chiral LLs are prominently contributing.  
Without tilt, the 3D Hall conductivity is an odd function of $\mu$ with $\sigma_{yz}(\mu)=-\sigma_{yz}(-\mu)$ inherited from $\sigma^{2D}_{yz}(k_x)$.
Likewise, in the parallel magnetic field case, the discontinuities in the slope of 3D Hall conductivity appear exactly at the peak of the DOS profile (see Figs.~\ref{fig:hallyz}(d)-(f)). As expected, with augmenting the tilt strength, the width of the $\mu$-linear zone reduces. The flat region in DOS shrinks, referring to the disintegration of chiral LLs into the bulk.

The behavior $\sigma_{yz} \sim \mu $ for $\mu \to 0$ is originated from the factor $f_{\alpha}-f_{\beta}$ while computed with the mid-gap chiral LLs. Notice that the energy spectrum $E_m(n< m,k_l)$  of mid-gap chiral LL varies linearly
with $k_l$ around the underlying WNs under the application of magnetic field $B_l$. As a result, the Fermi momentum, defined by $E_m(n<m,k_F)=\mu$ inside the bulk gap of the LL spectrum, is approximately linear in $\mu$. This might, in turn, lead to the linear behavior of $\mu$ for the 3D  Hall conductivity. Such linear behavior is more prominently visible for  $\sigma_{xy}$ than that for $\sigma_{yz}$ as the linearity of Fermi momentum with $\mu$ is more restricted for the case ${\mathbf B}\parallel x$. The linear variation of Hall conductivity as a function of $\mu$ can be regarded as a hallmark to distinguish type-I WSMs from type-II as 
this behavior is only observed for type-I WSM in the present case. Another vital point is that the slope of the $\mu$-linear region increases as the topological charge increases.
This pronounced response can be caused by the increasing number of chiral crossings within the bulk gap around $\mu=0$. The above findings are similar to that for parallel magnetic field ${\bm B}||z$. 

We now focus on the evolution of 2D Hall conductivity while tuning the amplitude of magnetic fields for a given value of tilt $t_0=0.5$ as shown in Figs.~\ref{fig:hallyzBvary} (a)-(c). The width of the quantized plateau increases because of the enhanced degeneracy of each LL, as shown at $k_x=3\pi/20$, also consistently reflected in the corresponding PDOS profiles. Notably, the monotonic pattern observed for triple WSM, around $\mu=0$, becomes less prominent with increasing $B$. The momentum integrated 3D Hall conductivity shows that the $\mu$-linear region gets broadened with increasing $B$ while the corresponding DOSs exhibit flat profiles (see Figs.~\ref{fig:hallyzBvary} (d1)-(f2)). Therefore, similar to the parallel magnetic field, one can also observe similar tendencies in the magnetoconductivity in the case of a perpendicular magnetic field.   

\section{Comparison with literature}\label{literature}

After extensively analyzing our results on the 2D sheet Hall conductivity $\sigma^{2D}_{ij}(k_l)$ and 3D Hall conductivity $\sigma_{ij}$, we here connect our findings with other relevant work in a similar direction. To begin with, we reiterate that the strong magnetic field essentially gaps out the WNs leading to a Fermi surface at $\mu=0$ with a finite value of  Fermi wave-vector $k_F$. The magnetic field-induced such a charge density wave of length $\lambda_F= 2\pi/k_F$ is analyzed for a single WSM in the context of QHE~\cite{Kai-Yu11}. Our results are consistent with the above study predicting $k_F=0$ and $\ne 0$ respectively for $ B \parallel x$ and $ B \parallel z$ where the WNs are located along $k_z$ without any magnetic field.
From the theoretical perspective, 3D WSMs exhibit quantized QHE investigated in \cite{Wang17,Li20}.
Interestingly, the quasi-quantization is experimentally observed in the presence of a magnetic field for the 3D QHE due to such charge density wave \cite{Galeski2021}. The quantization is investigated while varying the magnetic field for a fixed chemical potential. 
Such a quantized behavior can be anticipated from our analysis of $\sigma_{xy}$ being equivalent to the 2D sheet Hall conductivity $\sigma^{2D}_{xy}(k_F)$ for a given $k_z=k_F$ (see Fig.~\ref{fig:hallxykz}). We encounter the staircase profile for the quantization versus $\mu$ keeping $B$ fixed. 
Notice that LL energies increase with $B$, as shown in our analytical calculations.
This changes $k_F$ even when $\mu$ keeps fixed inside the bulk gap. One can hence expect that $\sigma^{2D}_{xy}(k_F)$ can, in principle, exhibit a staircase profile under the variation of $B$ as flat $k_x$-independent LLs cross a given $\mu$. One can obtain a staircase-like behavior of $\sigma^{2D}_{yz}(k_F)$ while varying $B$ for $\mu \ne 0$ residing in the bulk gap. It is thus evident that $\sum_{k_l}$ over the 2D sheet Hall conductivity
$\sigma^{2D}_{ij}(k_l)$ does not lead to a quantized plateau. Due to the limitation of our current framework, the 2D staircase sheet Hall conductivity as varying $B$ is beyond our scope, and we leave such a study for the future.

In the context of topological transport in WSM, the role of surface Fermi arc states is very important. It has been shown that the time taken by the electrons with velocity $v$ to execute the cyclic motion, i.e., magnetic cyclotron orbit, through the Fermi loop is divided into two parts such that $t=t_{\rm arc}+t_{\rm LL}$ where $t_{\rm arc} \approx k_0/(evB)$ with $k_0$ being the length of the Fermi arc, and $t_{\rm LL}\approx L/v$ denote time spent by electron on the surface and inside the bulk of the WSM  \cite{potter2014quantum}. 
The contribution from a chiral LL (i.e., bulk) dominates for $B>B_c= k_0/L$ while surface Fermi arc contribution prevails for magnetic field below such critical field strength. For a thick slab of WSM with $L\gg l_B$ and finite chemical potential such that $\mu$ intersects the bulk LLs, one expects the surface contribution to become insignificant. In the present case, we consider $B=2\pi n/L$ with $n$ being integer, and $L\gg k_0/ k_F^2$,  we find the chiral and non-chiral bulk LL are responsible for the conductivity. The oscillation in the 2D density of states, observed in Figs.~\ref{fig:hallxykz} and \ref{fig:hallyzkx}, as a function of $\mu$ is related to the quantum oscillations in terms of $1/B$. These oscillations, in our case, are governed by the bulk LLs, and hence we believe that the bulk conductivity $\sigma_{ij}=\sum_{k_l} \sigma^{2D}_{ij}(k_l)$ is maximally governed by the LLs. 
The magnetic field has to be perpendicular to the surface hosting the Fermi arc to receive the Fermi arc contribution. In our case, the magnetic field always lies parallel to the $xz$-surfaces, hosting the Fermi arcs,  as we do not have any $y$-component of ${\bf B}$. Therefore, the surface arc contribution under a perpendicular magnetic field is less than that of the bulk. Furthermore, to minimize the finite size effect, we consider the PBC along the $y$-direction so that the flat LLs become dispersionless. This might result in further reduction in the surface effect. However, the contribution of the surface Fermi arc is yet to be explored in more detail in future studies.

Finally, we discuss the effect of disorder on the 3D Hall conductivity. We believe that similar to the integer quantum Hall effect in 2D, the quantization in 2D sheet Hall conductivity remains unaffected in the presence of weak on-site random disorder $W_{\rm dis}$. For strong disorder, scattering between the localized edge states leads to the deviation from the quantization. This means that scattering between the $k_i$-dependent LLs can destroy the staircase profile of 2D Hall conductivity under magnetic field along the $i$-th direction.  
It has been shown that the quantized Hall conductivity, caused by the chiral zeroth LLs traversing
the gap, is robust against disorder scattering for an intermediate number of layers in the direction of the
magnetic field \cite{Ma21}. One can hence anticipate that 
the scattering between two opposite chiral LLs gets suppressed as long as the disorder is weak compared to energy scale $\Delta E \approx  v_F k_0$. Here, $v_F$ denotes the Fermi velocity and WNs appear at $(0,0,\pm k_0/2)$.  The  quantized 2D Hall conductivity is expected to be observed  (destroyed) for $W_{\rm dis}< \Delta E$ ($W_{\rm dis}> \Delta E$). We notice that the disordered 3D QHE is a completely new research direction, and it requires further investigations beyond the scope of the present study. The sample thickness and the mean free path caused by the disorder play interesting roles in quantizing 2D Hall conductivity for the disordered case.

\begin{table*}[ht]
\begin{center}
\begin{tabular}{| p{1.7cm}|p{2.6cm}|p{2cm}|p{4.3cm}|p{3cm}|p{3cm} |} 
\hline
 Topological charge $m$ & Number of chiral LL through WN  & Dispersive (flat) along  & Sheet Hall conductivity without tilt &  Steps in $\sigma_{xy}^{2D}(k_{z},\mu)$ for type-I & Steps in $\sigma_{xy}^{2D}(k_{z},\mu)$ for type-II  \\ 
\hline
 $1$   & 1 &  $k_z~ (k_x)$ & $\sigma_{xy}^{2D}(k_{z},\mu)\neq -\sigma_{xy}^{2D}(k_{z},-\mu)$ &1, monotonic & 1, monotonic   \\ 
\hline
 $2$   & 2 &  $k_z~ (k_x)$ &  $\sigma_{xy}^{2D}(k_{z},\mu)\neq -\sigma_{xy}^{2D}(k_{z},-\mu)$ & 1, monotonic & 1, monotonic   \\ 
\hline
$3$   & 3 &  $k_z~ (k_x)$ & $\sigma_{xy}^{2D}(k_{z},\mu)\neq -\sigma_{xy}^{2D}(k_{z},-\mu)$ & 1, monotonic &  1, monotonic   \\ 
\hline 
\end{tabular}
\end{center}
\caption{The main findings on LLs (Fig. ~\ref{fig:LLBz}) and 2D sheet Hall conductivity $\sigma_{xy}^{2D}(k_{z},\mu)$ (Fig. ~\ref{fig:hallxykz}) are presented in a table for $\bf {B} \parallel z$. The step represents the difference between two consecutive quantized plateaus in $\sigma_{xy}^{2D}(k_{z},\mu)$ in the unit of $e^2/\hbar$. The staircase profile of $\sigma_{xy}^{2D}(k_{z},\mu)$ remains unaltered with $\mu$ referring to the monotonic behavior.}
\label{tab:Bz} 
\end{table*}

\begin{table*}[ht]
\begin{center}
\begin{tabular}{| p{1.7cm}|p{2.6cm}|p{1.8cm}|p{4.3cm}|p{3.5cm}|p{3.cm} |} 
\hline
  Topological charge $m$ & Number of chiral LL through WN  & Dispersive (flat) along  & Sheet Hall conductivity without tilt &  Steps in $\sigma_{yz}^{2D}(k_{x},\mu)$ for type-I & Steps in $\sigma_{yz}^{2D}(k_{x},\mu)$ for type-II  \\ 
\hline
 $1$   & 2 &  $k_x~ (k_z)$  & $\sigma_{yz}^{2D}(k_{x},\mu) = -\sigma_{yz}^{2D}(k_{x},-\mu)$  & 1,2,uniform(mostly)  & non-uniform (mostly)   \\ 
\hline
 $2$   & 4 & $k_x~ (k_z)$ & $\sigma_{yz}^{2D}(k_{x},\mu)= -\sigma_{yz}^{2D}(k_{x},-\mu)$ &1,2,uniform(partially)  & irregular   \\ 
\hline
$3$   &6 &   $k_x~ (k_z)$ & $\sigma_{yz}^{2D}(k_{x},\mu)= -\sigma_{yz}^{2D}(k_{x},-\mu)$ &1,2,uniform(minimally) &irregular  \\ 
\hline 
\end{tabular}
\end{center}
\caption{The main findings on LLs (Fig. ~\ref{fig:LLBx}) and 2D sheet Hall conductivity $\sigma_{yz}^{2D}(k_{x},\mu)$ (Fig. ~\ref{fig:hallyzkx}) are presented in a table for $\bf {B} \parallel x$. The steps in $\sigma_{yz}^{2D}(k_{x},\mu)$ are not always found to be unity  indicating the deviation from uniform feature as described in Table~\ref{tab:Bz}. The profile also becomes irregular when the width of the plateaus are substantially different from each other. }
\label{tab:Bx} 
\end{table*}


\section{DISCUSSION and SUMMARY}\label{Summary}

Transport properties of topological systems have emerged as a central theme of recent research in condensed matter physics, with an inherent connection to the quantum Hall effect. Among them, 2D systems have been extensively studied theoretically early on~\cite{Laughlin81, Thouless82}. Interestingly, in recent experiments, ZrTe$_5$, HfTe$_5$, and Cd$_3$As$_2$ have been found to exhibit QHE in 3D \cite{zhang2019quantum,liang2018anomalous,tang2019three,Galeski2021,galeski2020unconventional}.
Therefore, in the present theoretical work on 3D WSMs in the quantum limit, we try to answer the following experimentally relevant questions: How does the Hall conductivity respond in generic, tilted mWSM models under different orientations of magnetic fields 
concerning the WN's separation? Further, we study 
how to distinguish type-I from type-II mWSMs following different magnetic field orientations. To answer these questions, we first analytically solve the LLs in a low-energy model consisting of two WNs at ${\bm k}^{\pm}_p=(0,0,\pm 1)$ and successfully depict the LLs in the lattice ones in the case of a parallel magnetic field $(0,0, B)$, namely, ${\bm B}$ aligns with the WN's separation. On the other hand, the LL spectrum under perpendicular magnetic field $(B,0,0)$, i.e., ${\bm B}$ being normal to the WN's separation, can be qualitatively explained by the low-energy model, describing a single WN. The sign and magnitude of higher topological charges imprint their signatures on the chirality (which is the slope of the mid-gap LLs) and the number of chiral LLs passing through a WN at $k_{z,p}=\pm\frac{\pi}{2}$ for ${\bm B} \parallel z$ (see Fig.~\ref{fig:LLBz}). In the case of ${\bm B} \parallel x$, the topological charge value can be obtained from the number of linear crossings of the mid-gap LLs at $k_x=0$ (see Fig.~\ref{fig:LLBx}), with two WNs of opposite chiralities projected simultaneously. In both the cases with ${\bm B}\parallel z$ and ${\bm B} \parallel x$, the bulk gap reduces with increasing the tilt strength. However, the chiral LLs continue to exist for the type-II phase only if ${\bm B} \parallel z$. Therefore, inspecting these distinct responses can simultaneously identify the type-I/type-II phase and topological charges.

The chiral structure of the LLs, dispersing along the magnetic field direction, essentially encrypts the quantization of the edge states. The magneto-Hall conductivity $\sigma_{xy}$ for $\bm B\parallel z$ and $\sigma_{yz}$ for $\bm B \parallel x$ are then the immediate measures to investigate the perceptible differences in terms of the tilt and topological charge. Notably, the WSMs can be envisioned as stacking Chern insulator plates along the direction of WN's separation. Our finding is, therefore, consistent with the fact that QHE in WSMs can only take place when Fermi arcs at opposite surfaces are connected through the bulk WNs to form the 
Fermi loop with a good quantum number \cite{Wang17}. 
This leads to the quantization in the 2D sheet Hall conductivity $\sigma^{2D}_{yz}(k_x)$ when ${\bm B}\parallel x$ such that the two Fermi arcs existing on the two opposite $yz$-surfaces, are strapped together via the WNs at ${\bm k}^{\pm}_p=(0,0,\pm \frac{\pi}{2})$ by considering $k_x$ as a good quantum number. Surprisingly, our findings indicate that the QHE can also be observed when $\bm B \parallel z$ such that both the WNs have an identical projection on the same Fermi point on the opposite $xy$-surfaces. This way, disconnected Fermi points can be coupled by considering $k_z$ as a good quantum number. This further leads to the staircase-like quantized behavior in $\sigma^{2D}_{xy}(k_z)$. To be more precise, the staircase-like structure in $\sigma^{2D}_{ij}(k_l)$ emerges from 3D WSMs under ${\bm B} \parallel l$ when electrons fill up the flat $k_p$-independent LLs serially with $l\neq p$ (see Fig.~\ref{fig:hallxykz} and \ref{fig:hallyzkx}).

Surprisingly, we find that $\sigma^{2D}_{xy}(k_z)$ always exhibits jumps by unity for both type-I and type-II phases. By contrast, the staircase pattern in $\sigma^{2D}_{yz}(k_x)$ and the double jump due to the crossing of the chiral LLs at $k_x=0$, 
observed for the type-I phase are maximally destroyed for the type-II phase. Apart from these distinct features with the tilt, the jump profiles between the adjacent quantized plateaus become different with the topological charge change. In the untilted case, $\sigma^{2D}_{yz}(k_x)$ becomes an odd function of $\mu$, unlike  $\sigma^{2D}_{xy}(k_z)$, as the flat LL spectrum is particle-hole symmetric irrespective of the topological charge for ${\bm B}\parallel x$. We also demonstrate the effect of an increasing magnetic field on the 2D Hall conductivity, where the width of the quantized plateau increases due to the degeneracy associated with each flat LL (see Fig.~\ref{fig:hallxyBvary} and \ref{fig:hallyzBvary}). In the end, we find linear dependence on $\mu$ for $\mu\to 0$ in the 3D Hall conductivities $\sigma_{ij}=\sum_{k_l}\sigma^{2D}_{ij}(k_l)/n_l$ irrespective of the direction of the magnetic fields (see Fig.~\ref{fig:hallxy} and \ref{fig:hallyz}). The mid-gap chiral LLs around $\mu=0$ are responsible for the above linear response that we also explain analytically by a plausible argument. Interestingly, the tilt reduces the width of the linear regime, eventually destroyed in the type-II phase. The slope associated with this $\mu$-linear region increases with the topological charge. The 
notable findings of our work are tabulated in Table~\ref{tab:Bz} and ~\ref{tab:Bx} for $\bf{B}\parallel z$ and $\bf{B}\parallel x$, respectively. 
Considering all the above theoretical predictions based on lattice models, we believe that our work is closely relevant in transport experiments with mWSMs.

Last, coming to a possible experimental investigation of our work, we can comment that building 3D systems by attaching electronic gates suitably in 2D systems or stacking the 2D materials \cite{Eisenstein86} appeared to fail as the resulting shape of the Fermi surface indicates its 2D nature. The 3D QHE was performed in ZrTe$_5$ with a magnetic field of around $2$ T and temperature around $0.6$ K, such that the lowest LL is only occupied in the extreme quantum limit \cite{tang2019three,zhang2019quantum, Galeski2021}.
The Hall resistivity plateau is proportional to half the Fermi wavelength along the magnetic field direction. The 3D QHE has been observed when the Fermi wavelength is much larger than the lattice constant. In our present case, a significant length of the Fermi arc (such as periodic boundary conditions can be imposed) is considered for the lattice along the magnetic-field direction. Thus, our results can be experimentally relevant under appropriate parameter regimes. However, we analyze our results with varying chemical potential to maintain the magnetic field commensurate with the sample size. Of course, it would be a practical topic to explore these effects in mWSMs in the context of the \textit{ab-initio} studies, e.g.,type-I (i.e., TaAs and NbAs) and type-II (i.e., MoTe$_2$, LaAlGe, and WTe$_2$, while varying magnetic field continuously. It could also be interesting to investigate 3D QHE in other topological semimetals, such as nodal-line and  Dirac semimetals.

\begin{acknowledgments}
TN thanks Tutul Biswas for valuable discussions. FX thanks A.L He for the discussion on Chern numbers on lattice models. One of the authors, FX, thanks the German Science Foundation (DFG) for support sincerely through RTG 1995 and RWTH0662 for granting computing time.
\end{acknowledgments}

\appendix

\section{Lattice Hamiltonian matrix}\label{LatticeHamil}

By employing a periodic boundary condition along $y$-direction, we can write down the Hamiltonian for single, double, and triple WSM described by Eqs.~(\ref{sWSM_lattice})-(\ref{tWSM_lattice}) in the tridiagonal matrix form.
These matrices for single, double, and triple WSM are as follows
\begin{equation}
H_m(\bm{k})= \begin{bmatrix}
h_{1} & h_{12} & h_{13} & \cdots & h_{1L_{y}-1} & h_{1L_{y}}\\
 & h_{2} & h_{23} & h_{24} & \cdots & h_{2L_{y}}\\
 &  & h_{3} & h_{34} & h_{35} & \vdots\\
\text{h.c.}&  &  & \ddots & \ddots & \vdots\\
 &  &  &  & h_{L_{y}-1} & h_{L_{y}-1L_{y}}\\
 &  &  &  &  & h_{L_{y}}
\end{bmatrix}\,,
\label{sWSMhamil_Lattice}
\end{equation}

where $h_{j_y}$ and $h_{j_{y}-1 j_y(j_{y}-2 j_y)}$ are the intra and interlayer (second nearest) hopping matrices. For single WSM, only the intra and interlayer contribute with
$h_{j_y} = \big( \cos(k_{z}-j_{y}B_{x})+2-\cos(k_{x}+j_{y}B_{z})\big)\sigma_{z} + \sin(k_{x}+j_{y}B_{z})\sigma_{x}+(t_{0} \cos (k_z-j_y B_x)-\mu)\sigma_{0}$ and $h_{j_{y-1}j_y}=\frac{1}{2i}\sigma_y-\frac{1}{2}\sigma_z$. Notice that for double and triple WSM, we need to take the second nearest interlayer hopping into account. In the case of double WSM, we have $h_{j_y}=\big(t_{z}\cos(k_{z}-j_{y}B_{x})+6 +\cos2(k_{x}+j_{y}B_{z})-4\cos(k_{x}+j_{y}B_{z})\big)\sigma_{z}+ \cos(k_{x}+j_{y}B_{z})\sigma_{x}+(t_{0} \cos (k_z-j_y B_x)-\mu)\sigma_{0}$, $h_{j_{y-1}j_y}=-\frac{1}{2}\sigma_x+\frac{1}{2i}\sin(k_x+j_y B_z)\sigma_y-2\sigma_z$ and $h_{j_{y-2}j_y}=\frac{1}{2}\sigma_z$. For triple WSM, these are given by  $h_{j_y}=\big(\cos(k_{z}-j_{y}B_{x})+6+\cos2(k_{x}+j_{y}B_{z})-4\cos(k_{x}+j_{y}B_{z})\big)\sigma_{z}+\sin(k_{x}+j_{y}B_{z})(-2-\cos(k_{x}+j_{y}B_{z}))\sigma_{x}+(t_{0} \cos (k_z-j_y B_x)-\mu)\sigma_{0}$, $h_{j_{y}-1 j_y}=\frac{3}{2}\sin(k_x+j_yB_z)\sigma_x-\frac{1}{2i}(-2+3\cos(k_x+j_y B_z))\sigma_y-2\sigma_z$ and $h_{j_{y}-2 j_y}=\frac{1}{4i}\sigma_y+\frac{1}{2}\sigma_z$.

For magnetic fields applied along with different directions ${\bm B}\parallel z$ or ${\bm B}\parallel x$, we only need to set either $B_x=0$ or $B_z=0$ in the matrices above.

\section{Lattice velocity matrix}\label{LatticeVelocity}

According to the definition of velocity $\mathcal{V}_{k_{i}}=\frac{\partial \mathcal{H}_{m}}{\partial k_{i}}$, the velocity can be also expressed by the matrix formula.

\subsection{$\bf{B}\parallel z$} \label{AppendBz}

In the case $\bm{B}$ along $z$ direction, the derivation of Hamiltion respective to $k_x$ and $k_y$ need to be considered. Then, we can write down the corresponding $\mathcal{V}_{k_x}$ and $\mathcal{V}_{y}$ for single, double, and triple WSM elaborately.

For $m=1$ single WSM, the corresponding matrices are following
\begin{equation}
\mathcal{V}_{k_x}= \begin{bmatrix}
V_{1} & 0 & \cdots & 0\\
 & V_{2} & \ddots & 0\\
 \text{h.c.}& & \ddots & \vdots\\
 &  &  & V_{L_{y}}
\end{bmatrix}\,,
\label{VxsWSM}
\end{equation} 
where onsite element $V_{j_y}=\cos\big(k_{x}+j_{y}B_{z}\big)\sigma_{x}+\sin\big(k_{x}+j_{y}B_{z}\big)\sigma_{z}$ and 
\begin{equation}
\mathcal{V}_{y}= \begin{bmatrix}
0 & V_{12} & \cdots & V_{1L_{y}}\\
 & 0 & V_{23} & 0\\
\text{h.c.} &  & \ddots & \vdots\\
 &  &  & 0
\end{bmatrix}\,,
\label{VysWSM}
\end{equation} 
where $V_{j_{y-1}j_y}=\frac{1}{2}\sigma^{y}+\frac{1}{2i}\sigma^{z}$ and the boundary connection $V_{1L_{y}}=V^\dagger_{12}$.

For $m=2$ double WSM, the velocity matrices are
\begin{equation}
\mathcal{V}_{k_x}= \begin{bmatrix}
V_1 & V_{12} & \cdots & V_{1L_{y}}\\
 & V_2 & V_{23} & 0\\
\text{h.c.} &  & \ddots & \vdots\\
 &  &  & V_{L_y}
\end{bmatrix}\,,
\label{VxdWSM}
\end{equation} 
where $V_{j_y}=-\sin\big(k_{x}+j_{y}B_{z}\big)\sigma_{x}+\big[-2\sin2\big(k_{x}+j_{y}B_{z}\big)+4\sin\big(k_{x}+j_{y}B_{z}\big)\big]\sigma_{z}$ and $V_{j_{y}-1 j_y}=\frac{1}{2i}\cos\big(k_{x}+j_{y}B_{z}\big)\sigma_{y}$.

\begin{equation}
\mathcal{V}_{y}= \begin{bmatrix}
0 & V_{12} & V_{13} & \cdots & V_{1L_{y}-1} & V_{1L_{y}}\\
 & 0 & V_{23} & V_{24} & \cdots & V_{2L_{y}}\\
 &  & 0 & V_{34} & V_{35} & \vdots\\
\text{h.c.}&  &  & \ddots & \ddots & \vdots\\
 &  &  &  & 0 & V_{L_{y}-1L_{y}}\\
 &  &  &  &  & 0
\end{bmatrix}\,,
\label{VydWSM}
\end{equation}
where $V_{j_{y}-1j_y}=\frac{1}{2i}\sigma_{x}+\frac{1}{2}\sin\big(k_{x}+j_{y}B_{z}\big)\sigma_{y}+\frac{2}{i}\sigma_{z}$ and $V_{j_{y}-2j_y}=-\frac{1}{i}\sigma^{z}$.

For $m=3$ triple WSM, the velocity matrices share the same structure as for $m=2$ in Eqs.~(\ref{VxdWSM}) and (\ref{VydWSM}). For $\mathcal{V}_{k_x}$, the nonzero elements in upper diagonal matrix $V_{j_y}=(-2\cos\big(k_{x}+j_{y}B_{z}\big)-\cos2\big(k_{x}+j_{y}B_{z}\big))\sigma_{x}+(-2\sin2\big(k_{x}+j_{y}B_{z}\big)+4\sin\big(k_{x}+j_{y}B_{z}\big))\sigma_{z}$ and $V_{j_{y}-1 j_y}=\frac{3}{2}\cos\big(k_{x}+j_{y}B_{z}\big)\sigma_{x}+\frac{3}{2i}\sin\big(k_{x}+j_{y}B_{z}\big)\sigma_{y}$. While for $\mathcal{V}_y$, the elements are $V_{j_{y}-1 j_y}=\frac{-3}{2i}\sin\big(k_{x}+j_{y}B_{z}\big)\sigma_{x}+(1-\frac{3}{2}\cos\big(k_{x}+j_{y}B_{z}\big))\sigma_{y}+\frac{2}{i}\sigma_{z}$ and $V_{j_{y}-2 j_y}=\frac{1}{2}\sigma^{y}-\frac{1}{i}\sigma^{z}$.

\subsection{$\bf{B}\parallel x$}

When applying a magnetic field in the $x$-direction, we need to consider the derivatives of Hamiltonian concerning the rest two variables $k_z$ and $k_y$. Since $k_z$ terms only appear in the onsite block, their velocity matrices $\mathcal{V}_{k_z}$ of single, double, and triple WSM are diagonal as given in Eq.~(\ref{VxsWSM}). Notice that  instead of filling with different values $V_{j_y}=-\sin(k_z-j_y B_x)(t_0\sigma_0+\sigma_{z})$ there we keep this same for all the three WSMs.
The $y$-direction velocity matrices are kept the same as what has been derived in Append.~\ref{AppendBz} with $k_x+j_y B_z$ replaced by $k_x$ as $B_z=0$ for ${\bm B}\parallel x$.

\bibliography{MWS.bib}

\begin{thebibliography}{94}%
\makeatletter
\providecommand \@ifxundefined [1]{%
 \@ifx{#1\undefined}
}%
\providecommand \@ifnum [1]{%
 \ifnum #1\expandafter \@firstoftwo
 \else \expandafter \@secondoftwo
 \fi
}%
\providecommand \@ifx [1]{%
 \ifx #1\expandafter \@firstoftwo
 \else \expandafter \@secondoftwo
 \fi
}%
\providecommand \natexlab [1]{#1}%
\providecommand \enquote  [1]{``#1''}%
\providecommand \bibnamefont  [1]{#1}%
\providecommand \bibfnamefont [1]{#1}%
\providecommand \citenamefont [1]{#1}%
\providecommand \href@noop [0]{\@secondoftwo}%
\providecommand \href [0]{\begingroup \@sanitize@url \@href}%
\providecommand \@href[1]{\@@startlink{#1}\@@href}%
\providecommand \@@href[1]{\endgroup#1\@@endlink}%
\providecommand \@sanitize@url [0]{\catcode `\\12\catcode `\$12\catcode
  `\&12\catcode `\#12\catcode `\^12\catcode `\_12\catcode `\%12\relax}%
\providecommand \@@startlink[1]{}%
\providecommand \@@endlink[0]{}%
\providecommand \url  [0]{\begingroup\@sanitize@url \@url }%
\providecommand \@url [1]{\endgroup\@href {#1}{\urlprefix }}%
\providecommand \urlprefix  [0]{URL }%
\providecommand \Eprint [0]{\href }%
\providecommand \doibase [0]{http://dx.doi.org/}%
\providecommand \selectlanguage [0]{\@gobble}%
\providecommand \bibinfo  [0]{\@secondoftwo}%
\providecommand \bibfield  [0]{\@secondoftwo}%
\providecommand \translation [1]{[#1]}%
\providecommand \BibitemOpen [0]{}%
\providecommand \bibitemStop [0]{}%
\providecommand \bibitemNoStop [0]{.\EOS\space}%
\providecommand \EOS [0]{\spacefactor3000\relax}%
\providecommand \BibitemShut  [1]{\csname bibitem#1\endcsname}%
\let\auto@bib@innerbib\@empty
\bibitem [{\citenamefont {Klitzing}\ \emph {et~al.}(1980)\citenamefont
  {Klitzing}, \citenamefont {Dorda},\ and\ \citenamefont
  {Pepper}}]{Klitzing80}%
  \BibitemOpen
  \bibfield  {author} {\bibinfo {author} {\bibfnamefont {K.~v.}\ \bibnamefont
  {Klitzing}}, \bibinfo {author} {\bibfnamefont {G.}~\bibnamefont {Dorda}}, \
  and\ \bibinfo {author} {\bibfnamefont {M.}~\bibnamefont {Pepper}},\ }\href
  {\doibase 10.1103/PhysRevLett.45.494} {\bibfield  {journal} {\bibinfo
  {journal} {Phys. Rev. Lett.}\ }\textbf {\bibinfo {volume} {45}},\ \bibinfo
  {pages} {494} (\bibinfo {year} {1980})}\BibitemShut {NoStop}%
\bibitem [{\citenamefont {Thouless}\ \emph {et~al.}(1982)\citenamefont
  {Thouless}, \citenamefont {Kohmoto}, \citenamefont {Nightingale},\ and\
  \citenamefont {den Nijs}}]{Thouless82}%
  \BibitemOpen
  \bibfield  {author} {\bibinfo {author} {\bibfnamefont {D.~J.}\ \bibnamefont
  {Thouless}}, \bibinfo {author} {\bibfnamefont {M.}~\bibnamefont {Kohmoto}},
  \bibinfo {author} {\bibfnamefont {M.~P.}\ \bibnamefont {Nightingale}}, \ and\
  \bibinfo {author} {\bibfnamefont {M.}~\bibnamefont {den Nijs}},\ }\href
  {\doibase 10.1103/PhysRevLett.49.405} {\bibfield  {journal} {\bibinfo
  {journal} {Phys. Rev. Lett.}\ }\textbf {\bibinfo {volume} {49}},\ \bibinfo
  {pages} {405} (\bibinfo {year} {1982})}\BibitemShut {NoStop}%
\bibitem [{\citenamefont {Zhang}\ \emph {et~al.}(2005)\citenamefont {Zhang},
  \citenamefont {Tan}, \citenamefont {Stormer},\ and\ \citenamefont
  {Kim}}]{zhang2005experimental}%
  \BibitemOpen
  \bibfield  {author} {\bibinfo {author} {\bibfnamefont {Y.}~\bibnamefont
  {Zhang}}, \bibinfo {author} {\bibfnamefont {Y.-W.}\ \bibnamefont {Tan}},
  \bibinfo {author} {\bibfnamefont {H.~L.}\ \bibnamefont {Stormer}}, \ and\
  \bibinfo {author} {\bibfnamefont {P.}~\bibnamefont {Kim}},\ }\href@noop {}
  {\bibfield  {journal} {\bibinfo  {journal} {nature}\ }\textbf {\bibinfo
  {volume} {438}},\ \bibinfo {pages} {201} (\bibinfo {year}
  {2005})}\BibitemShut {NoStop}%
\bibitem [{\citenamefont {Xu}\ \emph {et~al.}(2014)\citenamefont {Xu},
  \citenamefont {Miotkowski}, \citenamefont {Liu}, \citenamefont {Tian},
  \citenamefont {Nam}, \citenamefont {Alidoust}, \citenamefont {Hu},
  \citenamefont {Shih}, \citenamefont {Hasan},\ and\ \citenamefont
  {Chen}}]{xu2014observation}%
  \BibitemOpen
  \bibfield  {author} {\bibinfo {author} {\bibfnamefont {Y.}~\bibnamefont
  {Xu}}, \bibinfo {author} {\bibfnamefont {I.}~\bibnamefont {Miotkowski}},
  \bibinfo {author} {\bibfnamefont {C.}~\bibnamefont {Liu}}, \bibinfo {author}
  {\bibfnamefont {J.}~\bibnamefont {Tian}}, \bibinfo {author} {\bibfnamefont
  {H.}~\bibnamefont {Nam}}, \bibinfo {author} {\bibfnamefont {N.}~\bibnamefont
  {Alidoust}}, \bibinfo {author} {\bibfnamefont {J.}~\bibnamefont {Hu}},
  \bibinfo {author} {\bibfnamefont {C.-K.}\ \bibnamefont {Shih}}, \bibinfo
  {author} {\bibfnamefont {M.~Z.}\ \bibnamefont {Hasan}}, \ and\ \bibinfo
  {author} {\bibfnamefont {Y.~P.}\ \bibnamefont {Chen}},\ }\href@noop {}
  {\bibfield  {journal} {\bibinfo  {journal} {Nature Physics}\ }\textbf
  {\bibinfo {volume} {10}},\ \bibinfo {pages} {956} (\bibinfo {year}
  {2014})}\BibitemShut {NoStop}%
\bibitem [{\citenamefont {Halperin}(1987)}]{Halperin_1987}%
  \BibitemOpen
  \bibfield  {author} {\bibinfo {author} {\bibfnamefont {B.~I.}\ \bibnamefont
  {Halperin}},\ }\href {\doibase 10.7567/jjaps.26s3.1913} {\bibfield  {journal}
  {\bibinfo  {journal} {Japanese Journal of Applied Physics}\ }\textbf
  {\bibinfo {volume} {26}},\ \bibinfo {pages} {1913} (\bibinfo {year}
  {1987})}\BibitemShut {NoStop}%
\bibitem [{\citenamefont {Kohmoto}\ \emph {et~al.}(1992)\citenamefont
  {Kohmoto}, \citenamefont {Halperin},\ and\ \citenamefont {Wu}}]{Kohmoto92}%
  \BibitemOpen
  \bibfield  {author} {\bibinfo {author} {\bibfnamefont {M.}~\bibnamefont
  {Kohmoto}}, \bibinfo {author} {\bibfnamefont {B.~I.}\ \bibnamefont
  {Halperin}}, \ and\ \bibinfo {author} {\bibfnamefont {Y.-S.}\ \bibnamefont
  {Wu}},\ }\href {\doibase 10.1103/PhysRevB.45.13488} {\bibfield  {journal}
  {\bibinfo  {journal} {Phys. Rev. B}\ }\textbf {\bibinfo {volume} {45}},\
  \bibinfo {pages} {13488} (\bibinfo {year} {1992})}\BibitemShut {NoStop}%
\bibitem [{\citenamefont {Tang}\ \emph {et~al.}(2019)\citenamefont {Tang},
  \citenamefont {Ren}, \citenamefont {Wang}, \citenamefont {Zhong},
  \citenamefont {Schneeloch}, \citenamefont {Yang}, \citenamefont {Yang},
  \citenamefont {Lee}, \citenamefont {Gu}, \citenamefont {Qiao} \emph
  {et~al.}}]{tang2019three}%
  \BibitemOpen
  \bibfield  {author} {\bibinfo {author} {\bibfnamefont {F.}~\bibnamefont
  {Tang}}, \bibinfo {author} {\bibfnamefont {Y.}~\bibnamefont {Ren}}, \bibinfo
  {author} {\bibfnamefont {P.}~\bibnamefont {Wang}}, \bibinfo {author}
  {\bibfnamefont {R.}~\bibnamefont {Zhong}}, \bibinfo {author} {\bibfnamefont
  {J.}~\bibnamefont {Schneeloch}}, \bibinfo {author} {\bibfnamefont {S.~A.}\
  \bibnamefont {Yang}}, \bibinfo {author} {\bibfnamefont {K.}~\bibnamefont
  {Yang}}, \bibinfo {author} {\bibfnamefont {P.~A.}\ \bibnamefont {Lee}},
  \bibinfo {author} {\bibfnamefont {G.}~\bibnamefont {Gu}}, \bibinfo {author}
  {\bibfnamefont {Z.}~\bibnamefont {Qiao}},  \emph {et~al.},\ }\href@noop {}
  {\bibfield  {journal} {\bibinfo  {journal} {Nature}\ }\textbf {\bibinfo
  {volume} {569}},\ \bibinfo {pages} {537} (\bibinfo {year}
  {2019})}\BibitemShut {NoStop}%
\bibitem [{\citenamefont {Galeski}\ \emph {et~al.}(2021)\citenamefont
  {Galeski}, \citenamefont {Ehmcke}, \citenamefont {Wawrzy{\'{n}}czak},
  \citenamefont {Lozano}, \citenamefont {Cho}, \citenamefont {Sharma},
  \citenamefont {Das}, \citenamefont {K{\"u}ster}, \citenamefont {Sessi},
  \citenamefont {Brando}, \citenamefont {K{\"u}chler}, \citenamefont {Markou},
  \citenamefont {K{\"o}nig}, \citenamefont {Swekis}, \citenamefont {Felser},
  \citenamefont {Sassa}, \citenamefont {Li}, \citenamefont {Gu}, \citenamefont
  {Zimmermann}, \citenamefont {Ivashko}, \citenamefont {Gorbunov},
  \citenamefont {Zherlitsyn}, \citenamefont {F{\"o}rster}, \citenamefont
  {Parkin}, \citenamefont {Wosnitza}, \citenamefont {Meng},\ and\ \citenamefont
  {Gooth}}]{Galeski2021}%
  \BibitemOpen
  \bibfield  {author} {\bibinfo {author} {\bibfnamefont {S.}~\bibnamefont
  {Galeski}}, \bibinfo {author} {\bibfnamefont {T.}~\bibnamefont {Ehmcke}},
  \bibinfo {author} {\bibfnamefont {R.}~\bibnamefont {Wawrzy{\'{n}}czak}},
  \bibinfo {author} {\bibfnamefont {P.~M.}\ \bibnamefont {Lozano}}, \bibinfo
  {author} {\bibfnamefont {K.}~\bibnamefont {Cho}}, \bibinfo {author}
  {\bibfnamefont {A.}~\bibnamefont {Sharma}}, \bibinfo {author} {\bibfnamefont
  {S.}~\bibnamefont {Das}}, \bibinfo {author} {\bibfnamefont {F.}~\bibnamefont
  {K{\"u}ster}}, \bibinfo {author} {\bibfnamefont {P.}~\bibnamefont {Sessi}},
  \bibinfo {author} {\bibfnamefont {M.}~\bibnamefont {Brando}}, \bibinfo
  {author} {\bibfnamefont {R.}~\bibnamefont {K{\"u}chler}}, \bibinfo {author}
  {\bibfnamefont {A.}~\bibnamefont {Markou}}, \bibinfo {author} {\bibfnamefont
  {M.}~\bibnamefont {K{\"o}nig}}, \bibinfo {author} {\bibfnamefont
  {P.}~\bibnamefont {Swekis}}, \bibinfo {author} {\bibfnamefont
  {C.}~\bibnamefont {Felser}}, \bibinfo {author} {\bibfnamefont
  {Y.}~\bibnamefont {Sassa}}, \bibinfo {author} {\bibfnamefont
  {Q.}~\bibnamefont {Li}}, \bibinfo {author} {\bibfnamefont {G.}~\bibnamefont
  {Gu}}, \bibinfo {author} {\bibfnamefont {M.~V.}\ \bibnamefont {Zimmermann}},
  \bibinfo {author} {\bibfnamefont {O.}~\bibnamefont {Ivashko}}, \bibinfo
  {author} {\bibfnamefont {D.~I.}\ \bibnamefont {Gorbunov}}, \bibinfo {author}
  {\bibfnamefont {S.}~\bibnamefont {Zherlitsyn}}, \bibinfo {author}
  {\bibfnamefont {T.}~\bibnamefont {F{\"o}rster}}, \bibinfo {author}
  {\bibfnamefont {S.~S.~P.}\ \bibnamefont {Parkin}}, \bibinfo {author}
  {\bibfnamefont {J.}~\bibnamefont {Wosnitza}}, \bibinfo {author}
  {\bibfnamefont {T.}~\bibnamefont {Meng}}, \ and\ \bibinfo {author}
  {\bibfnamefont {J.}~\bibnamefont {Gooth}},\ }\href {\doibase
  10.1038/s41467-021-23435-y} {\bibfield  {journal} {\bibinfo  {journal}
  {Nature Communications}\ }\textbf {\bibinfo {volume} {12}},\ \bibinfo {pages}
  {3197} (\bibinfo {year} {2021})}\BibitemShut {NoStop}%
\bibitem [{\citenamefont {Wan}\ \emph {et~al.}(2011)\citenamefont {Wan},
  \citenamefont {Turner}, \citenamefont {Vishwanath},\ and\ \citenamefont
  {Savrasov}}]{Wan11}%
  \BibitemOpen
  \bibfield  {author} {\bibinfo {author} {\bibfnamefont {X.}~\bibnamefont
  {Wan}}, \bibinfo {author} {\bibfnamefont {A.~M.}\ \bibnamefont {Turner}},
  \bibinfo {author} {\bibfnamefont {A.}~\bibnamefont {Vishwanath}}, \ and\
  \bibinfo {author} {\bibfnamefont {S.~Y.}\ \bibnamefont {Savrasov}},\ }\href
  {\doibase 10.1103/PhysRevB.83.205101} {\bibfield  {journal} {\bibinfo
  {journal} {Phys. Rev. B}\ }\textbf {\bibinfo {volume} {83}},\ \bibinfo
  {pages} {205101} (\bibinfo {year} {2011})}\BibitemShut {NoStop}%
\bibitem [{\citenamefont {Burkov}\ and\ \citenamefont
  {Balents}(2011)}]{Burkov11}%
  \BibitemOpen
  \bibfield  {author} {\bibinfo {author} {\bibfnamefont {A.~A.}\ \bibnamefont
  {Burkov}}\ and\ \bibinfo {author} {\bibfnamefont {L.}~\bibnamefont
  {Balents}},\ }\href {\doibase 10.1103/PhysRevLett.107.127205} {\bibfield
  {journal} {\bibinfo  {journal} {Phys. Rev. Lett.}\ }\textbf {\bibinfo
  {volume} {107}},\ \bibinfo {pages} {127205} (\bibinfo {year}
  {2011})}\BibitemShut {NoStop}%
\bibitem [{\citenamefont {Aji}(2012)}]{Aji12}%
  \BibitemOpen
  \bibfield  {author} {\bibinfo {author} {\bibfnamefont {V.}~\bibnamefont
  {Aji}},\ }\href {\doibase 10.1103/PhysRevB.85.241101} {\bibfield  {journal}
  {\bibinfo  {journal} {Phys. Rev. B}\ }\textbf {\bibinfo {volume} {85}},\
  \bibinfo {pages} {241101} (\bibinfo {year} {2012})}\BibitemShut {NoStop}%
\bibitem [{\citenamefont {Zyuzin}\ and\ \citenamefont
  {Burkov}(2012)}]{Zyuzin12}%
  \BibitemOpen
  \bibfield  {author} {\bibinfo {author} {\bibfnamefont {A.~A.}\ \bibnamefont
  {Zyuzin}}\ and\ \bibinfo {author} {\bibfnamefont {A.~A.}\ \bibnamefont
  {Burkov}},\ }\href {\doibase 10.1103/PhysRevB.86.115133} {\bibfield
  {journal} {\bibinfo  {journal} {Phys. Rev. B}\ }\textbf {\bibinfo {volume}
  {86}},\ \bibinfo {pages} {115133} (\bibinfo {year} {2012})}\BibitemShut
  {NoStop}%
\bibitem [{\citenamefont {Huang}\ \emph {et~al.}(2015)\citenamefont {Huang},
  \citenamefont {Zhao}, \citenamefont {Long}, \citenamefont {Wang},
  \citenamefont {Chen}, \citenamefont {Yang}, \citenamefont {Liang},
  \citenamefont {Xue}, \citenamefont {Weng}, \citenamefont {Fang},
  \citenamefont {Dai},\ and\ \citenamefont {Chen}}]{Huang15}%
  \BibitemOpen
  \bibfield  {author} {\bibinfo {author} {\bibfnamefont {X.}~\bibnamefont
  {Huang}}, \bibinfo {author} {\bibfnamefont {L.}~\bibnamefont {Zhao}},
  \bibinfo {author} {\bibfnamefont {Y.}~\bibnamefont {Long}}, \bibinfo {author}
  {\bibfnamefont {P.}~\bibnamefont {Wang}}, \bibinfo {author} {\bibfnamefont
  {D.}~\bibnamefont {Chen}}, \bibinfo {author} {\bibfnamefont {Z.}~\bibnamefont
  {Yang}}, \bibinfo {author} {\bibfnamefont {H.}~\bibnamefont {Liang}},
  \bibinfo {author} {\bibfnamefont {M.}~\bibnamefont {Xue}}, \bibinfo {author}
  {\bibfnamefont {H.}~\bibnamefont {Weng}}, \bibinfo {author} {\bibfnamefont
  {Z.}~\bibnamefont {Fang}}, \bibinfo {author} {\bibfnamefont {X.}~\bibnamefont
  {Dai}}, \ and\ \bibinfo {author} {\bibfnamefont {G.}~\bibnamefont {Chen}},\
  }\href {\doibase 10.1103/PhysRevX.5.031023} {\bibfield  {journal} {\bibinfo
  {journal} {Phys. Rev. X}\ }\textbf {\bibinfo {volume} {5}},\ \bibinfo {pages}
  {031023} (\bibinfo {year} {2015})}\BibitemShut {NoStop}%
\bibitem [{\citenamefont {Yang}\ \emph
  {et~al.}(2011{\natexlab{a}})\citenamefont {Yang}, \citenamefont {Lu},\ and\
  \citenamefont {Ran}}]{Yang11}%
  \BibitemOpen
  \bibfield  {author} {\bibinfo {author} {\bibfnamefont {K.-Y.}\ \bibnamefont
  {Yang}}, \bibinfo {author} {\bibfnamefont {Y.-M.}\ \bibnamefont {Lu}}, \ and\
  \bibinfo {author} {\bibfnamefont {Y.}~\bibnamefont {Ran}},\ }\href {\doibase
  10.1103/PhysRevB.84.075129} {\bibfield  {journal} {\bibinfo  {journal} {Phys.
  Rev. B}\ }\textbf {\bibinfo {volume} {84}},\ \bibinfo {pages} {075129}
  (\bibinfo {year} {2011}{\natexlab{a}})}\BibitemShut {NoStop}%
\bibitem [{\citenamefont {Xu}\ \emph {et~al.}(2011)\citenamefont {Xu},
  \citenamefont {Weng}, \citenamefont {Wang}, \citenamefont {Dai},\ and\
  \citenamefont {Fang}}]{Xu11}%
  \BibitemOpen
  \bibfield  {author} {\bibinfo {author} {\bibfnamefont {G.}~\bibnamefont
  {Xu}}, \bibinfo {author} {\bibfnamefont {H.}~\bibnamefont {Weng}}, \bibinfo
  {author} {\bibfnamefont {Z.}~\bibnamefont {Wang}}, \bibinfo {author}
  {\bibfnamefont {X.}~\bibnamefont {Dai}}, \ and\ \bibinfo {author}
  {\bibfnamefont {Z.}~\bibnamefont {Fang}},\ }\href {\doibase
  10.1103/PhysRevLett.107.186806} {\bibfield  {journal} {\bibinfo  {journal}
  {Phys. Rev. Lett.}\ }\textbf {\bibinfo {volume} {107}},\ \bibinfo {pages}
  {186806} (\bibinfo {year} {2011})}\BibitemShut {NoStop}%
\bibitem [{\citenamefont {Parameswaran}\ \emph {et~al.}(2014)\citenamefont
  {Parameswaran}, \citenamefont {Grover}, \citenamefont {Abanin}, \citenamefont
  {Pesin},\ and\ \citenamefont {Vishwanath}}]{Parameswaran14}%
  \BibitemOpen
  \bibfield  {author} {\bibinfo {author} {\bibfnamefont {S.~A.}\ \bibnamefont
  {Parameswaran}}, \bibinfo {author} {\bibfnamefont {T.}~\bibnamefont
  {Grover}}, \bibinfo {author} {\bibfnamefont {D.~A.}\ \bibnamefont {Abanin}},
  \bibinfo {author} {\bibfnamefont {D.~A.}\ \bibnamefont {Pesin}}, \ and\
  \bibinfo {author} {\bibfnamefont {A.}~\bibnamefont {Vishwanath}},\ }\href
  {\doibase 10.1103/PhysRevX.4.031035} {\bibfield  {journal} {\bibinfo
  {journal} {Phys. Rev. X}\ }\textbf {\bibinfo {volume} {4}},\ \bibinfo {pages}
  {031035} (\bibinfo {year} {2014})}\BibitemShut {NoStop}%
\bibitem [{\citenamefont {Zhou}\ \emph {et~al.}(2015)\citenamefont {Zhou},
  \citenamefont {Chang},\ and\ \citenamefont {Xiao}}]{Zhou15}%
  \BibitemOpen
  \bibfield  {author} {\bibinfo {author} {\bibfnamefont {J.}~\bibnamefont
  {Zhou}}, \bibinfo {author} {\bibfnamefont {H.-R.}\ \bibnamefont {Chang}}, \
  and\ \bibinfo {author} {\bibfnamefont {D.}~\bibnamefont {Xiao}},\ }\href
  {\doibase 10.1103/PhysRevB.91.035114} {\bibfield  {journal} {\bibinfo
  {journal} {Phys. Rev. B}\ }\textbf {\bibinfo {volume} {91}},\ \bibinfo
  {pages} {035114} (\bibinfo {year} {2015})}\BibitemShut {NoStop}%
\bibitem [{\citenamefont {Son}\ and\ \citenamefont {Spivak}(2013)}]{Son13}%
  \BibitemOpen
  \bibfield  {author} {\bibinfo {author} {\bibfnamefont {D.~T.}\ \bibnamefont
  {Son}}\ and\ \bibinfo {author} {\bibfnamefont {B.~Z.}\ \bibnamefont
  {Spivak}},\ }\href {\doibase 10.1103/PhysRevB.88.104412} {\bibfield
  {journal} {\bibinfo  {journal} {Phys. Rev. B}\ }\textbf {\bibinfo {volume}
  {88}},\ \bibinfo {pages} {104412} (\bibinfo {year} {2013})}\BibitemShut
  {NoStop}%
\bibitem [{\citenamefont {Xu}\ \emph {et~al.}(2015{\natexlab{a}})\citenamefont
  {Xu}, \citenamefont {Zhang},\ and\ \citenamefont {Zhang}}]{YXu15}%
  \BibitemOpen
  \bibfield  {author} {\bibinfo {author} {\bibfnamefont {Y.}~\bibnamefont
  {Xu}}, \bibinfo {author} {\bibfnamefont {F.}~\bibnamefont {Zhang}}, \ and\
  \bibinfo {author} {\bibfnamefont {C.}~\bibnamefont {Zhang}},\ }\href
  {\doibase 10.1103/PhysRevLett.115.265304} {\bibfield  {journal} {\bibinfo
  {journal} {Phys. Rev. Lett.}\ }\textbf {\bibinfo {volume} {115}},\ \bibinfo
  {pages} {265304} (\bibinfo {year} {2015}{\natexlab{a}})}\BibitemShut
  {NoStop}%
\bibitem [{\citenamefont {Yan}\ and\ \citenamefont
  {Felser}(2017)}]{yan2017topological}%
  \BibitemOpen
  \bibfield  {author} {\bibinfo {author} {\bibfnamefont {B.}~\bibnamefont
  {Yan}}\ and\ \bibinfo {author} {\bibfnamefont {C.}~\bibnamefont {Felser}},\
  }\href@noop {} {\bibfield  {journal} {\bibinfo  {journal} {Annual Review of
  Condensed Matter Physics}\ }\textbf {\bibinfo {volume} {8}},\ \bibinfo
  {pages} {337} (\bibinfo {year} {2017})}\BibitemShut {NoStop}%
\bibitem [{\citenamefont {Soluyanov}\ \emph {et~al.}(2015)\citenamefont
  {Soluyanov}, \citenamefont {Gresch}, \citenamefont {Wang}, \citenamefont
  {Wu}, \citenamefont {Troyer}, \citenamefont {Dai},\ and\ \citenamefont
  {Bernevig}}]{soluyanov2015type}%
  \BibitemOpen
  \bibfield  {author} {\bibinfo {author} {\bibfnamefont {A.~A.}\ \bibnamefont
  {Soluyanov}}, \bibinfo {author} {\bibfnamefont {D.}~\bibnamefont {Gresch}},
  \bibinfo {author} {\bibfnamefont {Z.}~\bibnamefont {Wang}}, \bibinfo {author}
  {\bibfnamefont {Q.}~\bibnamefont {Wu}}, \bibinfo {author} {\bibfnamefont
  {M.}~\bibnamefont {Troyer}}, \bibinfo {author} {\bibfnamefont
  {X.}~\bibnamefont {Dai}}, \ and\ \bibinfo {author} {\bibfnamefont {B.~A.}\
  \bibnamefont {Bernevig}},\ }\href@noop {} {\bibfield  {journal} {\bibinfo
  {journal} {Nature}\ }\textbf {\bibinfo {volume} {527}},\ \bibinfo {pages}
  {495} (\bibinfo {year} {2015})}\BibitemShut {NoStop}%
\bibitem [{\citenamefont {Lv}\ \emph {et~al.}(2015)\citenamefont {Lv},
  \citenamefont {Xu}, \citenamefont {Weng}, \citenamefont {Ma}, \citenamefont
  {Richard}, \citenamefont {Huang}, \citenamefont {Zhao}, \citenamefont {Chen},
  \citenamefont {Matt}, \citenamefont {Bisti} \emph
  {et~al.}}]{lv2015observation}%
  \BibitemOpen
  \bibfield  {author} {\bibinfo {author} {\bibfnamefont {B.}~\bibnamefont
  {Lv}}, \bibinfo {author} {\bibfnamefont {N.}~\bibnamefont {Xu}}, \bibinfo
  {author} {\bibfnamefont {H.}~\bibnamefont {Weng}}, \bibinfo {author}
  {\bibfnamefont {J.}~\bibnamefont {Ma}}, \bibinfo {author} {\bibfnamefont
  {P.}~\bibnamefont {Richard}}, \bibinfo {author} {\bibfnamefont
  {X.}~\bibnamefont {Huang}}, \bibinfo {author} {\bibfnamefont
  {L.}~\bibnamefont {Zhao}}, \bibinfo {author} {\bibfnamefont {G.}~\bibnamefont
  {Chen}}, \bibinfo {author} {\bibfnamefont {C.}~\bibnamefont {Matt}}, \bibinfo
  {author} {\bibfnamefont {F.}~\bibnamefont {Bisti}},  \emph {et~al.},\
  }\href@noop {} {\bibfield  {journal} {\bibinfo  {journal} {Nature Physics}\
  }\textbf {\bibinfo {volume} {11}},\ \bibinfo {pages} {724} (\bibinfo {year}
  {2015})}\BibitemShut {NoStop}%
\bibitem [{\citenamefont {Xu}\ \emph {et~al.}(2015{\natexlab{b}})\citenamefont
  {Xu}, \citenamefont {Belopolski}, \citenamefont {Alidoust}, \citenamefont
  {Neupane}, \citenamefont {Bian}, \citenamefont {Zhang}, \citenamefont
  {Sankar}, \citenamefont {Chang}, \citenamefont {Yuan}, \citenamefont {Lee}
  \emph {et~al.}}]{xu2015discovery}%
  \BibitemOpen
  \bibfield  {author} {\bibinfo {author} {\bibfnamefont {S.-Y.}\ \bibnamefont
  {Xu}}, \bibinfo {author} {\bibfnamefont {I.}~\bibnamefont {Belopolski}},
  \bibinfo {author} {\bibfnamefont {N.}~\bibnamefont {Alidoust}}, \bibinfo
  {author} {\bibfnamefont {M.}~\bibnamefont {Neupane}}, \bibinfo {author}
  {\bibfnamefont {G.}~\bibnamefont {Bian}}, \bibinfo {author} {\bibfnamefont
  {C.}~\bibnamefont {Zhang}}, \bibinfo {author} {\bibfnamefont
  {R.}~\bibnamefont {Sankar}}, \bibinfo {author} {\bibfnamefont
  {G.}~\bibnamefont {Chang}}, \bibinfo {author} {\bibfnamefont
  {Z.}~\bibnamefont {Yuan}}, \bibinfo {author} {\bibfnamefont {C.-C.}\
  \bibnamefont {Lee}},  \emph {et~al.},\ }\href@noop {} {\bibfield  {journal}
  {\bibinfo  {journal} {Science}\ }\textbf {\bibinfo {volume} {349}},\ \bibinfo
  {pages} {613} (\bibinfo {year} {2015}{\natexlab{b}})}\BibitemShut {NoStop}%
\bibitem [{\citenamefont {Fang}\ \emph {et~al.}(2012)\citenamefont {Fang},
  \citenamefont {Gilbert}, \citenamefont {Dai},\ and\ \citenamefont
  {Bernevig}}]{Fang12}%
  \BibitemOpen
  \bibfield  {author} {\bibinfo {author} {\bibfnamefont {C.}~\bibnamefont
  {Fang}}, \bibinfo {author} {\bibfnamefont {M.~J.}\ \bibnamefont {Gilbert}},
  \bibinfo {author} {\bibfnamefont {X.}~\bibnamefont {Dai}}, \ and\ \bibinfo
  {author} {\bibfnamefont {B.~A.}\ \bibnamefont {Bernevig}},\ }\href {\doibase
  10.1103/PhysRevLett.108.266802} {\bibfield  {journal} {\bibinfo  {journal}
  {Phys. Rev. Lett.}\ }\textbf {\bibinfo {volume} {108}},\ \bibinfo {pages}
  {266802} (\bibinfo {year} {2012})}\BibitemShut {NoStop}%
\bibitem [{\citenamefont {Liu}\ and\ \citenamefont {Zunger}(2017)}]{Liu17}%
  \BibitemOpen
  \bibfield  {author} {\bibinfo {author} {\bibfnamefont {Q.}~\bibnamefont
  {Liu}}\ and\ \bibinfo {author} {\bibfnamefont {A.}~\bibnamefont {Zunger}},\
  }\href {\doibase 10.1103/PhysRevX.7.021019} {\bibfield  {journal} {\bibinfo
  {journal} {Phys. Rev. X}\ }\textbf {\bibinfo {volume} {7}},\ \bibinfo {pages}
  {021019} (\bibinfo {year} {2017})}\BibitemShut {NoStop}%
\bibitem [{\citenamefont {McCann}\ and\ \citenamefont
  {Fal'ko}(2006)}]{McCann06}%
  \BibitemOpen
  \bibfield  {author} {\bibinfo {author} {\bibfnamefont {E.}~\bibnamefont
  {McCann}}\ and\ \bibinfo {author} {\bibfnamefont {V.~I.}\ \bibnamefont
  {Fal'ko}},\ }\href {\doibase 10.1103/PhysRevLett.96.086805} {\bibfield
  {journal} {\bibinfo  {journal} {Phys. Rev. Lett.}\ }\textbf {\bibinfo
  {volume} {96}},\ \bibinfo {pages} {086805} (\bibinfo {year}
  {2006})}\BibitemShut {NoStop}%
\bibitem [{\citenamefont {Min}\ and\ \citenamefont {MacDonald}(2008)}]{Min08}%
  \BibitemOpen
  \bibfield  {author} {\bibinfo {author} {\bibfnamefont {H.}~\bibnamefont
  {Min}}\ and\ \bibinfo {author} {\bibfnamefont {A.~H.}\ \bibnamefont
  {MacDonald}},\ }\href {\doibase 10.1103/PhysRevB.77.155416} {\bibfield
  {journal} {\bibinfo  {journal} {Phys. Rev. B}\ }\textbf {\bibinfo {volume}
  {77}},\ \bibinfo {pages} {155416} (\bibinfo {year} {2008})}\BibitemShut
  {NoStop}%
\bibitem [{\citenamefont {Ahn}\ \emph {et~al.}(2017)\citenamefont {Ahn},
  \citenamefont {Mele},\ and\ \citenamefont {Min}}]{Ahn17}%
  \BibitemOpen
  \bibfield  {author} {\bibinfo {author} {\bibfnamefont {S.}~\bibnamefont
  {Ahn}}, \bibinfo {author} {\bibfnamefont {E.~J.}\ \bibnamefont {Mele}}, \
  and\ \bibinfo {author} {\bibfnamefont {H.}~\bibnamefont {Min}},\ }\href
  {\doibase 10.1103/PhysRevB.95.161112} {\bibfield  {journal} {\bibinfo
  {journal} {Phys. Rev. B}\ }\textbf {\bibinfo {volume} {95}},\ \bibinfo
  {pages} {161112} (\bibinfo {year} {2017})}\BibitemShut {NoStop}%
\bibitem [{\citenamefont {Roy}\ \emph {et~al.}(2017)\citenamefont {Roy},
  \citenamefont {Goswami},\ and\ \citenamefont {Juri\ifmmode \check{c}\else
  \v{c}\fi{}i\ifmmode~\acute{c}\else \'{c}\fi{}}}]{Roy17}%
  \BibitemOpen
  \bibfield  {author} {\bibinfo {author} {\bibfnamefont {B.}~\bibnamefont
  {Roy}}, \bibinfo {author} {\bibfnamefont {P.}~\bibnamefont {Goswami}}, \ and\
  \bibinfo {author} {\bibfnamefont {V.}~\bibnamefont {Juri\ifmmode
  \check{c}\else \v{c}\fi{}i\ifmmode~\acute{c}\else \'{c}\fi{}}},\ }\href
  {\doibase 10.1103/PhysRevB.95.201102} {\bibfield  {journal} {\bibinfo
  {journal} {Phys. Rev. B}\ }\textbf {\bibinfo {volume} {95}},\ \bibinfo
  {pages} {201102} (\bibinfo {year} {2017})}\BibitemShut {NoStop}%
\bibitem [{\citenamefont {Lundgren}\ \emph {et~al.}(2014)\citenamefont
  {Lundgren}, \citenamefont {Laurell},\ and\ \citenamefont
  {Fiete}}]{Lundgren14}%
  \BibitemOpen
  \bibfield  {author} {\bibinfo {author} {\bibfnamefont {R.}~\bibnamefont
  {Lundgren}}, \bibinfo {author} {\bibfnamefont {P.}~\bibnamefont {Laurell}}, \
  and\ \bibinfo {author} {\bibfnamefont {G.~A.}\ \bibnamefont {Fiete}},\ }\href
  {\doibase 10.1103/PhysRevB.90.165115} {\bibfield  {journal} {\bibinfo
  {journal} {Phys. Rev. B}\ }\textbf {\bibinfo {volume} {90}},\ \bibinfo
  {pages} {165115} (\bibinfo {year} {2014})}\BibitemShut {NoStop}%
\bibitem [{\citenamefont {Sharma}\ \emph {et~al.}(2016)\citenamefont {Sharma},
  \citenamefont {Goswami},\ and\ \citenamefont {Tewari}}]{Sharma16}%
  \BibitemOpen
  \bibfield  {author} {\bibinfo {author} {\bibfnamefont {G.}~\bibnamefont
  {Sharma}}, \bibinfo {author} {\bibfnamefont {P.}~\bibnamefont {Goswami}}, \
  and\ \bibinfo {author} {\bibfnamefont {S.}~\bibnamefont {Tewari}},\ }\href
  {\doibase 10.1103/PhysRevB.93.035116} {\bibfield  {journal} {\bibinfo
  {journal} {Phys. Rev. B}\ }\textbf {\bibinfo {volume} {93}},\ \bibinfo
  {pages} {035116} (\bibinfo {year} {2016})}\BibitemShut {NoStop}%
\bibitem [{\citenamefont {Spivak}\ and\ \citenamefont
  {Andreev}(2016)}]{Spivak16}%
  \BibitemOpen
  \bibfield  {author} {\bibinfo {author} {\bibfnamefont {B.~Z.}\ \bibnamefont
  {Spivak}}\ and\ \bibinfo {author} {\bibfnamefont {A.~V.}\ \bibnamefont
  {Andreev}},\ }\href {\doibase 10.1103/PhysRevB.93.085107} {\bibfield
  {journal} {\bibinfo  {journal} {Phys. Rev. B}\ }\textbf {\bibinfo {volume}
  {93}},\ \bibinfo {pages} {085107} (\bibinfo {year} {2016})}\BibitemShut
  {NoStop}%
\bibitem [{\citenamefont {Zyuzin}(2017)}]{Zyuzin17}%
  \BibitemOpen
  \bibfield  {author} {\bibinfo {author} {\bibfnamefont {V.~A.}\ \bibnamefont
  {Zyuzin}},\ }\href {\doibase 10.1103/PhysRevB.95.245128} {\bibfield
  {journal} {\bibinfo  {journal} {Phys. Rev. B}\ }\textbf {\bibinfo {volume}
  {95}},\ \bibinfo {pages} {245128} (\bibinfo {year} {2017})}\BibitemShut
  {NoStop}%
\bibitem [{\citenamefont {Nandy}\ \emph {et~al.}(2017)\citenamefont {Nandy},
  \citenamefont {Sharma}, \citenamefont {Taraphder},\ and\ \citenamefont
  {Tewari}}]{Nandy17}%
  \BibitemOpen
  \bibfield  {author} {\bibinfo {author} {\bibfnamefont {S.}~\bibnamefont
  {Nandy}}, \bibinfo {author} {\bibfnamefont {G.}~\bibnamefont {Sharma}},
  \bibinfo {author} {\bibfnamefont {A.}~\bibnamefont {Taraphder}}, \ and\
  \bibinfo {author} {\bibfnamefont {S.}~\bibnamefont {Tewari}},\ }\href
  {\doibase 10.1103/PhysRevLett.119.176804} {\bibfield  {journal} {\bibinfo
  {journal} {Phys. Rev. Lett.}\ }\textbf {\bibinfo {volume} {119}},\ \bibinfo
  {pages} {176804} (\bibinfo {year} {2017})}\BibitemShut {NoStop}%
\bibitem [{\citenamefont {Nandy}\ \emph {et~al.}(2019)\citenamefont {Nandy},
  \citenamefont {Taraphder},\ and\ \citenamefont {Tewari}}]{Nandy19}%
  \BibitemOpen
  \bibfield  {author} {\bibinfo {author} {\bibfnamefont {S.}~\bibnamefont
  {Nandy}}, \bibinfo {author} {\bibfnamefont {A.}~\bibnamefont {Taraphder}}, \
  and\ \bibinfo {author} {\bibfnamefont {S.}~\bibnamefont {Tewari}},\ }\href
  {\doibase 10.1103/PhysRevB.100.115139} {\bibfield  {journal} {\bibinfo
  {journal} {Phys. Rev. B}\ }\textbf {\bibinfo {volume} {100}},\ \bibinfo
  {pages} {115139} (\bibinfo {year} {2019})}\BibitemShut {NoStop}%
\bibitem [{\citenamefont {Chen}\ and\ \citenamefont {Fiete}(2016)}]{Chen16}%
  \BibitemOpen
  \bibfield  {author} {\bibinfo {author} {\bibfnamefont {Q.}~\bibnamefont
  {Chen}}\ and\ \bibinfo {author} {\bibfnamefont {G.~A.}\ \bibnamefont
  {Fiete}},\ }\href {\doibase 10.1103/PhysRevB.93.155125} {\bibfield  {journal}
  {\bibinfo  {journal} {Phys. Rev. B}\ }\textbf {\bibinfo {volume} {93}},\
  \bibinfo {pages} {155125} (\bibinfo {year} {2016})}\BibitemShut {NoStop}%
\bibitem [{\citenamefont {Park}\ \emph {et~al.}(2017)\citenamefont {Park},
  \citenamefont {Woo}, \citenamefont {Mele},\ and\ \citenamefont
  {Min}}]{Park17}%
  \BibitemOpen
  \bibfield  {author} {\bibinfo {author} {\bibfnamefont {S.}~\bibnamefont
  {Park}}, \bibinfo {author} {\bibfnamefont {S.}~\bibnamefont {Woo}}, \bibinfo
  {author} {\bibfnamefont {E.~J.}\ \bibnamefont {Mele}}, \ and\ \bibinfo
  {author} {\bibfnamefont {H.}~\bibnamefont {Min}},\ }\href {\doibase
  10.1103/PhysRevB.95.161113} {\bibfield  {journal} {\bibinfo  {journal} {Phys.
  Rev. B}\ }\textbf {\bibinfo {volume} {95}},\ \bibinfo {pages} {161113}
  (\bibinfo {year} {2017})}\BibitemShut {NoStop}%
\bibitem [{\citenamefont {Gorbar}\ \emph {et~al.}(2017)\citenamefont {Gorbar},
  \citenamefont {Miransky}, \citenamefont {Shovkovy},\ and\ \citenamefont
  {Sukhachov}}]{Gorbar17}%
  \BibitemOpen
  \bibfield  {author} {\bibinfo {author} {\bibfnamefont {E.~V.}\ \bibnamefont
  {Gorbar}}, \bibinfo {author} {\bibfnamefont {V.~A.}\ \bibnamefont
  {Miransky}}, \bibinfo {author} {\bibfnamefont {I.~A.}\ \bibnamefont
  {Shovkovy}}, \ and\ \bibinfo {author} {\bibfnamefont {P.~O.}\ \bibnamefont
  {Sukhachov}},\ }\href {\doibase 10.1103/PhysRevB.96.155138} {\bibfield
  {journal} {\bibinfo  {journal} {Phys. Rev. B}\ }\textbf {\bibinfo {volume}
  {96}},\ \bibinfo {pages} {155138} (\bibinfo {year} {2017})}\BibitemShut
  {NoStop}%
\bibitem [{\citenamefont {Dantas}\ \emph {et~al.}(2018)\citenamefont {Dantas},
  \citenamefont {Pe{\~n}a-Benitez}, \citenamefont {Roy},\ and\ \citenamefont
  {Sur{\'o}wka}}]{dantas2018magnetotransport}%
  \BibitemOpen
  \bibfield  {author} {\bibinfo {author} {\bibfnamefont {R.~M.}\ \bibnamefont
  {Dantas}}, \bibinfo {author} {\bibfnamefont {F.}~\bibnamefont
  {Pe{\~n}a-Benitez}}, \bibinfo {author} {\bibfnamefont {B.}~\bibnamefont
  {Roy}}, \ and\ \bibinfo {author} {\bibfnamefont {P.}~\bibnamefont
  {Sur{\'o}wka}},\ }\href@noop {} {\bibfield  {journal} {\bibinfo  {journal}
  {Journal of High Energy Physics}\ }\textbf {\bibinfo {volume} {2018}},\
  \bibinfo {pages} {69} (\bibinfo {year} {2018})}\BibitemShut {NoStop}%
\bibitem [{\citenamefont {Nag}\ and\ \citenamefont {Nandy}(2020)}]{Nag_2020a}%
  \BibitemOpen
  \bibfield  {author} {\bibinfo {author} {\bibfnamefont {T.}~\bibnamefont
  {Nag}}\ and\ \bibinfo {author} {\bibfnamefont {S.}~\bibnamefont {Nandy}},\
  }\href {\doibase 10.1088/1361-648x/abc310} {\bibfield  {journal} {\bibinfo
  {journal} {Journal of Physics: Condensed Matter}\ }\textbf {\bibinfo {volume}
  {33}},\ \bibinfo {pages} {075504} (\bibinfo {year} {2020})}\BibitemShut
  {NoStop}%
\bibitem [{\citenamefont {Das}\ \emph {et~al.}(2021)\citenamefont {Das},
  \citenamefont {Nag},\ and\ \citenamefont {Nandy}}]{Das21}%
  \BibitemOpen
  \bibfield  {author} {\bibinfo {author} {\bibfnamefont {S.~K.}\ \bibnamefont
  {Das}}, \bibinfo {author} {\bibfnamefont {T.}~\bibnamefont {Nag}}, \ and\
  \bibinfo {author} {\bibfnamefont {S.}~\bibnamefont {Nandy}},\ }\href
  {\doibase 10.1103/PhysRevB.104.115420} {\bibfield  {journal} {\bibinfo
  {journal} {Phys. Rev. B}\ }\textbf {\bibinfo {volume} {104}},\ \bibinfo
  {pages} {115420} (\bibinfo {year} {2021})}\BibitemShut {NoStop}%
\bibitem [{\citenamefont {Nag}\ and\ \citenamefont
  {Kennes}(2022)}]{nag2022distinct}%
  \BibitemOpen
  \bibfield  {author} {\bibinfo {author} {\bibfnamefont {T.}~\bibnamefont
  {Nag}}\ and\ \bibinfo {author} {\bibfnamefont {D.~M.}\ \bibnamefont
  {Kennes}},\ }\href@noop {} {\bibfield  {journal} {\bibinfo  {journal} {arXiv
  preprint arXiv:2201.11417}\ } (\bibinfo {year} {2022})}\BibitemShut {NoStop}%
\bibitem [{\citenamefont {Zyuzin}\ and\ \citenamefont
  {Tiwari}(2016)}]{zyuzin2016intrinsic}%
  \BibitemOpen
  \bibfield  {author} {\bibinfo {author} {\bibfnamefont {A.~A.}\ \bibnamefont
  {Zyuzin}}\ and\ \bibinfo {author} {\bibfnamefont {R.~P.}\ \bibnamefont
  {Tiwari}},\ }\href@noop {} {\bibfield  {journal} {\bibinfo  {journal} {JETP
  letters}\ }\textbf {\bibinfo {volume} {103}},\ \bibinfo {pages} {717}
  (\bibinfo {year} {2016})}\BibitemShut {NoStop}%
\bibitem [{\citenamefont {Ferreiros}\ \emph {et~al.}(2017)\citenamefont
  {Ferreiros}, \citenamefont {Zyuzin},\ and\ \citenamefont
  {Bardarson}}]{Ferreiros17}%
  \BibitemOpen
  \bibfield  {author} {\bibinfo {author} {\bibfnamefont {Y.}~\bibnamefont
  {Ferreiros}}, \bibinfo {author} {\bibfnamefont {A.~A.}\ \bibnamefont
  {Zyuzin}}, \ and\ \bibinfo {author} {\bibfnamefont {J.~H.}\ \bibnamefont
  {Bardarson}},\ }\href {\doibase 10.1103/PhysRevB.96.115202} {\bibfield
  {journal} {\bibinfo  {journal} {Phys. Rev. B}\ }\textbf {\bibinfo {volume}
  {96}},\ \bibinfo {pages} {115202} (\bibinfo {year} {2017})}\BibitemShut
  {NoStop}%
\bibitem [{\citenamefont {Mukherjee}\ and\ \citenamefont
  {Carbotte}(2017)}]{Mukherjee17}%
  \BibitemOpen
  \bibfield  {author} {\bibinfo {author} {\bibfnamefont {S.~P.}\ \bibnamefont
  {Mukherjee}}\ and\ \bibinfo {author} {\bibfnamefont {J.~P.}\ \bibnamefont
  {Carbotte}},\ }\href {\doibase 10.1103/PhysRevB.96.085114} {\bibfield
  {journal} {\bibinfo  {journal} {Phys. Rev. B}\ }\textbf {\bibinfo {volume}
  {96}},\ \bibinfo {pages} {085114} (\bibinfo {year} {2017})}\BibitemShut
  {NoStop}%
\bibitem [{\citenamefont {Tabert}\ and\ \citenamefont
  {Carbotte}(2016)}]{Tabert16}%
  \BibitemOpen
  \bibfield  {author} {\bibinfo {author} {\bibfnamefont {C.~J.}\ \bibnamefont
  {Tabert}}\ and\ \bibinfo {author} {\bibfnamefont {J.~P.}\ \bibnamefont
  {Carbotte}},\ }\href {\doibase 10.1103/PhysRevB.93.085442} {\bibfield
  {journal} {\bibinfo  {journal} {Phys. Rev. B}\ }\textbf {\bibinfo {volume}
  {93}},\ \bibinfo {pages} {085442} (\bibinfo {year} {2016})}\BibitemShut
  {NoStop}%
\bibitem [{\citenamefont {Menon}\ \emph {et~al.}(2018)\citenamefont {Menon},
  \citenamefont {Chowdhury},\ and\ \citenamefont {Basu}}]{Menon18}%
  \BibitemOpen
  \bibfield  {author} {\bibinfo {author} {\bibfnamefont {A.}~\bibnamefont
  {Menon}}, \bibinfo {author} {\bibfnamefont {D.}~\bibnamefont {Chowdhury}}, \
  and\ \bibinfo {author} {\bibfnamefont {B.}~\bibnamefont {Basu}},\ }\href
  {\doibase 10.1103/PhysRevB.98.205109} {\bibfield  {journal} {\bibinfo
  {journal} {Phys. Rev. B}\ }\textbf {\bibinfo {volume} {98}},\ \bibinfo
  {pages} {205109} (\bibinfo {year} {2018})}\BibitemShut {NoStop}%
\bibitem [{\citenamefont {Nag}\ \emph {et~al.}(2020)\citenamefont {Nag},
  \citenamefont {Menon},\ and\ \citenamefont {Basu}}]{Nag20}%
  \BibitemOpen
  \bibfield  {author} {\bibinfo {author} {\bibfnamefont {T.}~\bibnamefont
  {Nag}}, \bibinfo {author} {\bibfnamefont {A.}~\bibnamefont {Menon}}, \ and\
  \bibinfo {author} {\bibfnamefont {B.}~\bibnamefont {Basu}},\ }\href {\doibase
  10.1103/PhysRevB.102.014307} {\bibfield  {journal} {\bibinfo  {journal}
  {Phys. Rev. B}\ }\textbf {\bibinfo {volume} {102}},\ \bibinfo {pages}
  {014307} (\bibinfo {year} {2020})}\BibitemShut {NoStop}%
\bibitem [{\citenamefont {Menon}\ and\ \citenamefont
  {Basu}(2020)}]{Menon_2020}%
  \BibitemOpen
  \bibfield  {author} {\bibinfo {author} {\bibfnamefont {A.}~\bibnamefont
  {Menon}}\ and\ \bibinfo {author} {\bibfnamefont {B.}~\bibnamefont {Basu}},\
  }\href {\doibase 10.1088/1361-648x/abb9b8} {\bibfield  {journal} {\bibinfo
  {journal} {Journal of Physics: Condensed Matter}\ }\textbf {\bibinfo {volume}
  {33}},\ \bibinfo {pages} {045602} (\bibinfo {year} {2020})}\BibitemShut
  {NoStop}%
\bibitem [{\citenamefont {Sadhukhan}\ and\ \citenamefont
  {Nag}(2021{\natexlab{a}})}]{Sadhukhan21}%
  \BibitemOpen
  \bibfield  {author} {\bibinfo {author} {\bibfnamefont {B.}~\bibnamefont
  {Sadhukhan}}\ and\ \bibinfo {author} {\bibfnamefont {T.}~\bibnamefont
  {Nag}},\ }\href {\doibase 10.1103/PhysRevB.103.144308} {\bibfield  {journal}
  {\bibinfo  {journal} {Phys. Rev. B}\ }\textbf {\bibinfo {volume} {103}},\
  \bibinfo {pages} {144308} (\bibinfo {year} {2021}{\natexlab{a}})}\BibitemShut
  {NoStop}%
\bibitem [{\citenamefont {Sadhukhan}\ and\ \citenamefont
  {Nag}(2021{\natexlab{b}})}]{Sadhukhan21b}%
  \BibitemOpen
  \bibfield  {author} {\bibinfo {author} {\bibfnamefont {B.}~\bibnamefont
  {Sadhukhan}}\ and\ \bibinfo {author} {\bibfnamefont {T.}~\bibnamefont
  {Nag}},\ }\href {\doibase 10.1103/PhysRevB.104.245122} {\bibfield  {journal}
  {\bibinfo  {journal} {Phys. Rev. B}\ }\textbf {\bibinfo {volume} {104}},\
  \bibinfo {pages} {245122} (\bibinfo {year} {2021}{\natexlab{b}})}\BibitemShut
  {NoStop}%
\bibitem [{\citenamefont {Zeng}\ \emph {et~al.}(2021)\citenamefont {Zeng},
  \citenamefont {Nandy},\ and\ \citenamefont {Tewari}}]{Zeng21}%
  \BibitemOpen
  \bibfield  {author} {\bibinfo {author} {\bibfnamefont {C.}~\bibnamefont
  {Zeng}}, \bibinfo {author} {\bibfnamefont {S.}~\bibnamefont {Nandy}}, \ and\
  \bibinfo {author} {\bibfnamefont {S.}~\bibnamefont {Tewari}},\ }\href
  {\doibase 10.1103/PhysRevB.103.245119} {\bibfield  {journal} {\bibinfo
  {journal} {Phys. Rev. B}\ }\textbf {\bibinfo {volume} {103}},\ \bibinfo
  {pages} {245119} (\bibinfo {year} {2021})}\BibitemShut {NoStop}%
\bibitem [{\citenamefont {Avery}\ \emph {et~al.}(2012)\citenamefont {Avery},
  \citenamefont {Pufall},\ and\ \citenamefont {Zink}}]{Avery12}%
  \BibitemOpen
  \bibfield  {author} {\bibinfo {author} {\bibfnamefont {A.~D.}\ \bibnamefont
  {Avery}}, \bibinfo {author} {\bibfnamefont {M.~R.}\ \bibnamefont {Pufall}}, \
  and\ \bibinfo {author} {\bibfnamefont {B.~L.}\ \bibnamefont {Zink}},\ }\href
  {\doibase 10.1103/PhysRevLett.109.196602} {\bibfield  {journal} {\bibinfo
  {journal} {Phys. Rev. Lett.}\ }\textbf {\bibinfo {volume} {109}},\ \bibinfo
  {pages} {196602} (\bibinfo {year} {2012})}\BibitemShut {NoStop}%
\bibitem [{\citenamefont {Li}\ \emph {et~al.}(2016{\natexlab{a}})\citenamefont
  {Li}, \citenamefont {Kharzeev}, \citenamefont {Zhang}, \citenamefont {Huang},
  \citenamefont {Pletikosi{\'c}}, \citenamefont {Fedorov}, \citenamefont
  {Zhong}, \citenamefont {Schneeloch}, \citenamefont {Gu},\ and\ \citenamefont
  {Valla}}]{li2016chiral}%
  \BibitemOpen
  \bibfield  {author} {\bibinfo {author} {\bibfnamefont {Q.}~\bibnamefont
  {Li}}, \bibinfo {author} {\bibfnamefont {D.~E.}\ \bibnamefont {Kharzeev}},
  \bibinfo {author} {\bibfnamefont {C.}~\bibnamefont {Zhang}}, \bibinfo
  {author} {\bibfnamefont {Y.}~\bibnamefont {Huang}}, \bibinfo {author}
  {\bibfnamefont {I.}~\bibnamefont {Pletikosi{\'c}}}, \bibinfo {author}
  {\bibfnamefont {A.}~\bibnamefont {Fedorov}}, \bibinfo {author} {\bibfnamefont
  {R.}~\bibnamefont {Zhong}}, \bibinfo {author} {\bibfnamefont
  {J.}~\bibnamefont {Schneeloch}}, \bibinfo {author} {\bibfnamefont
  {G.}~\bibnamefont {Gu}}, \ and\ \bibinfo {author} {\bibfnamefont
  {T.}~\bibnamefont {Valla}},\ }\href@noop {} {\bibfield  {journal} {\bibinfo
  {journal} {Nature Physics}\ }\textbf {\bibinfo {volume} {12}},\ \bibinfo
  {pages} {550} (\bibinfo {year} {2016}{\natexlab{a}})}\BibitemShut {NoStop}%
\bibitem [{\citenamefont {Liang}\ \emph {et~al.}(2017)\citenamefont {Liang},
  \citenamefont {Lin}, \citenamefont {Gibson}, \citenamefont {Gao},
  \citenamefont {Hirschberger}, \citenamefont {Liu}, \citenamefont {Cava},\
  and\ \citenamefont {Ong}}]{Liang17}%
  \BibitemOpen
  \bibfield  {author} {\bibinfo {author} {\bibfnamefont {T.}~\bibnamefont
  {Liang}}, \bibinfo {author} {\bibfnamefont {J.}~\bibnamefont {Lin}}, \bibinfo
  {author} {\bibfnamefont {Q.}~\bibnamefont {Gibson}}, \bibinfo {author}
  {\bibfnamefont {T.}~\bibnamefont {Gao}}, \bibinfo {author} {\bibfnamefont
  {M.}~\bibnamefont {Hirschberger}}, \bibinfo {author} {\bibfnamefont
  {M.}~\bibnamefont {Liu}}, \bibinfo {author} {\bibfnamefont {R.~J.}\
  \bibnamefont {Cava}}, \ and\ \bibinfo {author} {\bibfnamefont {N.~P.}\
  \bibnamefont {Ong}},\ }\href {\doibase 10.1103/PhysRevLett.118.136601}
  {\bibfield  {journal} {\bibinfo  {journal} {Phys. Rev. Lett.}\ }\textbf
  {\bibinfo {volume} {118}},\ \bibinfo {pages} {136601} (\bibinfo {year}
  {2017})}\BibitemShut {NoStop}%
\bibitem [{\citenamefont {Hirschberger}\ \emph {et~al.}(2016)\citenamefont
  {Hirschberger}, \citenamefont {Kushwaha}, \citenamefont {Wang}, \citenamefont
  {Gibson}, \citenamefont {Liang}, \citenamefont {Belvin}, \citenamefont
  {Bernevig}, \citenamefont {Cava},\ and\ \citenamefont
  {Ong}}]{hirschberger2016chiral}%
  \BibitemOpen
  \bibfield  {author} {\bibinfo {author} {\bibfnamefont {M.}~\bibnamefont
  {Hirschberger}}, \bibinfo {author} {\bibfnamefont {S.}~\bibnamefont
  {Kushwaha}}, \bibinfo {author} {\bibfnamefont {Z.}~\bibnamefont {Wang}},
  \bibinfo {author} {\bibfnamefont {Q.}~\bibnamefont {Gibson}}, \bibinfo
  {author} {\bibfnamefont {S.}~\bibnamefont {Liang}}, \bibinfo {author}
  {\bibfnamefont {C.~A.}\ \bibnamefont {Belvin}}, \bibinfo {author}
  {\bibfnamefont {B.~A.}\ \bibnamefont {Bernevig}}, \bibinfo {author}
  {\bibfnamefont {R.~J.}\ \bibnamefont {Cava}}, \ and\ \bibinfo {author}
  {\bibfnamefont {N.~P.}\ \bibnamefont {Ong}},\ }\href@noop {} {\bibfield
  {journal} {\bibinfo  {journal} {Nature materials}\ }\textbf {\bibinfo
  {volume} {15}},\ \bibinfo {pages} {1161} (\bibinfo {year}
  {2016})}\BibitemShut {NoStop}%
\bibitem [{\citenamefont {Watzman}\ \emph {et~al.}(2018)\citenamefont
  {Watzman}, \citenamefont {McCormick}, \citenamefont {Shekhar}, \citenamefont
  {Wu}, \citenamefont {Sun}, \citenamefont {Prakash}, \citenamefont {Felser},
  \citenamefont {Trivedi},\ and\ \citenamefont {Heremans}}]{Watzman18}%
  \BibitemOpen
  \bibfield  {author} {\bibinfo {author} {\bibfnamefont {S.~J.}\ \bibnamefont
  {Watzman}}, \bibinfo {author} {\bibfnamefont {T.~M.}\ \bibnamefont
  {McCormick}}, \bibinfo {author} {\bibfnamefont {C.}~\bibnamefont {Shekhar}},
  \bibinfo {author} {\bibfnamefont {S.-C.}\ \bibnamefont {Wu}}, \bibinfo
  {author} {\bibfnamefont {Y.}~\bibnamefont {Sun}}, \bibinfo {author}
  {\bibfnamefont {A.}~\bibnamefont {Prakash}}, \bibinfo {author} {\bibfnamefont
  {C.}~\bibnamefont {Felser}}, \bibinfo {author} {\bibfnamefont
  {N.}~\bibnamefont {Trivedi}}, \ and\ \bibinfo {author} {\bibfnamefont
  {J.~P.}\ \bibnamefont {Heremans}},\ }\href {\doibase
  10.1103/PhysRevB.97.161404} {\bibfield  {journal} {\bibinfo  {journal} {Phys.
  Rev. B}\ }\textbf {\bibinfo {volume} {97}},\ \bibinfo {pages} {161404}
  (\bibinfo {year} {2018})}\BibitemShut {NoStop}%
\bibitem [{\citenamefont {Lu}\ \emph {et~al.}(2015)\citenamefont {Lu},
  \citenamefont {Zhang},\ and\ \citenamefont {Shen}}]{H-ZLu15}%
  \BibitemOpen
  \bibfield  {author} {\bibinfo {author} {\bibfnamefont {H.-Z.}\ \bibnamefont
  {Lu}}, \bibinfo {author} {\bibfnamefont {S.-B.}\ \bibnamefont {Zhang}}, \
  and\ \bibinfo {author} {\bibfnamefont {S.-Q.}\ \bibnamefont {Shen}},\ }\href
  {\doibase 10.1103/PhysRevB.92.045203} {\bibfield  {journal} {\bibinfo
  {journal} {Phys. Rev. B}\ }\textbf {\bibinfo {volume} {92}},\ \bibinfo
  {pages} {045203} (\bibinfo {year} {2015})}\BibitemShut {NoStop}%
\bibitem [{\citenamefont {Li}\ \emph {et~al.}(2016{\natexlab{b}})\citenamefont
  {Li}, \citenamefont {Roy},\ and\ \citenamefont {Das~Sarma}}]{Li16}%
  \BibitemOpen
  \bibfield  {author} {\bibinfo {author} {\bibfnamefont {X.}~\bibnamefont
  {Li}}, \bibinfo {author} {\bibfnamefont {B.}~\bibnamefont {Roy}}, \ and\
  \bibinfo {author} {\bibfnamefont {S.}~\bibnamefont {Das~Sarma}},\ }\href
  {\doibase 10.1103/PhysRevB.94.195144} {\bibfield  {journal} {\bibinfo
  {journal} {Phys. Rev. B}\ }\textbf {\bibinfo {volume} {94}},\ \bibinfo
  {pages} {195144} (\bibinfo {year} {2016}{\natexlab{b}})}\BibitemShut
  {NoStop}%
\bibitem [{\citenamefont {Wang}\ \emph {et~al.}(2017)\citenamefont {Wang},
  \citenamefont {Sun}, \citenamefont {Lu},\ and\ \citenamefont {Xie}}]{Wang17}%
  \BibitemOpen
  \bibfield  {author} {\bibinfo {author} {\bibfnamefont {C.~M.}\ \bibnamefont
  {Wang}}, \bibinfo {author} {\bibfnamefont {H.-P.}\ \bibnamefont {Sun}},
  \bibinfo {author} {\bibfnamefont {H.-Z.}\ \bibnamefont {Lu}}, \ and\ \bibinfo
  {author} {\bibfnamefont {X.~C.}\ \bibnamefont {Xie}},\ }\href {\doibase
  10.1103/PhysRevLett.119.136806} {\bibfield  {journal} {\bibinfo  {journal}
  {Phys. Rev. Lett.}\ }\textbf {\bibinfo {volume} {119}},\ \bibinfo {pages}
  {136806} (\bibinfo {year} {2017})}\BibitemShut {NoStop}%
\bibitem [{\citenamefont {Li}\ \emph {et~al.}(2020)\citenamefont {Li},
  \citenamefont {Liu}, \citenamefont {Jiang},\ and\ \citenamefont
  {Xie}}]{Li20}%
  \BibitemOpen
  \bibfield  {author} {\bibinfo {author} {\bibfnamefont {H.}~\bibnamefont
  {Li}}, \bibinfo {author} {\bibfnamefont {H.}~\bibnamefont {Liu}}, \bibinfo
  {author} {\bibfnamefont {H.}~\bibnamefont {Jiang}}, \ and\ \bibinfo {author}
  {\bibfnamefont {X.~C.}\ \bibnamefont {Xie}},\ }\href {\doibase
  10.1103/PhysRevLett.125.036602} {\bibfield  {journal} {\bibinfo  {journal}
  {Phys. Rev. Lett.}\ }\textbf {\bibinfo {volume} {125}},\ \bibinfo {pages}
  {036602} (\bibinfo {year} {2020})}\BibitemShut {NoStop}%
\bibitem [{\citenamefont {Udagawa}\ and\ \citenamefont
  {Bergholtz}(2016)}]{Udagawa16}%
  \BibitemOpen
  \bibfield  {author} {\bibinfo {author} {\bibfnamefont {M.}~\bibnamefont
  {Udagawa}}\ and\ \bibinfo {author} {\bibfnamefont {E.~J.}\ \bibnamefont
  {Bergholtz}},\ }\href {\doibase 10.1103/PhysRevLett.117.086401} {\bibfield
  {journal} {\bibinfo  {journal} {Phys. Rev. Lett.}\ }\textbf {\bibinfo
  {volume} {117}},\ \bibinfo {pages} {086401} (\bibinfo {year}
  {2016})}\BibitemShut {NoStop}%
\bibitem [{\citenamefont {Ma}\ \emph {et~al.}(2021)\citenamefont {Ma},
  \citenamefont {Sheng},\ and\ \citenamefont {Sheng}}]{Ma21}%
  \BibitemOpen
  \bibfield  {author} {\bibinfo {author} {\bibfnamefont {R.}~\bibnamefont
  {Ma}}, \bibinfo {author} {\bibfnamefont {D.~N.}\ \bibnamefont {Sheng}}, \
  and\ \bibinfo {author} {\bibfnamefont {L.}~\bibnamefont {Sheng}},\ }\href
  {\doibase 10.1103/PhysRevB.104.075425} {\bibfield  {journal} {\bibinfo
  {journal} {Phys. Rev. B}\ }\textbf {\bibinfo {volume} {104}},\ \bibinfo
  {pages} {075425} (\bibinfo {year} {2021})}\BibitemShut {NoStop}%
\bibitem [{\citenamefont {Chang}\ and\ \citenamefont {Sheng}(2021)}]{Chang21a}%
  \BibitemOpen
  \bibfield  {author} {\bibinfo {author} {\bibfnamefont {M.}~\bibnamefont
  {Chang}}\ and\ \bibinfo {author} {\bibfnamefont {L.}~\bibnamefont {Sheng}},\
  }\href {\doibase 10.1103/PhysRevB.103.245409} {\bibfield  {journal} {\bibinfo
   {journal} {Phys. Rev. B}\ }\textbf {\bibinfo {volume} {103}},\ \bibinfo
  {pages} {245409} (\bibinfo {year} {2021})}\BibitemShut {NoStop}%
\bibitem [{\citenamefont {Chang}\ \emph {et~al.}(2021)\citenamefont {Chang},
  \citenamefont {Geng}, \citenamefont {Sheng},\ and\ \citenamefont
  {Xing}}]{Chang21b}%
  \BibitemOpen
  \bibfield  {author} {\bibinfo {author} {\bibfnamefont {M.}~\bibnamefont
  {Chang}}, \bibinfo {author} {\bibfnamefont {H.}~\bibnamefont {Geng}},
  \bibinfo {author} {\bibfnamefont {L.}~\bibnamefont {Sheng}}, \ and\ \bibinfo
  {author} {\bibfnamefont {D.~Y.}\ \bibnamefont {Xing}},\ }\href {\doibase
  10.1103/PhysRevB.103.245434} {\bibfield  {journal} {\bibinfo  {journal}
  {Phys. Rev. B}\ }\textbf {\bibinfo {volume} {103}},\ \bibinfo {pages}
  {245434} (\bibinfo {year} {2021})}\BibitemShut {NoStop}%
\bibitem [{\citenamefont {Jeon}\ \emph {et~al.}(2014)\citenamefont {Jeon},
  \citenamefont {Zhou}, \citenamefont {Gyenis}, \citenamefont {Feldman},
  \citenamefont {Kimchi}, \citenamefont {Potter}, \citenamefont {Gibson},
  \citenamefont {Cava}, \citenamefont {Vishwanath},\ and\ \citenamefont
  {Yazdani}}]{jeon2014landau}%
  \BibitemOpen
  \bibfield  {author} {\bibinfo {author} {\bibfnamefont {S.}~\bibnamefont
  {Jeon}}, \bibinfo {author} {\bibfnamefont {B.~B.}\ \bibnamefont {Zhou}},
  \bibinfo {author} {\bibfnamefont {A.}~\bibnamefont {Gyenis}}, \bibinfo
  {author} {\bibfnamefont {B.~E.}\ \bibnamefont {Feldman}}, \bibinfo {author}
  {\bibfnamefont {I.}~\bibnamefont {Kimchi}}, \bibinfo {author} {\bibfnamefont
  {A.~C.}\ \bibnamefont {Potter}}, \bibinfo {author} {\bibfnamefont {Q.~D.}\
  \bibnamefont {Gibson}}, \bibinfo {author} {\bibfnamefont {R.~J.}\
  \bibnamefont {Cava}}, \bibinfo {author} {\bibfnamefont {A.}~\bibnamefont
  {Vishwanath}}, \ and\ \bibinfo {author} {\bibfnamefont {A.}~\bibnamefont
  {Yazdani}},\ }\href@noop {} {\bibfield  {journal} {\bibinfo  {journal}
  {Nature materials}\ }\textbf {\bibinfo {volume} {13}},\ \bibinfo {pages}
  {851} (\bibinfo {year} {2014})}\BibitemShut {NoStop}%
\bibitem [{\citenamefont {Novak}\ \emph {et~al.}(2015)\citenamefont {Novak},
  \citenamefont {Sasaki}, \citenamefont {Segawa},\ and\ \citenamefont
  {Ando}}]{Novak15}%
  \BibitemOpen
  \bibfield  {author} {\bibinfo {author} {\bibfnamefont {M.}~\bibnamefont
  {Novak}}, \bibinfo {author} {\bibfnamefont {S.}~\bibnamefont {Sasaki}},
  \bibinfo {author} {\bibfnamefont {K.}~\bibnamefont {Segawa}}, \ and\ \bibinfo
  {author} {\bibfnamefont {Y.}~\bibnamefont {Ando}},\ }\href {\doibase
  10.1103/PhysRevB.91.041203} {\bibfield  {journal} {\bibinfo  {journal} {Phys.
  Rev. B}\ }\textbf {\bibinfo {volume} {91}},\ \bibinfo {pages} {041203}
  (\bibinfo {year} {2015})}\BibitemShut {NoStop}%
\bibitem [{\citenamefont {Uchida}\ \emph {et~al.}(2017)\citenamefont {Uchida},
  \citenamefont {Nakazawa}, \citenamefont {Nishihaya}, \citenamefont {Akiba},
  \citenamefont {Kriener}, \citenamefont {Kozuka}, \citenamefont {Miyake},
  \citenamefont {Taguchi}, \citenamefont {Tokunaga}, \citenamefont {Nagaosa}
  \emph {et~al.}}]{uchida2017quantum}%
  \BibitemOpen
  \bibfield  {author} {\bibinfo {author} {\bibfnamefont {M.}~\bibnamefont
  {Uchida}}, \bibinfo {author} {\bibfnamefont {Y.}~\bibnamefont {Nakazawa}},
  \bibinfo {author} {\bibfnamefont {S.}~\bibnamefont {Nishihaya}}, \bibinfo
  {author} {\bibfnamefont {K.}~\bibnamefont {Akiba}}, \bibinfo {author}
  {\bibfnamefont {M.}~\bibnamefont {Kriener}}, \bibinfo {author} {\bibfnamefont
  {Y.}~\bibnamefont {Kozuka}}, \bibinfo {author} {\bibfnamefont
  {A.}~\bibnamefont {Miyake}}, \bibinfo {author} {\bibfnamefont
  {Y.}~\bibnamefont {Taguchi}}, \bibinfo {author} {\bibfnamefont
  {M.}~\bibnamefont {Tokunaga}}, \bibinfo {author} {\bibfnamefont
  {N.}~\bibnamefont {Nagaosa}},  \emph {et~al.},\ }\href@noop {} {\bibfield
  {journal} {\bibinfo  {journal} {Nature communications}\ }\textbf {\bibinfo
  {volume} {8}},\ \bibinfo {pages} {2274} (\bibinfo {year} {2017})}\BibitemShut
  {NoStop}%
\bibitem [{\citenamefont {Schumann}\ \emph {et~al.}(2018)\citenamefont
  {Schumann}, \citenamefont {Galletti}, \citenamefont {Kealhofer},
  \citenamefont {Kim}, \citenamefont {Goyal},\ and\ \citenamefont
  {Stemmer}}]{Schumann18}%
  \BibitemOpen
  \bibfield  {author} {\bibinfo {author} {\bibfnamefont {T.}~\bibnamefont
  {Schumann}}, \bibinfo {author} {\bibfnamefont {L.}~\bibnamefont {Galletti}},
  \bibinfo {author} {\bibfnamefont {D.~A.}\ \bibnamefont {Kealhofer}}, \bibinfo
  {author} {\bibfnamefont {H.}~\bibnamefont {Kim}}, \bibinfo {author}
  {\bibfnamefont {M.}~\bibnamefont {Goyal}}, \ and\ \bibinfo {author}
  {\bibfnamefont {S.}~\bibnamefont {Stemmer}},\ }\href {\doibase
  10.1103/PhysRevLett.120.016801} {\bibfield  {journal} {\bibinfo  {journal}
  {Phys. Rev. Lett.}\ }\textbf {\bibinfo {volume} {120}},\ \bibinfo {pages}
  {016801} (\bibinfo {year} {2018})}\BibitemShut {NoStop}%
\bibitem [{\citenamefont {Zhang}\ \emph {et~al.}(2019)\citenamefont {Zhang},
  \citenamefont {Zhang}, \citenamefont {Yuan}, \citenamefont {Lu},
  \citenamefont {Zhang}, \citenamefont {Narayan}, \citenamefont {Liu},
  \citenamefont {Zhang}, \citenamefont {Ni}, \citenamefont {Liu} \emph
  {et~al.}}]{zhang2019quantum}%
  \BibitemOpen
  \bibfield  {author} {\bibinfo {author} {\bibfnamefont {C.}~\bibnamefont
  {Zhang}}, \bibinfo {author} {\bibfnamefont {Y.}~\bibnamefont {Zhang}},
  \bibinfo {author} {\bibfnamefont {X.}~\bibnamefont {Yuan}}, \bibinfo {author}
  {\bibfnamefont {S.}~\bibnamefont {Lu}}, \bibinfo {author} {\bibfnamefont
  {J.}~\bibnamefont {Zhang}}, \bibinfo {author} {\bibfnamefont
  {A.}~\bibnamefont {Narayan}}, \bibinfo {author} {\bibfnamefont
  {Y.}~\bibnamefont {Liu}}, \bibinfo {author} {\bibfnamefont {H.}~\bibnamefont
  {Zhang}}, \bibinfo {author} {\bibfnamefont {Z.}~\bibnamefont {Ni}}, \bibinfo
  {author} {\bibfnamefont {R.}~\bibnamefont {Liu}},  \emph {et~al.},\
  }\href@noop {} {\bibfield  {journal} {\bibinfo  {journal} {Nature}\ }\textbf
  {\bibinfo {volume} {565}},\ \bibinfo {pages} {331} (\bibinfo {year}
  {2019})}\BibitemShut {NoStop}%
\bibitem [{\citenamefont {Qin}\ \emph {et~al.}(2020)\citenamefont {Qin},
  \citenamefont {Li}, \citenamefont {Du}, \citenamefont {Wang}, \citenamefont
  {Zhang}, \citenamefont {Yu}, \citenamefont {Lu},\ and\ \citenamefont
  {Xie}}]{Qin20}%
  \BibitemOpen
  \bibfield  {author} {\bibinfo {author} {\bibfnamefont {F.}~\bibnamefont
  {Qin}}, \bibinfo {author} {\bibfnamefont {S.}~\bibnamefont {Li}}, \bibinfo
  {author} {\bibfnamefont {Z.~Z.}\ \bibnamefont {Du}}, \bibinfo {author}
  {\bibfnamefont {C.~M.}\ \bibnamefont {Wang}}, \bibinfo {author}
  {\bibfnamefont {W.}~\bibnamefont {Zhang}}, \bibinfo {author} {\bibfnamefont
  {D.}~\bibnamefont {Yu}}, \bibinfo {author} {\bibfnamefont {H.-Z.}\
  \bibnamefont {Lu}}, \ and\ \bibinfo {author} {\bibfnamefont {X.~C.}\
  \bibnamefont {Xie}},\ }\href {\doibase 10.1103/PhysRevLett.125.206601}
  {\bibfield  {journal} {\bibinfo  {journal} {Phys. Rev. Lett.}\ }\textbf
  {\bibinfo {volume} {125}},\ \bibinfo {pages} {206601} (\bibinfo {year}
  {2020})}\BibitemShut {NoStop}%
\bibitem [{\citenamefont {Chen}\ \emph {et~al.}(2021)\citenamefont {Chen},
  \citenamefont {Wang}, \citenamefont {Liu}, \citenamefont {Lu},\ and\
  \citenamefont {Xie}}]{RuiChen21}%
  \BibitemOpen
  \bibfield  {author} {\bibinfo {author} {\bibfnamefont {R.}~\bibnamefont
  {Chen}}, \bibinfo {author} {\bibfnamefont {C.~M.}\ \bibnamefont {Wang}},
  \bibinfo {author} {\bibfnamefont {T.}~\bibnamefont {Liu}}, \bibinfo {author}
  {\bibfnamefont {H.-Z.}\ \bibnamefont {Lu}}, \ and\ \bibinfo {author}
  {\bibfnamefont {X.~C.}\ \bibnamefont {Xie}},\ }\href {\doibase
  10.1103/PhysRevResearch.3.033227} {\bibfield  {journal} {\bibinfo  {journal}
  {Phys. Rev. Research}\ }\textbf {\bibinfo {volume} {3}},\ \bibinfo {pages}
  {033227} (\bibinfo {year} {2021})}\BibitemShut {NoStop}%
\bibitem [{\citenamefont {Zhao}\ \emph {et~al.}(2021)\citenamefont {Zhao},
  \citenamefont {Lu},\ and\ \citenamefont {Xie}}]{Peng-Lu21}%
  \BibitemOpen
  \bibfield  {author} {\bibinfo {author} {\bibfnamefont {P.-L.}\ \bibnamefont
  {Zhao}}, \bibinfo {author} {\bibfnamefont {H.-Z.}\ \bibnamefont {Lu}}, \ and\
  \bibinfo {author} {\bibfnamefont {X.~C.}\ \bibnamefont {Xie}},\ }\href
  {\doibase 10.1103/PhysRevLett.127.046602} {\bibfield  {journal} {\bibinfo
  {journal} {Phys. Rev. Lett.}\ }\textbf {\bibinfo {volume} {127}},\ \bibinfo
  {pages} {046602} (\bibinfo {year} {2021})}\BibitemShut {NoStop}%
\bibitem [{\citenamefont {McCormick}\ \emph {et~al.}(2017)\citenamefont
  {McCormick}, \citenamefont {Kimchi},\ and\ \citenamefont
  {Trivedi}}]{McCormick17}%
  \BibitemOpen
  \bibfield  {author} {\bibinfo {author} {\bibfnamefont {T.~M.}\ \bibnamefont
  {McCormick}}, \bibinfo {author} {\bibfnamefont {I.}~\bibnamefont {Kimchi}}, \
  and\ \bibinfo {author} {\bibfnamefont {N.}~\bibnamefont {Trivedi}},\ }\href
  {\doibase 10.1103/PhysRevB.95.075133} {\bibfield  {journal} {\bibinfo
  {journal} {Phys. Rev. B}\ }\textbf {\bibinfo {volume} {95}},\ \bibinfo
  {pages} {075133} (\bibinfo {year} {2017})}\BibitemShut {NoStop}%
\bibitem [{\citenamefont {Yang}\ and\ \citenamefont
  {Nagaosa}(2014)}]{yang2014classification}%
  \BibitemOpen
  \bibfield  {author} {\bibinfo {author} {\bibfnamefont {B.-J.}\ \bibnamefont
  {Yang}}\ and\ \bibinfo {author} {\bibfnamefont {N.}~\bibnamefont {Nagaosa}},\
  }\href@noop {} {\bibfield  {journal} {\bibinfo  {journal} {Nature
  communications}\ }\textbf {\bibinfo {volume} {5}},\ \bibinfo {pages} {4898}
  (\bibinfo {year} {2014})}\BibitemShut {NoStop}%
\bibitem [{\citenamefont {Slager}\ \emph {et~al.}(2017)\citenamefont {Slager},
  \citenamefont {Juri\ifmmode \check{c}\else \v{c}\fi{}i\ifmmode~\acute{c}\else
  \'{c}\fi{}},\ and\ \citenamefont {Roy}}]{Slager17}%
  \BibitemOpen
  \bibfield  {author} {\bibinfo {author} {\bibfnamefont {R.-J.}\ \bibnamefont
  {Slager}}, \bibinfo {author} {\bibfnamefont {V.}~\bibnamefont {Juri\ifmmode
  \check{c}\else \v{c}\fi{}i\ifmmode~\acute{c}\else \'{c}\fi{}}}, \ and\
  \bibinfo {author} {\bibfnamefont {B.}~\bibnamefont {Roy}},\ }\href {\doibase
  10.1103/PhysRevB.96.201401} {\bibfield  {journal} {\bibinfo  {journal} {Phys.
  Rev. B}\ }\textbf {\bibinfo {volume} {96}},\ \bibinfo {pages} {201401}
  (\bibinfo {year} {2017})}\BibitemShut {NoStop}%
\bibitem [{\citenamefont {Dantas}\ \emph {et~al.}(2020)\citenamefont {Dantas},
  \citenamefont {Pe\~na Benitez}, \citenamefont {Roy},\ and\ \citenamefont
  {Sur\'owka}}]{Dantas20}%
  \BibitemOpen
  \bibfield  {author} {\bibinfo {author} {\bibfnamefont {R.~M.~A.}\
  \bibnamefont {Dantas}}, \bibinfo {author} {\bibfnamefont {F.}~\bibnamefont
  {Pe\~na Benitez}}, \bibinfo {author} {\bibfnamefont {B.}~\bibnamefont {Roy}},
  \ and\ \bibinfo {author} {\bibfnamefont {P.}~\bibnamefont {Sur\'owka}},\
  }\href {\doibase 10.1103/PhysRevResearch.2.013007} {\bibfield  {journal}
  {\bibinfo  {journal} {Phys. Rev. Research}\ }\textbf {\bibinfo {volume}
  {2}},\ \bibinfo {pages} {013007} (\bibinfo {year} {2020})}\BibitemShut
  {NoStop}%
\bibitem [{\citenamefont {Yin}\ \emph {et~al.}(2017)\citenamefont {Yin},
  \citenamefont {Bai}, \citenamefont {Wang}, \citenamefont {Li}, \citenamefont
  {Zhang},\ and\ \citenamefont {He}}]{yin2017landau}%
  \BibitemOpen
  \bibfield  {author} {\bibinfo {author} {\bibfnamefont {L.-J.}\ \bibnamefont
  {Yin}}, \bibinfo {author} {\bibfnamefont {K.-K.}\ \bibnamefont {Bai}},
  \bibinfo {author} {\bibfnamefont {W.-X.}\ \bibnamefont {Wang}}, \bibinfo
  {author} {\bibfnamefont {S.-Y.}\ \bibnamefont {Li}}, \bibinfo {author}
  {\bibfnamefont {Y.}~\bibnamefont {Zhang}}, \ and\ \bibinfo {author}
  {\bibfnamefont {L.}~\bibnamefont {He}},\ }\href@noop {} {\bibfield  {journal}
  {\bibinfo  {journal} {Frontiers of Physics}\ }\textbf {\bibinfo {volume}
  {12}},\ \bibinfo {pages} {127208} (\bibinfo {year} {2017})}\BibitemShut
  {NoStop}%
\bibitem [{\citenamefont {Islam}\ and\ \citenamefont
  {Jayannavar}(2017)}]{Islam17}%
  \BibitemOpen
  \bibfield  {author} {\bibinfo {author} {\bibfnamefont {S.~F.}\ \bibnamefont
  {Islam}}\ and\ \bibinfo {author} {\bibfnamefont {A.~M.}\ \bibnamefont
  {Jayannavar}},\ }\href {\doibase 10.1103/PhysRevB.96.235405} {\bibfield
  {journal} {\bibinfo  {journal} {Phys. Rev. B}\ }\textbf {\bibinfo {volume}
  {96}},\ \bibinfo {pages} {235405} (\bibinfo {year} {2017})}\BibitemShut
  {NoStop}%
\bibitem [{\citenamefont {Lukose}\ \emph {et~al.}(2007)\citenamefont {Lukose},
  \citenamefont {Shankar},\ and\ \citenamefont {Baskaran}}]{Lukose07}%
  \BibitemOpen
  \bibfield  {author} {\bibinfo {author} {\bibfnamefont {V.}~\bibnamefont
  {Lukose}}, \bibinfo {author} {\bibfnamefont {R.}~\bibnamefont {Shankar}}, \
  and\ \bibinfo {author} {\bibfnamefont {G.}~\bibnamefont {Baskaran}},\ }\href
  {\doibase 10.1103/PhysRevLett.98.116802} {\bibfield  {journal} {\bibinfo
  {journal} {Phys. Rev. Lett.}\ }\textbf {\bibinfo {volume} {98}},\ \bibinfo
  {pages} {116802} (\bibinfo {year} {2007})}\BibitemShut {NoStop}%
\bibitem [{\citenamefont {Peres}\ and\ \citenamefont
  {Castro}(2007)}]{peres2007algebraic}%
  \BibitemOpen
  \bibfield  {author} {\bibinfo {author} {\bibfnamefont {N.}~\bibnamefont
  {Peres}}\ and\ \bibinfo {author} {\bibfnamefont {E.~V.}\ \bibnamefont
  {Castro}},\ }\href@noop {} {\bibfield  {journal} {\bibinfo  {journal}
  {Journal of Physics: Condensed Matter}\ }\textbf {\bibinfo {volume} {19}},\
  \bibinfo {pages} {406231} (\bibinfo {year} {2007})}\BibitemShut {NoStop}%
\bibitem [{\citenamefont {Abdulla}\ \emph {et~al.}(2021)\citenamefont
  {Abdulla}, \citenamefont {Das}, \citenamefont {Rao},\ and\ \citenamefont
  {Murthy}}]{abdulla2021time}%
  \BibitemOpen
  \bibfield  {author} {\bibinfo {author} {\bibfnamefont {F.}~\bibnamefont
  {Abdulla}}, \bibinfo {author} {\bibfnamefont {A.}~\bibnamefont {Das}},
  \bibinfo {author} {\bibfnamefont {S.}~\bibnamefont {Rao}}, \ and\ \bibinfo
  {author} {\bibfnamefont {G.}~\bibnamefont {Murthy}},\ }\href@noop {}
  {\bibfield  {journal} {\bibinfo  {journal} {arXiv preprint arXiv:2108.03196}\
  } (\bibinfo {year} {2021})}\BibitemShut {NoStop}%
\bibitem [{\citenamefont {Peeters}\ and\ \citenamefont
  {Vasilopoulos}(1992)}]{Peeters92}%
  \BibitemOpen
  \bibfield  {author} {\bibinfo {author} {\bibfnamefont {F.~M.}\ \bibnamefont
  {Peeters}}\ and\ \bibinfo {author} {\bibfnamefont {P.}~\bibnamefont
  {Vasilopoulos}},\ }\href {\doibase 10.1103/PhysRevB.46.4667} {\bibfield
  {journal} {\bibinfo  {journal} {Phys. Rev. B}\ }\textbf {\bibinfo {volume}
  {46}},\ \bibinfo {pages} {4667} (\bibinfo {year} {1992})}\BibitemShut
  {NoStop}%
\bibitem [{\citenamefont {Charbonneau}\ \emph {et~al.}(1982)\citenamefont
  {Charbonneau}, \citenamefont {van Vliet},\ and\ \citenamefont
  {Vasilopoulos}}]{Charbonneau82}%
  \BibitemOpen
  \bibfield  {author} {\bibinfo {author} {\bibfnamefont {M.}~\bibnamefont
  {Charbonneau}}, \bibinfo {author} {\bibfnamefont {K.~M.}\ \bibnamefont {van
  Vliet}}, \ and\ \bibinfo {author} {\bibfnamefont {P.}~\bibnamefont
  {Vasilopoulos}},\ }\href {\doibase 10.1063/1.525355} {\bibfield  {journal}
  {\bibinfo  {journal} {Journal of Mathematical Physics}\ }\textbf {\bibinfo
  {volume} {23}},\ \bibinfo {pages} {318} (\bibinfo {year} {1982})},\ \Eprint
  {http://arxiv.org/abs/https://doi.org/10.1063/1.525355}
  {https://doi.org/10.1063/1.525355} \BibitemShut {NoStop}%
\bibitem [{\citenamefont {Krstaji\ifmmode~\acute{c}\else \'{c}\fi{}}\ and\
  \citenamefont {Vasilopoulos}(2012)}]{Vasilopoulos12}%
  \BibitemOpen
  \bibfield  {author} {\bibinfo {author} {\bibfnamefont {P.~M.}\ \bibnamefont
  {Krstaji\ifmmode~\acute{c}\else \'{c}\fi{}}}\ and\ \bibinfo {author}
  {\bibfnamefont {P.}~\bibnamefont {Vasilopoulos}},\ }\href {\doibase
  10.1103/PhysRevB.86.115432} {\bibfield  {journal} {\bibinfo  {journal} {Phys.
  Rev. B}\ }\textbf {\bibinfo {volume} {86}},\ \bibinfo {pages} {115432}
  (\bibinfo {year} {2012})}\BibitemShut {NoStop}%
\bibitem [{\citenamefont {Islam}(2018)}]{Islam_2018}%
  \BibitemOpen
  \bibfield  {author} {\bibinfo {author} {\bibfnamefont {S.~F.}\ \bibnamefont
  {Islam}},\ }\href {\doibase 10.1088/1361-648x/aac8b3} {\bibfield  {journal}
  {\bibinfo  {journal} {Journal of Physics: Condensed Matter}\ }\textbf
  {\bibinfo {volume} {30}},\ \bibinfo {pages} {275301} (\bibinfo {year}
  {2018})}\BibitemShut {NoStop}%
\bibitem [{\citenamefont {Steiner}\ \emph {et~al.}(2017)\citenamefont
  {Steiner}, \citenamefont {Andreev},\ and\ \citenamefont {Pesin}}]{Steiner17}%
  \BibitemOpen
  \bibfield  {author} {\bibinfo {author} {\bibfnamefont {J.~F.}\ \bibnamefont
  {Steiner}}, \bibinfo {author} {\bibfnamefont {A.~V.}\ \bibnamefont
  {Andreev}}, \ and\ \bibinfo {author} {\bibfnamefont {D.~A.}\ \bibnamefont
  {Pesin}},\ }\href {\doibase 10.1103/PhysRevLett.119.036601} {\bibfield
  {journal} {\bibinfo  {journal} {Phys. Rev. Lett.}\ }\textbf {\bibinfo
  {volume} {119}},\ \bibinfo {pages} {036601} (\bibinfo {year}
  {2017})}\BibitemShut {NoStop}%
\bibitem [{\citenamefont {Burkov}(2014)}]{Burkov14_PRL}%
  \BibitemOpen
  \bibfield  {author} {\bibinfo {author} {\bibfnamefont {A.~A.}\ \bibnamefont
  {Burkov}},\ }\href {\doibase 10.1103/PhysRevLett.113.187202} {\bibfield
  {journal} {\bibinfo  {journal} {Phys. Rev. Lett.}\ }\textbf {\bibinfo
  {volume} {113}},\ \bibinfo {pages} {187202} (\bibinfo {year}
  {2014})}\BibitemShut {NoStop}%
\bibitem [{\citenamefont {Yang}\ \emph
  {et~al.}(2011{\natexlab{b}})\citenamefont {Yang}, \citenamefont {Lu},\ and\
  \citenamefont {Ran}}]{Kai-Yu11}%
  \BibitemOpen
  \bibfield  {author} {\bibinfo {author} {\bibfnamefont {K.-Y.}\ \bibnamefont
  {Yang}}, \bibinfo {author} {\bibfnamefont {Y.-M.}\ \bibnamefont {Lu}}, \ and\
  \bibinfo {author} {\bibfnamefont {Y.}~\bibnamefont {Ran}},\ }\href {\doibase
  10.1103/PhysRevB.84.075129} {\bibfield  {journal} {\bibinfo  {journal} {Phys.
  Rev. B}\ }\textbf {\bibinfo {volume} {84}},\ \bibinfo {pages} {075129}
  (\bibinfo {year} {2011}{\natexlab{b}})}\BibitemShut {NoStop}%
\bibitem [{\citenamefont {Potter}\ \emph {et~al.}(2014)\citenamefont {Potter},
  \citenamefont {Kimchi},\ and\ \citenamefont
  {Vishwanath}}]{potter2014quantum}%
  \BibitemOpen
  \bibfield  {author} {\bibinfo {author} {\bibfnamefont {A.~C.}\ \bibnamefont
  {Potter}}, \bibinfo {author} {\bibfnamefont {I.}~\bibnamefont {Kimchi}}, \
  and\ \bibinfo {author} {\bibfnamefont {A.}~\bibnamefont {Vishwanath}},\
  }\href@noop {} {\bibfield  {journal} {\bibinfo  {journal} {Nature
  communications}\ }\textbf {\bibinfo {volume} {5}},\ \bibinfo {pages} {5161}
  (\bibinfo {year} {2014})}\BibitemShut {NoStop}%
\bibitem [{\citenamefont {Laughlin}(1981)}]{Laughlin81}%
  \BibitemOpen
  \bibfield  {author} {\bibinfo {author} {\bibfnamefont {R.~B.}\ \bibnamefont
  {Laughlin}},\ }\href {\doibase 10.1103/PhysRevB.23.5632} {\bibfield
  {journal} {\bibinfo  {journal} {Phys. Rev. B}\ }\textbf {\bibinfo {volume}
  {23}},\ \bibinfo {pages} {5632} (\bibinfo {year} {1981})}\BibitemShut
  {NoStop}%
\bibitem [{\citenamefont {Liang}\ \emph {et~al.}(2018)\citenamefont {Liang},
  \citenamefont {Lin}, \citenamefont {Gibson}, \citenamefont {Kushwaha},
  \citenamefont {Liu}, \citenamefont {Wang}, \citenamefont {Xiong},
  \citenamefont {Sobota}, \citenamefont {Hashimoto}, \citenamefont {Kirchmann}
  \emph {et~al.}}]{liang2018anomalous}%
  \BibitemOpen
  \bibfield  {author} {\bibinfo {author} {\bibfnamefont {T.}~\bibnamefont
  {Liang}}, \bibinfo {author} {\bibfnamefont {J.}~\bibnamefont {Lin}}, \bibinfo
  {author} {\bibfnamefont {Q.}~\bibnamefont {Gibson}}, \bibinfo {author}
  {\bibfnamefont {S.}~\bibnamefont {Kushwaha}}, \bibinfo {author}
  {\bibfnamefont {M.}~\bibnamefont {Liu}}, \bibinfo {author} {\bibfnamefont
  {W.}~\bibnamefont {Wang}}, \bibinfo {author} {\bibfnamefont {H.}~\bibnamefont
  {Xiong}}, \bibinfo {author} {\bibfnamefont {J.~A.}\ \bibnamefont {Sobota}},
  \bibinfo {author} {\bibfnamefont {M.}~\bibnamefont {Hashimoto}}, \bibinfo
  {author} {\bibfnamefont {P.~S.}\ \bibnamefont {Kirchmann}},  \emph {et~al.},\
  }\href@noop {} {\bibfield  {journal} {\bibinfo  {journal} {Nature Physics}\
  }\textbf {\bibinfo {volume} {14}},\ \bibinfo {pages} {451} (\bibinfo {year}
  {2018})}\BibitemShut {NoStop}%
\bibitem [{\citenamefont {Galeski}\ \emph {et~al.}(2020)\citenamefont
  {Galeski}, \citenamefont {Zhao}, \citenamefont {Wawrzy{\'n}czak},
  \citenamefont {Meng}, \citenamefont {F{\"o}rster}, \citenamefont {Lozano},
  \citenamefont {Honnali}, \citenamefont {Lamba}, \citenamefont {Ehmcke},
  \citenamefont {Markou} \emph {et~al.}}]{galeski2020unconventional}%
  \BibitemOpen
  \bibfield  {author} {\bibinfo {author} {\bibfnamefont {S.}~\bibnamefont
  {Galeski}}, \bibinfo {author} {\bibfnamefont {X.}~\bibnamefont {Zhao}},
  \bibinfo {author} {\bibfnamefont {R.}~\bibnamefont {Wawrzy{\'n}czak}},
  \bibinfo {author} {\bibfnamefont {T.}~\bibnamefont {Meng}}, \bibinfo {author}
  {\bibfnamefont {T.}~\bibnamefont {F{\"o}rster}}, \bibinfo {author}
  {\bibfnamefont {P.}~\bibnamefont {Lozano}}, \bibinfo {author} {\bibfnamefont
  {S.}~\bibnamefont {Honnali}}, \bibinfo {author} {\bibfnamefont
  {N.}~\bibnamefont {Lamba}}, \bibinfo {author} {\bibfnamefont
  {T.}~\bibnamefont {Ehmcke}}, \bibinfo {author} {\bibfnamefont
  {A.}~\bibnamefont {Markou}},  \emph {et~al.},\ }\href@noop {} {\bibfield
  {journal} {\bibinfo  {journal} {Nature communications}\ }\textbf {\bibinfo
  {volume} {11}},\ \bibinfo {pages} {5926} (\bibinfo {year}
  {2020})}\BibitemShut {NoStop}%
\bibitem [{\citenamefont {St\"ormer}\ \emph {et~al.}(1986)\citenamefont
  {St\"ormer}, \citenamefont {Eisenstein}, \citenamefont {Gossard},
  \citenamefont {Wiegmann},\ and\ \citenamefont {Baldwin}}]{Eisenstein86}%
  \BibitemOpen
  \bibfield  {author} {\bibinfo {author} {\bibfnamefont {H.~L.}\ \bibnamefont
  {St\"ormer}}, \bibinfo {author} {\bibfnamefont {J.~P.}\ \bibnamefont
  {Eisenstein}}, \bibinfo {author} {\bibfnamefont {A.~C.}\ \bibnamefont
  {Gossard}}, \bibinfo {author} {\bibfnamefont {W.}~\bibnamefont {Wiegmann}}, \
  and\ \bibinfo {author} {\bibfnamefont {K.}~\bibnamefont {Baldwin}},\ }\href
  {\doibase 10.1103/PhysRevLett.56.85} {\bibfield  {journal} {\bibinfo
  {journal} {Phys. Rev. Lett.}\ }\textbf {\bibinfo {volume} {56}},\ \bibinfo
  {pages} {85} (\bibinfo {year} {1986})}\BibitemShut {NoStop}%
\end{thebibliography}%
\end{document}